\documentclass{aa}  

\usepackage{graphicx}
\usepackage{txfonts}
\DeclareMathOperator{\arctanh}{arctanh}
\usepackage{siunitx}
\usepackage{natbib,twoopt}
\usepackage[breaklinks=true]{hyperref} 
\hypersetup{breaklinks=true} 
\bibpunct{(}{)}{;}{a}{}{,}             
\makeatletter
  \newcommandtwoopt{\citeads}[3][][]{\href{http://adsabs.harvard.edu/abs/#3}%
    {\def\hyper@linkstart##1##2{}%
     \let\hyper@linkend\@empty\citealp[#1][#2]{#3}}}
  \newcommandtwoopt{\citepads}[3][][]{\href{http://adsabs.harvard.edu/abs/#3}%
    {\def\hyper@linkstart##1##2{}%
     \let\hyper@linkend\@empty\citep[#1][#2]{#3}}}
  \newcommandtwoopt{\citetads}[3][][]{\href{http://adsabs.harvard.edu/abs/#3}%
    {\def\hyper@linkstart##1##2{}%
     \let\hyper@linkend\@empty\citet[#1][#2]{#3}}}
  \newcommandtwoopt{\citeyearads}[3][][]%
    {\href{http://adsabs.harvard.edu/abs/#3}
    {\def\hyper@linkstart##1##2{}%
     \let\hyper@linkend\@empty\citeyear[#1][#2]{#3}}}
\makeatother
\usepackage{tabularx}
\usepackage{tablefootnote}
\usepackage{array}
\usepackage{subcaption} 
\usepackage{caption} 
\usepackage{float} 
\usepackage[hang,flushmargin]{footmisc} 

\usepackage{soul}

\begin{document}

   \title{The SRG/eROSITA All-Sky Survey}
   \subtitle{$J$-Band Follow-Up Observations for Selected High-Redshift Galaxy Cluster Candidates}


   \author{N. Zimmermann \inst{1}
            \and
            M. Kluge \inst{2}            
            \and
            S. Grandis \inst{1}
            \and
            T. Schrabback \inst{1}
            \and
            F. Balzer \inst{2}
            \and
            E. Bulbul \inst{2}
            \and
            J. Comparat \inst{2}
            \and
            B. Csizi \inst{1}
            \and
            V. Ghirardini \inst{3}
            \and
            H. Jansen \inst{1}
            \and
            F. Kleinebreil \inst{1}
            \and
            A. Liu \inst{2}
            \and
            A. Merloni \inst{2}
            \and
            M. E. Ramos-Ceja \inst{2}
            \and
            J. Sanders \inst{2}
            \and
            X. Zhang \inst{2}
            \and
            P. Aschenbrenner \inst{1}
            \and
            F. Enescu \inst{1}
            \and
            S. Keiler \inst{1}
            \and
            M. Märk \inst{1}
            \and
            M. Rinner \inst{1}
            \and
            P. Schweitzer \inst{1}
            \and
            E. Silvestre-Rosello \inst{1}
            \and
            L. Stepman \inst{1}
          }

   \institute{$^1$University of Innsbruck, Institute for Astro- and Particle Physics, Technikerstraße 25, 6020 Innsbruck, Austria\\
              $^2$Max Planck Institute for Extraterrestrial Physics, Giessenbachstrasse 1, 85748 Garching, Germany\\
              $^3$INAF, Osservatorio di Astrofisica e Scienza dello Spazio, via Piero Gobetti 93/3, I-40129 Bologna, Italy\\
            }

   \date{Received ; accepted }

 
  \abstract{We select galaxy cluster candidates from the high-redshift ($\mathrm{BEST\_Z}>0.9$) end of the first SRG (Spectrum Roentgen Gamma)/eROSITA All-Sky Survey (eRASS1) galaxy cluster catalogue, for which we obtain moderately deep $J$-band imaging data with the OMEGA2000 camera at the $\SI{3.5}{m}$ telescope of the Calar Alto Observatory. We include $J$-band data of four additional targets obtained with the three-channel camera at the $\SI{2}{m}$ Fraunhofer telescope at the Wendelstein Observatory. We complement the new $J$-band photometric catalogue with forced photometry in the $i$- and $z$-bands of the tenth data release of the Legacy Survey (LS DR10) to derive the radial colour distribution around the eROSITA/eRASS1 clusters. Without assuming a priori to find a cluster red sequence at a specific colour, we try to find a radially weighted colour over-density to confirm the presence of high-redshift optical counterparts for the X-ray emission. We compare our confirmation with optical properties derived in earlier works based on LS DR10 data to refine the existing high-redshift cluster confirmation of eROSITA-selected clusters. We attempt to calibrate the colour-redshift-relation including the new $J$-band data by comparing our obtained photometric redshift estimate with the spectroscopic redshift of a confirmed, optically selected, high-redshift galaxy cluster. We confirm 9 out of 18 of the selected galaxy cluster candidates with a radial over-density of similar coloured galaxies for which we provide a photometric redshift estimate. We can report an increase in the relative colour measurement precision from $8\%$ to $4\%$ when including $J$-band data. In conclusion, our findings indicate a not insignificant spurious contaminant fraction at the high-redshift end ($\mathrm{BEST}\_\mathrm{Z}>0.9$) of the eROSITA/eRASS1 galaxy cluster catalogue, as well as it underlines the necessity for wide and deep near infrared imaging data for confirmation and characterisation of high-$z$ galaxy clusters.}




   \titlerunning{$J$-Band Observations for Selected High-Redshift eROSITA/eRASS1 Galaxy Cluster Candidates}
   \authorrunning{}
   \maketitle
%
\section{Introduction}
The distribution of galaxy clusters in mass and redshift is a probe of structure growth in the Universe \citep{Allen2011}, as well as its expansion history \citep{Weinberg2013}.
Simulations show that the population of galaxy clusters changes drastically at higher redshifts when using different cosmological models.
Thus, extending cluster samples to higher redshifts (\mbox{$z\gtrsim 1$}) yields powerful cosmological constraining power.
For a review, see e.g. \citet{Borgani2011}.\\
Galaxy clusters are observable in different wavelength regimes via different physical effects.
As the name suggests, galaxy clusters have first been observed as projected galaxy over-densities in the optical \citep{Zwicky1968}, which is still used to detect and characterise them in catalogues \citep{Rykoff2016,Maturi2019}.
Additionally, they are traceable in the millimeter and X-ray regime, as hot plasma accumulates at the core of the cluster (intracluster medium, ICM) and 1) leads to a characteristic spectral distortion of cosmic microwave background (CMB) photons and 2) emits X-ray photons due to thermal Bremsstrahlung.
The former is known as the thermal Sunyaev-Zeldovich (tSZ, \citealt{SunyaevZeldovich1970}) effect and can be observed in CMB maps due to the shift of the CMB spectrum in the direction of galaxy clusters, providing the means to create tSZ-selected galaxy cluster catalogues (e.g. \citealt{Hilton2021, Bleem2024}).
X-ray surveys delivered a variety of catalogues of galaxy clusters with an ever increasing redshift limit.
While the northern and southern ROSAT samples \citep{Boehringer2000,Ebeling2001,Boehringer2004} are limited to relatively nearby clusters ($z<0.7$), deeper X-ray-based galaxy cluster samples covering smaller sky areas, such as the XMM-Newton XXL sample \citep{Pacauad2016}, already reach out to redshift unity.
However, the new cluster catalogue published by the eROSITA (extended ROentgen Survey with an Imaging Telescope Array, \citealt{Predehl2021}, on board the Spectrum Roentgen Gama Mission, \citealt{Sunyaev2021}) collaboration, which is based on the first pass of eROSITA of the full sky (eRASS1), detected galaxy clusters up until a redshift of $z\sim1.3$ and covers the whole western hemisphere \citep{Bulbul2024}.\\
Two key characteristics of galaxy clusters that carry cosmological information are their mass and their redshift.
The former is not directly observable, thus different mass proxies have been introduced, studied, and discussed extensively elsewhere (e.g. \citealt{Pratt2019}).
The latter is in principle directly observable by measuring the spectrum of cluster member galaxies.
However, taking spectra of galaxies is a time-intensive endeavour, and additionally, the sample of luminous red galaxies in large-area spectroscopic surveys such as DESI \citep{DESI2022} is limited to redshifts \mbox{$z\lesssim 1$} \citep{Zhou2023}.\\
We therefore resort to photometry as the observational method of choice for cluster redshift estimation.
\citet{Bower1992} have observed a tight correlation between colour and magnitude in early-type elliptical galaxies in the Coma and Virgo cluster.
Subsequent work has shown that this relation is due to the evolution of these galaxies being monotonic with redshift and consistent with the passive ageing of stellar populations formed before $z \sim 2$ \citep{AragonEllisCouchCarter1993,BenderZieglerBruzual1996,KodamaArimoto1997}.
This implies that the colour-magnitude relation of cluster ellipticals (cluster red-sequence) is a probe for galaxy cluster detection and/or confirmation, as well as being sensitive to the cluster redshift and can thus be used to obtain a photometric redshift ($z_{\mathrm{ph}}$) estimate.
Photometric redshift estimates have been shown to be reasonably reliable for lower redshifts ($z_{\mathrm{ph}}<1$, e.g. $\delta z = 0.005/(1+z)$ in \citealt{Kluge2024}, see also \citealt{Klein2024}).
When extended to higher redshifts, the optical \emph{griz}-bands no longer encompass the $\SI{4000}{\angstrom}$ break in the observed (redshifted) spectral energy distribution (SED) of passive early-type galaxies, thus the photometric redshift estimation becomes uncertain or not possible at all.
In addition, as the redshift increases, passive red galaxies become fainter in spectral bands below $\SI{1}{micron}$, making photometric measurements also more noisy.\\
The goal of this work is to confirm and characterise some of the most distant X-ray-luminous clusters present in eRASS1.
Toward this goal, we obtain good-resolution imaging data red-wards of the $\SI{4000}{\angstrom}$ break in the form of moderately deep \emph{J}-band observations of selected cluster candidates aiming to detect and probe the cluster red-sequence.
We do not a priori assume to find such a red-sequence of a certain colour, but fit for radial colour over-densites in the vicinity of a supposed cluster centre for just one colour index at once.
We motivate this by the fact that even though fully developed cluster red-sequences at high-redshifts ($z>1$) have been found \citep{Menci2008,Cerulo2016,Strazzullo2019}, but whether or not they undergo any sort of evolution when compared to low redshift clusters is still not fully understood.
Furthermore, the lack of both extensive spectroscopic observations, as well as near-infrared (NIR) imaging data of high-redshift cluster red-sequences for eROSITA clusters so far prevented a robust calibration of any colour-to-redshift relation, as has been found and used for lower redshift objects.
It can be expected that the plethora of NIR data taken as part of the ongoing {\it Euclid} surveys \citep{Mellier2025} will push the optical high-$z$ galaxy clusters detection, confirmation and characterisation to new limits.
Thus, targeted follow-up observations as presented in this work have to fill the gap until {\it Euclid} data is available for large fractions of the extragalactic sky.\\
This paper is structured as follows: Section \ref{chap:Data} provides a description of the analysed data and image processing steps, as well as a short overview of the telescopes and instruments used for acquiring the data.
Section \ref{chap:Methods} introduces the method we applied to extract cluster red-sequences and how we obtain photometric redshift estimates.
Section \ref{chap:Results} presents the results of this work, which are consequently discussed in Sect. \ref{chap:Discussion}, and wrapped up in a conclusion in Sect. \ref{chap:Conclusion}.


\section{Data}\label{chap:Data}
In this work we make use of \emph{J}-band imaging data obtained with the CAHA3.5/Omega2000 NIR imager at the Calar Alto Astronomical Observatory (CAHA), as well as with the WST2.1/3KK at the Observatorium Wendelstein (WST) in dedicated observations of a eRASS1 galaxy cluster sub-sample.
We complement this analysis with \emph{i}- and \emph{z}-band imaging data of the tenth data release of the Legacy Survey (LSDR10, \citealt{Dey2019}).
While we use the \emph{J}-band data for both detection and photometry, the LSDR10 data is solely used for photometric purposes.

    \subsection{Parent sample: eROSITA/eRASS1}\label{SubSec:eRASS1}
    As already mentioned above, the cluster candidates investigated in this work originate from the eRASS1 galaxy cluster catalogue, published by the eROSITA collaboration.
    eROSITA is the primary instrument on-board the Russian-German "Spectrum-Roentgen-Gamma" (SRG) observatory launched in July of $2019$.
    It is a wide-field X-ray telescope capable of delivering deep images in the energy range $\sim \SIrange{0.2}{8}{keV}$.
    Since December $2019$, SRG/eROSITA has been surveying the whole sky, in which the entire celestial sphere is mapped once every six months, before the observations were halted in 2022.
    The first eROSITA All-Sky Survey (eRASS1, \citealt{Merloni2024, Bulbul2024}) encompasses the first six months of data, listing extended X-ray sources attributed to galaxy clusters and/or galaxy groups in the Western Galactic Hemisphere.
    The combination of X-ray with optical and NIR data facilitates the enhancement of the sample purity compared to X-ray-extent-only selected samples, as well as an identification and characterisation of galaxy cluster candidates up to \mbox{$z\sim 0.9$--$1$} with the use of the \emph{grzw}1-bands recorded ($grz$) and utilised ($w$1) in the Legacy Survey \citep{Kluge2024}.
    Even though the eRASS1 cluster catalogue finds clusters until $z=1.3$, Fig. 3 of \citet{Kluge2024} shows how the depth of the Legacy Surveys rapidly limits the usable survey area beyond $z=1$.
    They make use of the \texttt{eROMaPPer} algorithm based on the red-sequence matched-filter Probabilistic Percolation cluster finder \texttt{redMaPPer} \citep{Rykoff2014,Rykoff2016}, optimised to identify X-ray bright clusters and groups, and scanning for an over-density of passive red galaxies around the detected X-ray centres.\newline
    \noindent    
    However, eROSITA is expected to detect X-ray luminous galaxy clusters up to redshifts as high as $z \sim 1.3$ \citep{Grandis2019}, which was confirmed by \citet{Bulbul2024, Kluge2024}.
    These extended sources can not be identified and characterised properly at these high redshifts using optical imaging only.
    That is, because the $\SI{4000}{\angstrom}$ break is redshifted to longer wavelengths, out of the available $griz$-bands.
    The use of the IR bands obtained by the Near-Earth Object Wide-field Infrared Survey Explorer (NEOWISE, \citealt{NEOWISE2014}) does not improve the constraints sufficiently due to their limited depth, poor resolution, and resulting severe blending.
    In order to compensate for the lack of suitable NIR imaging redwards of the $z$ band, we obtained targeted $J$-band follow-up observations for a subset of high-$z$ eRASS1 cluster candidates.
    We show the list of the selected targets in Table \ref{Tab:TargetList}, as well as the results for BEST\_Z, EXT\_LIKE, and PCONT reported in the latest optical cluster catalogue\footnote{\href{https://erosita.mpe.mpg.de/dr1/AllSkySurveyData_dr1/Catalogues_dr1/}{https://erosita.mpe.mpg.de/dr1/AllSkySurveyData\_dr1/Catalogues\_dr1/}}.
    For a more detailed discussion of the individual parameters, please see \citet{Kluge2024, Ghirardini2024, Brunner2022, Seppi2022}, as we only provide a brief explanation.
    \citet{Kluge2024} report a "best" redshift BEST\_Z, which corresponds  to the first available of the following redshift estimates: spectroscopic redshift $z_{\mathrm{spec}}$ from at least three spectroscopic cluster members,  spectroscopic redshift $z_{\mathrm{spec,cg}}$ of at least one galaxy located in the optical cluster centre, photometric redshifts $z_{\lambda}$ (if within calibrated redshift range; the subscript $\lambda$ is used in their work to indicate photometric redshifts), or from literature redshifts $z_{\mathrm{lit}}$.
    We show the value of the obtained BEST\_Z redshift value in Table \ref{Tab:TargetList}.
    For all targets of our follow-up programme, the "best" redshift estimate corresponds to a photometric redshift.
    The EXT\_LIKE parameter quantifies the likelihood difference of a detected source to be of extended nature compared to a point source \citep{Brunner2022}.
    To estimate the overall contamination of the eROSITA optical cluster catalogue, \citet{Ghirardini2024} use the probability density funtion of the count rate and sky position of the active galactic nuclei (AGN) and random source (RS) populations obtained from eRASS1 simulations presented in \citet{Seppi2022, Comparat2020}.
    For this, \citet{Kluge2024} run the optical cluster finder \texttt{eROMaPPer} at positions of eRASS1 point sources as well as at locations of random lines of sight.
    Based on this analysis they estimate a contamination probability PCONT to any source in the catalogue.\\    
    Our targets were selected according to the following criteria:
    \begin{itemize}
    \setlength\itemsep{0.5em}
        \item likely to be high-redshift: $\mathrm{BEST\_Z} > 0.9$
        \item good visibility at the Calar Alto Observatory: $\delta > -20^{\circ}$
        \item showing extended X-ray emission: EXT\_LIKE $>5$
    \end{itemize}
    We note that one of our targets listed in Table \ref{Tab:TargetList} (1eRASS $J140945.2$-$130101$) was observed with both the CAHA3.5/Omega2000 and the WST2.1/3KK.
    We use this opportunity to compare the findings from the two telescopes and see if the results are consistent with each other.

    \begin{table*}
    \caption{
                Sub-selection of the eRASS1 cluster catalogue following the constraints described in Sect.\ref{SubSec:eRASS1}.
            }
    \label{Tab:TargetList}
    \resizebox{\linewidth}{!}{
        \begin{tabular}{l>{\raggedleft\arraybackslash}p{1.5cm}>{\raggedleft\arraybackslash}p{1.5cm}>{\raggedleft\arraybackslash}p{1.5cm}>{\raggedleft\arraybackslash}p{1.5cm}>{\raggedleft\arraybackslash}p{2.5cm}>{\raggedleft\arraybackslash}p{1.75cm}>{\raggedleft\arraybackslash}p{1.5cm}}
            \hline
            \noalign{\smallskip}
            CLUSTER ID & RA\tablefootmark{(a)} & DEC & RA$_{\mathrm{OPT}}$\tablefootmark{(b)} & DEC$_{\mathrm{OPT}}$ & BEST\_Z\tablefootmark{(c)} & EXT\_LIKE\tablefootmark{(d)} & PCONT\tablefootmark{(e)} \\
            \noalign{\smallskip}
            & [$^{\circ}$] & [$^{\circ}$] & [$^{\circ}$] & [$^{\circ}$] &  &  & \\
            \noalign{\smallskip}
            \hline
            \noalign{\smallskip}
            \multicolumn{8}{c}{CAHA/Omega2000}\\
            \hline
            \noalign{\smallskip}
            1eRASS $J$025044.2$-$044309 & 42.68 & $-4.72$ & 42.69 & $-4.72$ & 1.02 $\pm$ 0.02 & 7.35 & 0.00 \\
            1eRASS $J$031350.5$-$000546 & 48.46 & $-0.10$ & 48.44 & $-0.09$ & 1.15 $\pm$ 0.02 & 5.06 & 0.10 \\
            1eRASS $J$033041.6$-$053608 & 52.67 & $-5.60$ & 52.67 & $-5.60$ & 1.00 $\pm$ 0.02 & 5.99 & 0.31 \\
            1eRASS $J$042710.8$-$155324 & 66.80 & $-15.89$ & 66.81 & $-15.89$ & 1.18 $\pm$ 0.03 & 5.85 & 0.01 \\
            1eRASS $J$043901.1$-$022852 & 69.75 & $-2.48$ & 69.75 & $-2.48$ & 1.07 $\pm$ 0.03 & 5.25 & 0.60 \\
            1eRASS $J$053425.5$-$183444 & 83.61 & $-18.58$ & 83.60 & $-18.59$ & 1.00 $\pm$ 0.03 & 6.91 & 0.92 \\
            1eRASS $J$085742.5$-$054517 & 134.43 & $-5.75$ & 134.44 & $-5.75$ & 1.17 $\pm$ 0.02 & 7.00 & 0.00 \\
            1eRASS $J$091715.4$-$102343 & 139.31 & $-10.40$ & 139.32 & $-10.40$ & 0.96 $\pm$ 0.02 & 5.77 & 0.06 \\
            1eRASS $J$094854.3$+$133740 & 147.23 & $13.63$ & 147.23 & $13.63$ & 1.08 $\pm$ 0.03 & 9.33 & 0.01 \\
            1eRASS $J$102148.4$+$225133 & 155.45 & $22.86$ & 155.46 & $22.86$ & 1.12 $\pm$ 0.03 & 5.99 & 0.00 \\
            1eRASS $J$113751.5$+$072839 & 174.46 & $7.48$ & 174.46 & $7.47$ & 0.94 $\pm$ 0.02 & 7.74 & 0.04 \\
            1eRASS $J$124634.6$+$252236 & 191.64 & $25.38$ & 191.64 & $25.38$ & 1.07 $\pm$ 0.02 & 5.32 & 0.00 \\
            1eRASS $J$133333.8$+$062920 & 203.39 & $6.49$ & 203.38 & $6.48$ & 1.14 $\pm$ 0.02 & 6.57 & 0.00 \\
            1eRASS $J$133801.7$+$175957 & 204.51 & $18.00$ & 204.51 & $17.99$ & 1.21 $\pm$ 0.03 & 5.29 & 0.00 \\
            1eRASS $J$140945.2$-$130101 & 212.44 & $-13.02$ & 212.42 & $-13.02$ & 0.99 $\pm$ 0.02 & 6.49 & 0.00 \\
            1eRASS $J$142000.6$+$095651 & 215.00 & $9.95$ & 215.01 & $9.94$ & 1.19 $\pm$ 0.02 & 5.74 & 0.00 \\
            1eRASS $J$145552.2$-$030618 & 223.97 & $-3.11$ & 223.96 & $-3.09$ & 1.19 $\pm$ 0.05 & 8.44 & 0.00 \\
            \noalign{\smallskip}
            \hline
            \noalign{\smallskip}
            \multicolumn{8}{c}{WST/3KK}\\
            \hline
            \noalign{\smallskip}
            1eRASS $J$051425.6$-$093653 & 78.61 & -9.61 & 78.61 & -9.62 & 0.72 $\pm$ 0.02 & 15.47 & 0.00 \\
            1eRASS $J$115512.7$+$125759 & 178.80 & 12.97 & 178.80 & 12.97 & 1.02 $\pm$ 0.02 & 3.24 & 0.00 \\
            1eRASS $J$121051.6$+$315520 & 182.72 & 31.92 & 182.73 & 31.91 & 0.97 $\pm$ 0.01 & 3.92 & 0.00 \\
            1eRASS $J$140945.2$-$130101 & 212.44 & -13.02 & 212.42 & -13.02 & 0.99 $\pm$ 0.02 & 6.49 & 0.00 \\
            \hline
            \noalign{\smallskip}
        \end{tabular}
    }
    \tablefoottext{a}{X-Ray centre of the respective cluster.}\\
    \tablefoottext{b}{Optical cluster centre found by \citet{Kluge2024}.}\\
    \tablefoottext{c}{"Best" redshift estimate assigned to the cluster in \citet{Kluge2024} (see text).}\\
    \tablefoottext{d}{Likelihood difference of the source to be extended compared to being a point source (see Sect. \ref{SubSec:eRASS1} and \citealt{Brunner2022}).}\\
    \tablefoottext{e}{Estimate of the detection to be a contaminant due to noise, projection effects and point sources.}\\
    \end{table*}
    
    \subsection{CAHA}
    The selected high-\emph{z} galaxy cluster candidates have been observed with with Omega2000 on the CAHA3.5m as part of observing programmes 23B043 and 24B015 of the optical telescope calls of the Opticon RadioNet Pilot (ORP) Trans-National Access programme.
    The observations were conducted in visitor mode during the nights starting on Dec 20 and 21, 2023, Apr 17 and 18, 2024, as well as Nov 16 and 17, 2024.
    Omega2000 is a NIR imager equipped with a single HAWAII-2 HgCdTe detector, which covers a field of view of $15\farcm5 \times 15\farcm4$ with a pixel scale of $0\farcs4487/\mathrm{pixel}$.
    The \emph{J}-band images were obtained by taking $45 \times \SI{60}{s}$ individual exposures (totalling to $\SI{2700}{s}$) for each target, excluding 1eRASS $J113751.5$+$072839$, which was exposed in total for $\SI{3840}{s}$.
    Each of the individual $\SI{60}{s}$ exposures was split into $6 \times \SI{10}{s}$ sub-exposures to avoid background saturation.
    We dither the individual exposures following the standard pattern described in the CAHA3.5/Omega2000 manual\footnote{\href{https://w3.caha.es/CAHA/Instruments/O2000/OMEGA2000_manual.pdf}{https://w3.caha.es/CAHA/Instruments/O2000/ OMEGA2000\_manual.pdf}}.
    Every full observing block ($45 \times 6 \times \SI{10}{s}$) is split into three sub-blocks ($15 \times 6 \times \SI{10}{s}$) in between which we add an additional dither offset of $\SI{15}{''}$ going East, North or West, depending on the current dither position.\\
    \noindent
    The data was reduced using the end-to-end pipeline \texttt{THELI} \citep{Erben2005, Schirmer2013}, encompassing several reduction steps.
    They can be largely split into calibration tasks like overscan correction, dark subtraction and background modelling/subtraction, and the co-addition tasks like the generation of global and individual weight maps, computing an astrometric solution, a preliminary relative and absolute photometric calibration, and finally the stacking of the individual exposures.\\
    The background is modelled in two-pass mode, first removing the bulk background by median-combination without object detection and masking.
    A second more refined step then includes object masking.
    The median-combination usually incorporates 4--6 individual exposures, resulting in a background model that averages over 4--\SI{6}{min}.
    A global weight map based on the master flat field serves as base for the individual weight maps for each individual exposure.
    For the latter, the global weight map can be modified according to bad pixels, cosmetics in the FoV, and dynamic range thresholding of the pixel values in each individual exposure.
    The astrometry is determined by comparing the catalogue retrieved with \texttt{SExtractor} \citep{Bertin1996} with the corresponding Two Micron All Sky Survey (2MASS) counterpart \citep{Skrutskie2006}, the astrometric solution is then computed within \texttt{THELI} using \texttt{Scamp} \citep{Bertin2006}.
    Here we require the standard deviation of the position residuals of the matched sources to be less than $\sim 1/10$ of a pixel in both dimensions.
    Image resampling and co-addition in \texttt{THELI} is performed with \texttt{SWarp} \citep{Bertin2002} using a Lanczos3 kernel.    
    
    \subsection{Wendelstein}\label{SubSec:DataWST}
    We complement the observations obtained with the CAHA/Omega2000 with additional \emph{J}-band data obtained with the three-channel camera \citep[3KK;][]{Lang-Bardl2010,Lang-Bardl2016} at the $\SI{2}{m}$ Fraunhofer telescope at the Wendelstein Observatory.
    The field of view in the infrared channel is $8\arcmin\times8\arcmin$ with a pixel scale of $0\farcs24/\mathrm{pixel}$.
    Even though the observations were taken simultaneously in the $r'$, $z'$, and $J$ bands, we only used the $J$-band observations in this analysis, since they are least sensitive to increased sky backgrounds caused by the partially bright observing conditions.
    The observations were taken between April 2023 and April 2024 during a mixture of bright and dark observing conditions.
    The total integration times are 3.7 hours for 1eRASS $J$051425.6$-$093653, 6.9 hours for 1eRASS $J$115512.7$+$125759, 5.5 hours for 1eRASS $J$121051.6$+$315520, and 6.2 hours for 1eRASS $J$140945.2$-$130101.
    The observations were split into single exposures with 60\,s exposure time, taken in a dithered configuration with a step size of 50\arcsec.\\    
    The data were reduced using the pipeline detailed in \cite{Obermeier2020}.
    It includes bias and dark subtraction, flat fielding, hot pixel masking, astrometric calibration, crosstalk correction, and nonlinearity correction.
    Sky subtraction was separately performed using the night-sky-flat procedure described in \cite{Kluge2020} based on the target images themselves.
    Photometric zero points were calibrated to the 2MASS $J$-band magnitudes of matched stars.
    The magnitudes of the stars in the Wendelstein images are measured in circular apertures with a diameter of 4\arcsec.
    
    \subsection{Legacy Surveys}
    We complement the \emph{J}-band data with with forced photometry measurements in the \emph{i}- and \emph{z}-band.
    For this we use stacked image data from the tenth data release of the Legacy Surveys.
    This survey is built up from several individual surveys (Beijing-Arizona Sky Survey, BASS, \citealt{BASS2017}, Dark Energy Camera Legacy Survey, DECaLS, Mayall $z$-band Legacy Survey, MzLS, \citealt{DECaLS2019}) in the optical, complemented with four infrared bands from the NEOWISE survey.
    They were observed using different instruments like the 90Prime camera at the prime focus of the Bok2.3m telescope at the Kitt Peak National Observatory (BASS), the Dark Energy Camera (DECam) on the Blanco4m telescope located at the Cerro Tololo Inter-American Observatory (DECaLS), the MOSAIC-3 camera at the prime focus of the 4-meter Mayall telescope, adjacent to the Bok2.3m telescope at the Kitt Peak National Observatory, and lastly the four infrared detectors on board the space-based Wide-Field Infrared Survey Explorer (WISE) whose first re-activation served the observation of the NEOWISE survey.
    Together, they encompassed $~14000\ \mathrm{deg}^2$ of the extra-galactic sky visible from the northern hemisphere.
    As of the tenth data release (used in this work), this survey is expanded further into the southern hemisphere, encompassing in total >$20000\ \mathrm{deg}^2$ by including supplementary Dark Energy Camera (DECam) data from the NOIRLab\footnote{\href{https://noirlab.edu/public/projects/astrodataarchive/}{https://noirlab.edu/public/projects/astrodataarchive/}}, comprising now also $i$-band data. 
    The northern (southern) segment of LSDR10 has a defined location which is at $\delta \geq 32\fdg375$ ($\delta < 32\fdg375$) and (or) north (south) of the galactic plane.
    The reached depths in LSDR10 for the different bands depending on the increasing number of exposures are summarised on their catalogue description website\footnote{\href{https://www.legacysurvey.org/dr10/description/}{https://www.legacysurvey.org/dr10/description/}}.
    For a more detailed overview, please see \citet{Dey2019}.
    In this work we use exclusively data observed with the DECam which has a field of view of $3^{\circ} \times 3^{\circ}$ and a pixel scale of $0\farcs262/\mathrm{pixel}$.\\

    \subsection{Preparation and calibration}\label{subsec:PreCal}
    We used forced aperture photometry to obtain colour measurements for our retrieved source catalogues around the proposed galaxy cluster centres.
    We choose an aperture diameter of $2\farcs5$ to be adequate for typical cluster galaxies and run \texttt{SExtractor} in its dual image mode.
    We measure the flux in fixed circular apertures in both the \emph{J}- and \emph{i}- (\emph{z}-) band at the location of \emph{J}-band-detected sources.
    In order to obtain reliable colour measurements, we need to ensure, in each band, to always measure the flux in the same intrinsic part of the galaxy in question.
    Thus, we perform a homogenization of the point spread functions (PSFs) of the individual instruments.
    For this we extract \texttt{SExtractor}'s FLUX\_RADIUS parameter, which quantifies the size of a source containing a specified fraction (in this work $50\%$) of the total flux associated with that source.
    We use the median $50\%$ FLUX\_RADIUS of non-saturated stellar sources to estimate the image quality for each target in every band.
    We select stellar sources by matching our extracted catalogue with the corresponding Gaia \citep{Gaia2016,Gaia2018} counterpart, choosing only sources with a parallax measurement and a re-normalised unit weight error (RUWE) of $\leq 1.2$, which is indicating a good astrometric solution \citep{Lindgren2018}.
    In order to avoid including a selection-biased population of stars for calibration, we only use stellar sources whose magnitude is smaller than the mode of the magnitude distribution of all detected stellar sources in the FoV.
    This approach is exerted throughout this work in order to identify stellar sources in any of the used imaging data.
    For each target, we then homogenise the resolution of all bands,  matching that of the band with the poorest image quality.
    For this we first resample each set of three images ($i$-, $z$- and $J$-band) to the same pixel scale.
    The pixel scale of the CAHA/Omega2000 $J$-band images ($0\farcs4487/\mathrm{pixel}$) is larger than for the DECam $i$-, and $z$-band images ($0\farcs262/\mathrm{pixel}$), therefore we resample to the $J$-band pixel scale for all targets observed at CAHA.
    The pixel scale of the WST/3KK $J$-band images ($0\farcs24/\mathrm{pixel}$) is smaller than for the DECam $i$-, and $z$-band images, therefore we resample to the DECam pixel scale for all targets observed at Wendelstein.
    Afterwards we apply a \emph{Gauss} convolution with a kernel of adaptable size to match the $50\%$ FLUX\_RADIUS parameter of non-saturated stellar sources within $5\%$ after convolution.
    We report the maximum retrieved flux radius per-cent deviation with respect to the median flux radius in the $J$-band data of that target as $\Delta r_{\mathrm{flux}}$ in Table \ref{Tab:TargetListIQ}.
    \noindent
    
\begin{figure}
        \centering
        \includegraphics[width=0.5\textwidth]{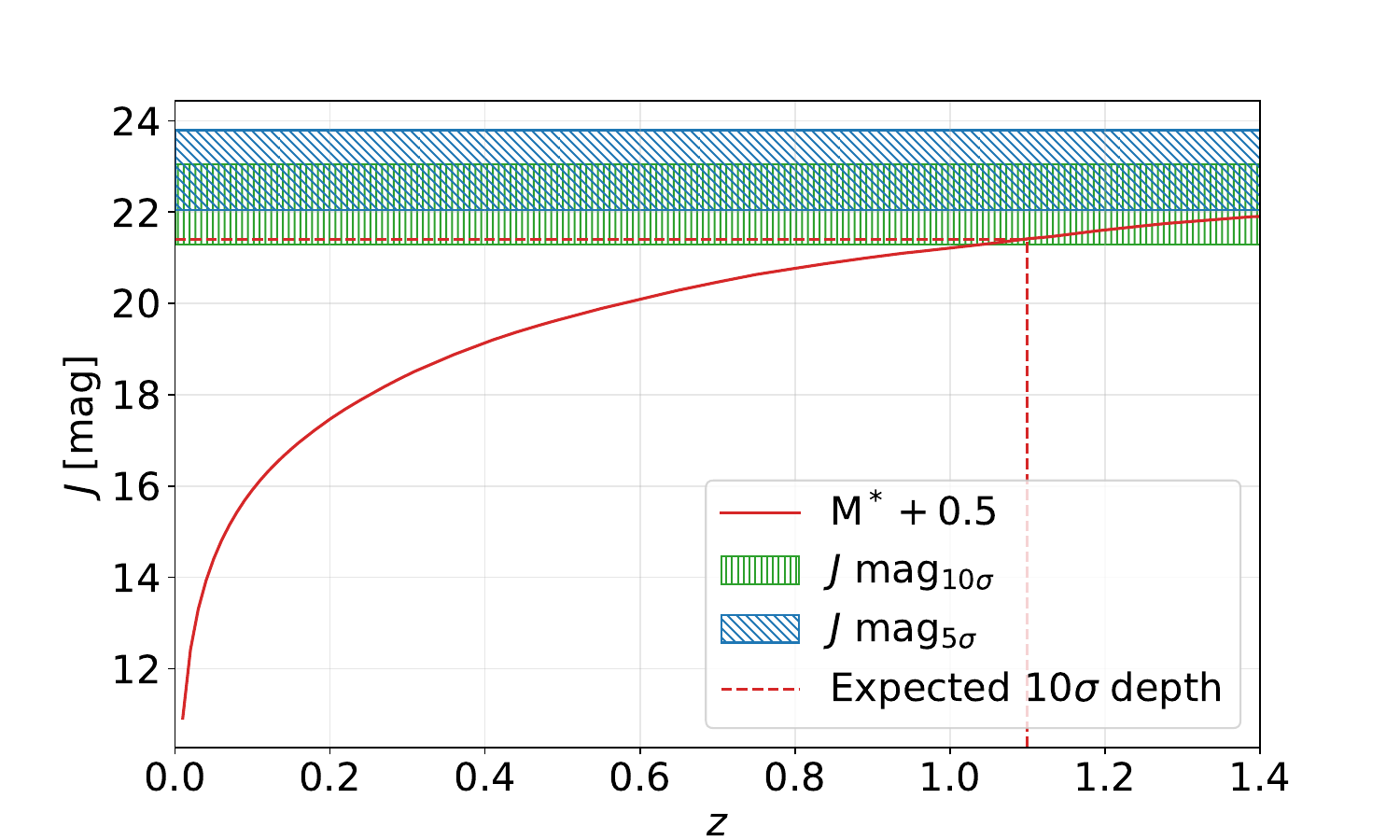}
        \caption{Comparison of the reached depth in the observed $J$-band images with the expected          CAHA $J$-band magnitude of early type ellipticals, depending on their redshift.
                 We plot the expected CAHA $J$-band magnitude of early type ellipticals against the redshift corresponding to $M^* + 0.5$ (solid, red line), and the nominal magnitude of such a galaxy at redshift \mbox{$z\sim 1.1$} (dashed, red line).
                 The shaded areas are the acquired $10\sigma$ (vertical, green), and $5\sigma$ (diagonal, blue) depth ranges of all targets.}
        \label{Fig:Mstar}
    \end{figure}

\begin{figure}
        \centering
        \includegraphics[width=0.5\textwidth]{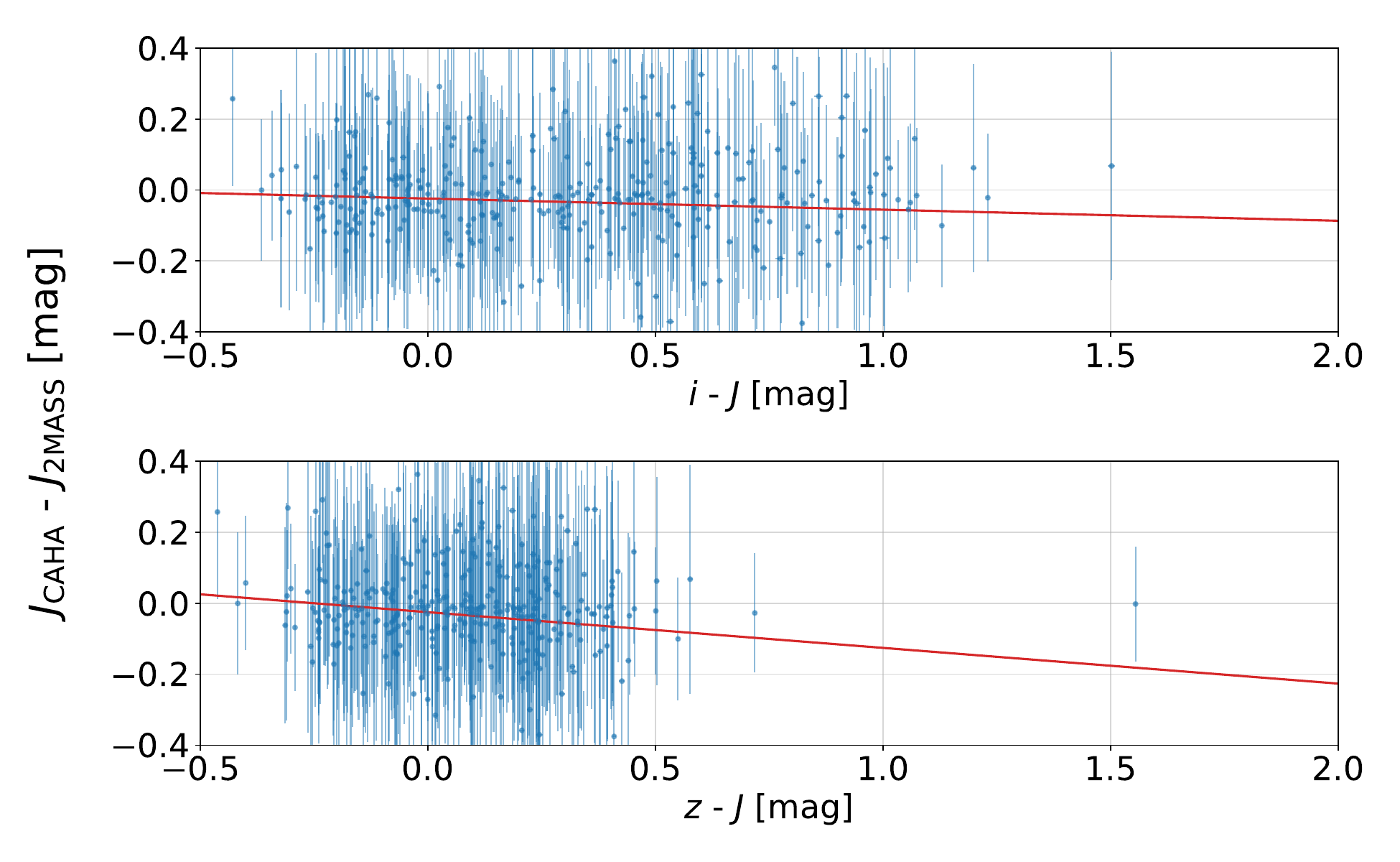}
        \caption{Linear fit of the colour term for stellar sources identified by matching with Gaia (further selection criteria regarding Gaia sources in the text) when calibrating the CAHA $J$-band against 2MASS $J$-band. The best-fit parameters are reported in the text.}
          \label{Fig:ColorTerm}
    \end{figure}

    We estimate the image depth of the stacks used in this work by measuring the magnitude as to which we would obtain an aperture flux measurement with $5\sigma$ significance for an individual source with respect to the background.
    For this we estimate the background signal in each image by placing $1000$ apertures with diameter $2\farcs5$ at random positions in the image, which are not attributed to a detected source using \texttt{SExtractor}'s SEGMENTATION MAP.
    We sum the flux in each aperture and use the standard deviation of the flux sums in all apertures as a measure of the sky background ($\sigma_{\mathrm{sky}}$) in each stacked exposure \citep{Klein2018}.
    The $5\sigma$ limiting magnitude is thus computed as
    \begin{equation}
        \mathrm{mag}_{5\sigma} = -2.5 \log_{10}\left( 5\sigma_{\mathrm{sky}} \right) + \mathrm{ZP},
    \end{equation}
    where ZP is the respective zeropoint of the stacked image.
    \noindent
    We perform a photometric calibration by comparing the extracted source magnitudes (MAG\_AUTO) of non-saturated sellar sources with the literature values reported in the 2MASS catalogue (\emph{J}-band) and the LSDR10 catalogues (\emph{i}- and \emph{z}-band), respectively.
    For the \emph{J}-band calibration, we convert the reported Vega magnitudes in the 2MASS catalogues to the AB magnitude system used in this work, following the Cambridge Astronomical Survey Unit\footnote{\href{http://casu.ast.cam.ac.uk/surveys-projects/vista/technical/filter-set}{http://casu.ast.cam.ac.uk/surveys-projects/vista/technical/filter-set}} (compare also with \citealt{Blanton2007}), by adding a constant offset.
    We choose only non-saturated stellar sources to calculate the photometric offset.
    Additionally we correct for galactic extinction using data from NASA/IPAC's Infrared Science Archive (IRSA)\footnote{\href{https://irsa.ipac.caltech.edu/applications/DUST/}{https://irsa.ipac.caltech.edu/applications/DUST/}}.
    For the further analysis we cleaned the source sample from stellar sources by excluding any Gaia-matched sources from our source catalogue.
    We apply magnitude cuts to ensure a minimum signal-to-noise (S/N) ratio of $5\sigma$ in the \emph{J}-band by cutting our source catalogue at the $5\sigma$ depth estimate explained above. 
    The resulting effective depth, as well as the $10\sigma$ depths are plotted in Fig. \ref{Fig:Mstar}.
    Along the obtained depths we show the expected magnitude for early type, passively evolving, elliptical galaxies seen through the CAHA3.5/Omega2000 $J$-band at redshift \mbox{$z\sim 1.1$}, detected with a S/N of $10\sigma$.
    As a benchmark depth we use  $M^* + 0.5$ (for a discussion on $M^*$ see, e.g., \citealt{To2020}).
    $M^* + 0.5$ is the magnitude corresponding to $0.6 L^*$, where $L^*$ is the characteristic luminosity in the modified Schechter luminosity function \citep{Schechter1976}, describing the luminosity distribution of galaxies in a given luminosity interval.
    This benchmark estimates the faint magnitude threshold for sources still detectable with a specific instrument.
    Furthermore, we fit for a colour term correction when calibrating against the 2MASS Vega magnitudes.
    The resulting linear colour correction term in both $i-J$ and $z-J$ is presented in Fig. \ref{Fig:ColorTerm}.
    The line is a linear fit of the residual between the calibrated CAHA $J$-band magnitudes and the corresponding reference 2MASS $J$-band magnitudes.
    The best fit parameter values for $i-J$ [$z-J$] are $10^{-2} \times (-2.67 \pm 0.02)$ [$10^{-2}\times(-4.23 \pm 0.03$)] for the slope and $10^{-2}\times(-2.709 \pm 0.005$) [$10^{-2}\times(-3.131 \pm 0.003$)] for the offset.
    We apply this colour term correction in the subsequent analysis on all source magnitudes left after the aforementioned selection cuts.
    
    \begin{table*}
    \caption{
                Image quality and depth of the \emph{J}-, \emph{i}-, and \emph{z}-band imaging data used in this work. Quantities shown are after PSF homogenization.
            }
    \label{Tab:TargetListIQ}
        \resizebox{\linewidth}{!}{
            \begin{tabular}{l>{\raggedleft\arraybackslash}p{1.5cm}>{\raggedleft\arraybackslash}p{1.5cm}>{\raggedleft\arraybackslash}p{1.5cm}>{\raggedleft\arraybackslash}p{1.5cm}>{\raggedleft\arraybackslash}p{1.5cm}>{\raggedleft\arraybackslash}p{1.5cm}>{\raggedleft\arraybackslash}p{1.5cm}}
                \hline
                \noalign{\smallskip}
                CLUSTER ID & $r_{\mathrm{flux},J}$\tablefootmark{(a)} & mag$_{5\sigma, J}$\tablefootmark{(b)} & $r_{\mathrm{flux},i}$ & mag$_{5\sigma, i}$ & $r_{\mathrm{flux},z}$ & mag$_{5\sigma, z}$ & $\Delta r_{\mathrm{flux}}$\tablefootmark{(c)} \\
                \noalign{\smallskip}
                & [''] & [mag$_{\mathrm{AB}}$] & [''] & [mag$_{\mathrm{AB}}$] & [''] & [mag$_{\mathrm{AB}}$] & [\%] \\
                \noalign{\smallskip}
                \hline
                \noalign{\smallskip}
                \multicolumn{8}{c}{CAHA/Omega2000}\\
                \hline
                \noalign{\smallskip}
                1eRASS $J$025044.2$-$044309 & 0.58 & 22.26 & 0.57 & 23.77 & 0.59 & 23.23 & 2.24 \\
                1eRASS $J$031350.5$-$000546 & 0.64 & 22.45 & - & - & 0.63 & 22.84 & 2.51 \\
                1eRASS $J$033041.6$-$053608 & 0.65 & 22.50 & 0.69 & 22.46 & 0.67 & 23.15 & 2.30\tablefootmark{(d)} \\
                1eRASS $J$042710.8$-$155324 & 0.72 & 22.37 & 0.74 & 23.40 & 0.73 & 23.11 & 1.74 \\
                1eRASS $J$043901.1$-$022852 & 0.73 & 22.25 & 0.70 & 23.60 & 0.70 & 23.26 & 3.46 \\
                1eRASS $J$053425.5$-$183444 & 0.81 & 22.11 & 0.79 & 24.17 & 0.79 & 23.59 & 2.72 \\
                1eRASS $J$085742.5$-$054517 & 0.97 & 22.21 & 0.94 & 23.70 & 0.93 & 23.18 & 4.29 \\
                1eRASS $J$091715.4$-$102343 & 0.67 & 22.18 & 0.69 & 23.01 & 0.68 & 22.93 & 4.18 \\
                1eRASS $J$094854.3$+$133740 & 0.94 & 22.37 & 0.90 & 23.82 & 0.92 & 23.15 & 3.49 \\
                1eRASS $J$102148.4$+$225133 & 0.90 & 22.41 & 0.88 & 23.33 & 0.88 & 23.18 & 2.88 \\
                1eRASS $J$113751.5$+$072839 & 0.66 & 22.48 & 0.66 & 23.55 & 0.68 & 22.77 & 3.51 \\
                1eRASS $J$124634.6$+$252236 & 0.83 & 22.46 & 0.82 & 23.71 & 0.82 & 23.22 & 1.35 \\
                1eRASS $J$133333.8$+$062920 & 0.73 & 22.48 & 0.72 & 23.58 & 0.73 & 22.75 & 1.72 \\
                1eRASS $J$133801.7$+$175957 & 0.67 & 22.04 & 0.65 & 23.17 & 0.67 & 22.86 & 2.67 \\
                1eRASS $J$140945.2$-$130101 & 1.02 & 22.34 & 0.98 & 23.76 & 0.98 & 23.14 & 3.85 \\
                1eRASS $J$142000.6$+$095651 & 0.79 & 22.55 & 0.78 & 23.57 & 0.78 & 23.00 & 1.25 \\
                1eRASS $J$145552.2$-$030618 & 0.77 & 22.45 & 0.77 & 23.45 & 0.78 & 23.18 & 1.11 \\
                MOO $J$0319\tablefootmark{(e)} & 0.60 & 22.07 & 0.65 & 23.17 & 0.63 & 23.08 & 4.68\tablefootmark{(d)} \\
                \noalign{\smallskip}
                \hline
                \noalign{\smallskip}
                \multicolumn{8}{c}{WST/3KK}\\
                \hline
                \noalign{\smallskip}
                1eRASS $J$051425.6$-$093653 & 0.79 & 22.59 & 0.79 & 24.38 & 0.79 & 23.22 & 0.23 \\
                1eRASS $J$115512.7$+$125759 & 0.61 & 23.31 & 0.62 & 24.18 & 0.62 & 23.49 & 2.02 \\
                1eRASS $J$121051.6$+$315520 & 0.79 & 23.80 & - & - & 0.82 & 23.54 & 3.80 \\
                1eRASS $J$140945.2$-$130101 & 0.63 & 22.34 & 0.63 & 23.76 & 0.63 & 23.14 & 0.41 \\
                \hline
                \noalign{\smallskip}
            \end{tabular}
        }
        \tablefoottext{a}{$\SI{50}{\%}$ flux radius of stellar sources in the field of view; used as an image quality estimate.}\\
        \tablefoottext{b}{$\SI{5}{\sigma}$ depth estimate for the individual bands (see text).}\\
        \tablefoottext{c}{Quality assessment of the performed PSF homogenization by Gauss convolution (see text).}\\
        \tablefoottext{d}{Due to tiling effects in the worst band ($i$-band), we here use the ratio to the $z$-band. The deviations to the $J$-band are $5.8\%$ (1eRASS $J$033041.6$-$053608), and $7.7\%$ (MOO $J$0319).}\\
        \tablefoottext{e}{Compared to Table \ref{Tab:TargetList} we added this target for calibration (see further down in the text).}
    \end{table*}


\section{Methods}\label{chap:Methods}
   
    In this section we explain the strategy employed in this work to extract possible cluster red-sequences around selected extended X-ray sources in the eRASS1 survey and how we estimate a photometric redshift for significant detections.
    First, we describe the two-component mixture model applied to find a radially weighted colour excess around the supposed cluster centre.
    Second, we use two methods of Bayesian inference, Expectation-Maximization (EM) and Markov Chain Monte Carlo (MCMC) sampling, to obtain estimates of the values and uncertainties for the parameters of the used two-component mixture model.
    Third, we compare the found mean colour of the colour excess to a red-sequence model \citep{Kluge2024} to obtain a photo-$z$ estimate.

    \subsection{Radial colour distribution}\label{Sec:RadColDist}
        
        The probability density function (pdf) $p(c_i\ |\ R_i)$ for a source $i$ at a given angular separation $R_i$ from the supposed cluster centre to have a certain colour $c_i$ is modelled by a two-component mixture model.
        One component is the random distribution of field galaxy colours, $p_{\mathrm{field}}(c_i)$, the other is the galaxy cluster member population, $p_{\mathrm{cl}}(c_i)$.
        The latter is assumed to be a Gaussian modelling a colour excess at a given mean colour $\mu$ and standard deviation $\sigma$.
        The model takes the form
        \begin{equation}\label{Eq:p_cGr}
            p(c_i\ |\ R_i) = f(A,R_i) p_{\mathrm{cl}}(c_i) + [1-f(A,R_i)] p_{\mathrm{field}}(c_i),
        \end{equation}
        where
        \begin{equation}\label{Eq:clustercomponent}
            p_{\mathrm{cl}}(c_i)  =    p_{\mathrm{cl}}(c_i\ |\ \mu,\sigma^2)   =   \mathcal{N}(c_i\ |\ \mu,\sigma^2).
        \end{equation}
        Here, $c_i$ and $R_i$ denote the colour and angular separation from the supposed cluster centre of the individual sources in the image, respectively.
        Following Appendix B in \citet{Grandis2024}, the two components are radially weighted by a weight function $f(A,R_i)$, assuming the spatial mass distribution of the galaxy cluster follows a Navarro-Frenk-White (NFW) profile \citep{NFW1996}.
        The weights have the form
        \begin{equation}\label{Eq:Frac}
            f(A,R_i)  =   \frac{A\ \Tilde{\Sigma}_{\mathrm{NFW}}(R_i)}{1+A\ \Tilde{\Sigma}_{\mathrm{NFW}}(R_i)}.
        \end{equation}
        Here, $\Tilde{\Sigma}_{\mathrm{NFW}}(R_i)$ is the corresponding surface mass density of a NFW galaxy cluster \citep{Wright2000}.
        If one introduces a dimensionless radial distance $x=R/R_\mathrm{s}$ for some scaling radius $R_\mathrm{s}$, it can be written as
        \begin{equation}
            \Tilde{\Sigma}_{\mathrm{NFW}}(x)  =
                \begin{cases}
                    \frac{2}{(x^2-1)} \left[ 1-\frac{2}{\sqrt{1-x^2}} \arctanh \left( \frac{1-x}{1+x} \right) \right], &\text{x $< 1$}\\
                    \frac{2}{3}, &\text{x $= 1$}\\
                    \frac{2}{(x^2-1)} \left[ 1-\frac{2}{\sqrt{x^2-1}} \arctan \left( \frac{x-1}{1+x} \right) \right], &\text{x $> 1$}.
                \end{cases}
        \end{equation}
        With this definition, $\Sigma_{\mathrm{NFW}}(x) = A\ \Tilde{\Sigma}_{\mathrm{NFW}}(x)$ is the same as in Eq. (11) in \citet{Wright2000}, and $A$ corresponds to the amplitude of the NFW-profile.
        The shape of $p_{\mathrm{field}}(c_i)$ is estimated by a kernel density estimation (KDE) using the implementation of \texttt{scikit-learn} \citep{scikit-learn2011}.
        We apply a Gaussian kernel and fit for the optimal bandwidth by splitting the field colours ($R_i > 3'$) into train and test sub-samples and use the value which maximises the log-likelihood function.\\
        For this work we fix the value for the scaling radius to $R_\mathrm{s}=0\farcm2$, which would correspond to a galaxy cluster of $M_{200\mathrm{c}} \sim 1e14 M_\odot$ with concentration $c_{200\mathrm{c}} \sim 3$ at redshift $z = 1$, assuming Planck18 cosmology \citep{Planck2020}.
        This leaves the amplitude $A$ of the radial profile, the mean colour of the excess $\mu$, and its standard deviation $\sigma$ free to vary.
        We initialise our mixture model with some initial guesses for the free parameters $\{A, \mu, \sigma\}_{\mathrm{init}} = \{1.0, \mu_\mathrm{init}, 0.05\}$, where $\mu_\mathrm{init}$ is computed for each target individually by setting it to the mode of the colour distribution red-wards of the mean colour of field galaxies.
        The latter are identified by only choosing sources which have an angular separation to the cluster centre of $> \SI{3}{\arcmin}$.
        While the contribution of the colour excess is suppressed by the radial weighting for larger angular separations, and thus is dominated by the field colour distribution towards the outer radial bins, it becomes quite prominent in the inner regions around the cluster centre.
        Following the definition in Eq. (\ref{Eq:p_cGr}), the members of our galaxy catalogue can either be associated with cluster member galaxies (clm), or field galaxies somewhere along the line of sight.
        Thus, the probability density $p_{\mathrm{clm}}^i$ for a a source $i$ to be a cluster member galaxy given a certain colour $c_i$ and angular separation from the cluster centre $R_i$ is given by the weighted cluster component (Eq. \ref{Eq:clustercomponent}), normalised by the whole model pdf (Eq. \ref{Eq:p_cGr}).
        It takes the form
        \begin{equation}\label{Eq:clmprob}
            p_{\mathrm{clm}}^i    =   p(\mathrm{clm}\ |\ c_i, R_i) =   \frac{f(A,R_i)\ p_{\mathrm{cl}}(c_i\ |\ \mu, \sigma)}{p(c_i\ |\ R_i)}.
        \end{equation}
        $p_\mathrm{clm}^i$ is sometimes also referred to as responsibility of a component, quantifying the probability for a data point $i$ to be sampled from that component.
        We can update the mean $\mu^\mathrm{new}$ and the standard deviation $\sigma^\mathrm{new}$ of the Gaussian modelling the colour excess by taking the gradient of the total log-likelihood function,
        \begin{equation}\label{Eq:totlogLike}
            \ln \left( {\mathcal{L}_{\mathrm{tot}}} \right) =   \sum_i \ln \left[ p(c_i\ |\ R_i) \right],
        \end{equation}
        with respect to the corresponding parameter and setting it to $0$.
        The estimators take the form
        \begin{eqnarray}
            \mu^{\mathrm{new}}  &=&   \frac{1}{\sum_i p_{\mathrm{clm}}^i} \ \sum_i p_{\mathrm{clm}}^i \ c_i     \label{Eq:MU}\\
            \sigma^{\mathrm{new}}  &=&   \frac{1}{\sum_i p_{\mathrm{clm}}^i} \ \sum_i p_{\mathrm{clm}}^i \ (c_i - \mu^{\mathrm{new}})^2.    \label{Eq:Sig}
        \end{eqnarray}
        For a proof of Eqs. (\ref{Eq:clmprob}), (\ref{Eq:MU}), and (\ref{Eq:Sig}), please see e.g. chapter 11 of \citet{Deisenroth2020}.
        The $i^\mathrm{th}$ source with colour $c_i$ and radial distance $R_i$ is either classified as a cluster member galaxy, or as a field galaxy.
        The frequency, here $f(A, R_i)$, to be in one of the two classes is given by a Bernoulli distribution, where the probability mass function to be classified as a cluster member is given by $p_\mathrm{clm}^i$, and $1 - p_\mathrm{clm}^i$ to be classified as a field galaxy.
        We estimate the amplitude $A$ by maximizing the Bernoulli log-likelihood function,
        \begin{equation}\label{Eq:logLike_A}
            \ln{\mathcal{L}(A)}    =   \sum_i p_{\mathrm{clm}}^i \ \ln{A} - \sum_i \ln \left( 1+A \ \Tilde{\Sigma}_{\mathrm{NFW}}^i \right),
        \end{equation}
        for each galaxy, such that $A^{\mathrm{new}} = \mathrm{argmax}_{A} \left[\ln{\mathcal{L}(A)}\right]$, i.e. the updated value for $A$ is the one which maximises Eq. (\ref{Eq:logLike_A}).
        We obtain the expression in Eq. (\ref{Eq:logLike_A}) by inserting Eq. (\ref{Eq:Frac}) into the general form of a Bernoulli log-likelihood function.
        Now we can update the initial guesses $\{A, \mu, \sigma\}_{\mathrm{init}}$ with the newly found parameter values (Expectation Step), only to estimate again the parameter values with the newly computed $p_{\mathrm{clm}}$ (Maximization Step).
        In order to find the maximum likelihood estimates of the parameters in Eqs. (\ref{Eq:MU} -- \ref{Eq:logLike_A}), we compute the total log-likelihood (Eq. \ref{Eq:totlogLike}) and iterate between the Expectation and the Maximization step until we reach convergence.
        As a convergence criterion we demand $\ln \left( {\mathcal{L}_{\mathrm{tot, new}}} \right) - \ln \left( {\mathcal{L}_{\mathrm{tot, old}}} \right) < \epsilon = \num{1e-10}$.

        We now initialise an MCMC run around the maximum likelihood estimators found by the EM algorithm to obtain an estimate for the uncertainty for each of the parameters.
        For this we use the \texttt{emcee} implementation in \texttt{python} presented in \citet{emceePython}.
        The log-likelihood function for this is the same as in Eq. (\ref{Eq:totlogLike}).
        To assure positiveness for the amplitude $A$ of the cluster profile, and the width of the red-sequence $\sigma$, we sample both of the parameters in log-space.
        We apply moderately wide and flat priors to all free parameters reading \{$\mathrm{ln}\left(0.001\right) < \mathrm{ln}\left(A\right) < \mathrm{ln}\left(20\right)$, $\mu_{\mathrm{low}} < \mu < 3$, $\mathrm{ln}\left(0.01\right) < \mathrm{ln}\left(\sigma\right) < \mathrm{ln}\left(0.2\right)$\}.
        The lower limit for $\sigma$ is chosen following the findings of \citet{Bower1992}, while the upper limit on $\sigma$ is used to exclude field galaxies during the sampling.
        The lower limit for the mean of the colour over-density, $\mu_{\mathrm{low}}$, is computed for each cluster candidate separately to avoid fitting for the field colour distribution,
        \begin{equation}
            \mu_{\mathrm{low}}  =   \mu_{\mathrm{field}} + \frac{1}{2} \left(\mu_{\mathrm{cen}} - \mu_{\mathrm{field}}\right),
        \end{equation}
        with $\mu_{\mathrm{field}}$ being the mean colour of the field galaxies and $\mu_{\mathrm{cen}}$ the mean colour of the central region (within a radius of $30''$ around the cluster centre).
        If the EM algorithm does not find an additional cluster component red-wards of the mean field colour, we take as an initial guess for $\mu$ the colour of the central region and for $\sigma$ the typical width of a red-sequence of $\SI{0.05}{mag}$ \citep{Bower1992}.
        We estimate the detection significance SIGN by comparing the total likelihood of a model assuming no cluster (no additional colour over-density, i.e. $A=0$), with the total likelihood of the two-component mixture model:
        \begin{equation}
            \mathrm{SIGN} = \sqrt{-2\ \left( \ln{\mathcal{L_{\mathrm{tot, none}}}} - \ln{\mathcal{L_{\mathrm{tot}}}} \right)},
        \end{equation}
        where $\mathcal{L_{\mathrm{tot, none}}}$ is the total log-likelihood assuming no additional cluster component.
        Here, we follow the formalism for a one dimensional Gaussian likelihood, where $\ln{\mathcal{L_{\mathrm{Gauss, 1D}}}} = -\frac{1}{2} \left(\frac{\Delta x}{\sigma}\right)^2$ and $\ln{\mathcal{L_{\mathrm{Gauss, 1D}}^\mathrm{ref}}} = \ln{\mathcal{L_{\mathrm{Gauss, 1D}}}} \left( \Delta x = 0 \right)$.
        Thus we define this quantity in terms of $\sigma$.
        Only sources with a detection significance $\mathrm{SIGN} \geq 3.0\sigma$ are treated as significant detections.
        In the case of a clear peak in the posterior distribution, we estimate the uncertainty for each parameter by taking the symmetric $68\%$ interval in parameter space around the retrieved maximum log-likelihood value.
        If $\mathrm{SIGN}<3\sigma$, we take the $95\%$-percentile of the retrieved posterior distribution as an upper limit on the respective parameter.
        If the peak of the posterior distribution is within $1\sigma$ of the prior boundary, we report the $5\%$-percentile as a lower boundary (if within $1\sigma$ of the upper prior boundary), and the $95\%$-percentile as an upper boundary (if within $1\sigma$ of the lower prior boundary).
        We validate the inference framework on mock catalogues, discussed in more detail in Appendix \ref{app:MockVal}
        
    \subsection{Red-sequence models}\label{Sec:RedSeqModel}

        \begin{figure}
            \centering
            \includegraphics[width=0.5\textwidth]{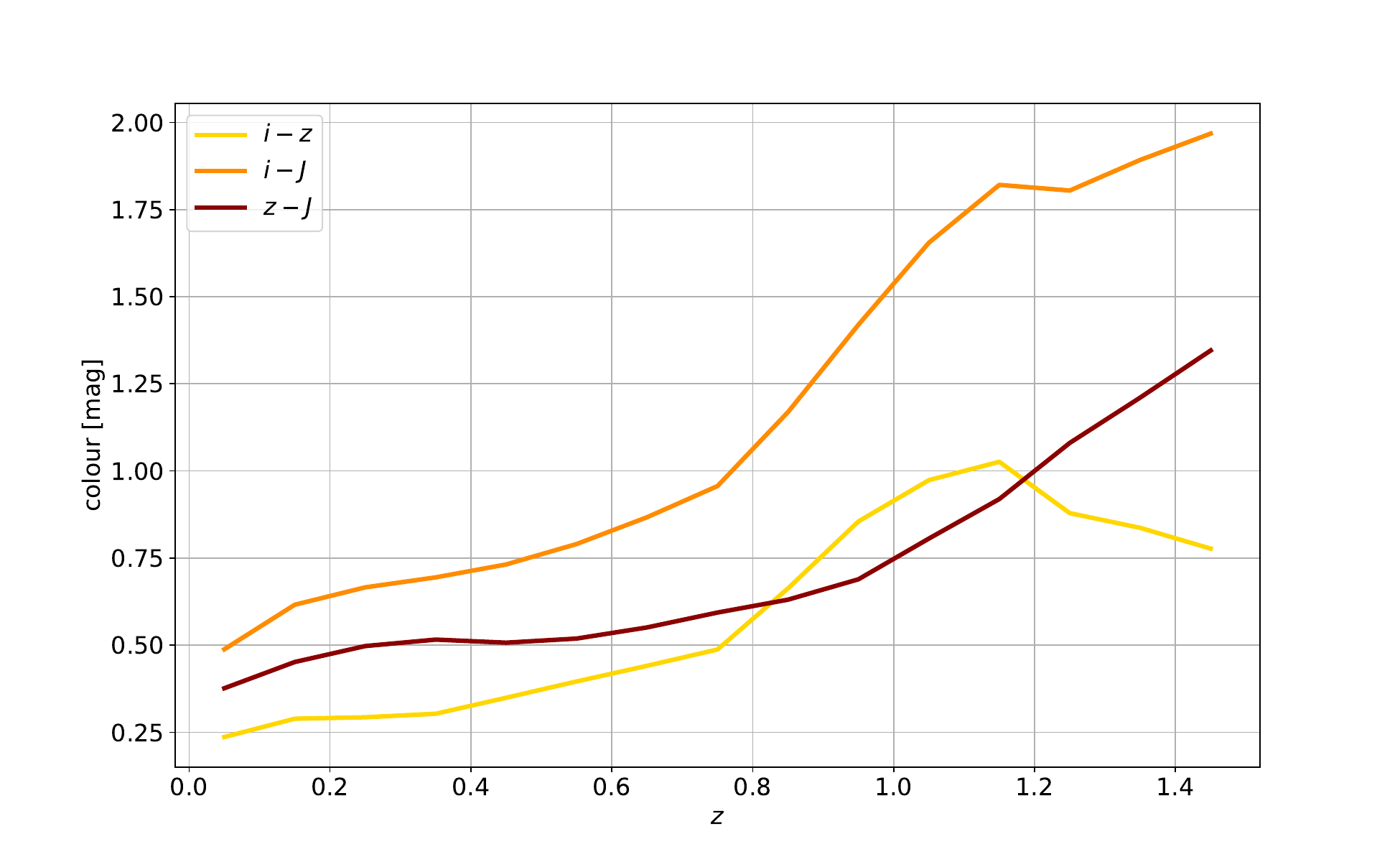}
            \caption{The red-sequence models for the colour indices $i-z$ (yellow), $i-J$ (orange), and $z-J$ (red) used in this work.}
            \label{Fig:RedSeqModel}
        \end{figure}

        The red-sequence models used in this work (see Fig. \ref{Fig:RedSeqModel}) have been obtained similarly to the model in \citet{Kluge2024}, which is based on the procedure described in \citet{Rykoff2014}.
        We generated predictions for the spectral energy distribution of a passively evolving stellar population model from \citet{Bruzual2003}.
        These were integrated within the Calar Alto and Wendelstein $J$ filter bands and Legacy Surveys $i$ and $z$ filter bands.
        By varying the age of the stellar population, we predict the optical and near-infrared colours on a grid in redshift from $z=0$ to $z=1.5$.
        Note the changing sensitivity of the different models in different redshift domains.
        While $i-z$ is fairly sensitive between redshift $0.8<z<1.0$, $i-J$ extends to redshifts until $z\sim1.1$.
        On the other hand, $z-J$ only starts becoming sensitive above $z>1.0$, making it especially interesting for higher redshift objects.
        Different to \cite{Kluge2024}, we did not calibrate these models using observed colours of galaxies with known redshifts.
        Such a sufficiently large data set does not exist for the instruments used in this work.
        The impact is on the order of $\lesssim0.1$\,mag for the models in \cite{Kluge2024} based on Legacy Surveys data, which corresponds to a redshift uncertainty of $\delta z_\lambda\approx0.03$.
        However, we were able to observe MOO $J$0319$-$0025, an optically selected galaxy cluster with a spectroscopic redshift of $z = 1.194$ presented in Table 5 of \citet{Gonzales2019}.
        We can use this target to do a $0^{\mathrm{th}}$-order re-calibration of the red-sequence model including $J$-band imaging data.
        Additional calibration uncertainties of the photometric zero-points can arise from the unknown conversion uncertainty from 2MASS Vega magnitudes (to which the images were calibrated) to AB magnitudes in the Calar Alto and Wendelstein filter bands.\\
        We define sensitive regimes in each red-sequence-model in which we use this model to compute a photometric redshift estimate from a retrieved mean colour $\mu$: $\mu_{i - z} \in \{0.44, 0.89\}$, $\mu_{i - J} \in \{0.87, 1.81\}$, and $\mu_{z - J} \in \{0.59, 1.35\}$.
        These correspond to redshift regimes of $z_{\mathrm{phot},i - z} \in \{0.65, 1.25\}$, $z_{\mathrm{phot},i - J} \in \{0.65, 1.25\}$, and $z_{\mathrm{phot},z - J} \in \{0.75, 1.45\}$ (compare with Fig. \ref{Fig:RedSeqModel}).
        

\section{Results}\label{chap:Results}

    \begin{figure*}
        \centering
        \includegraphics[width=0.95\textwidth]{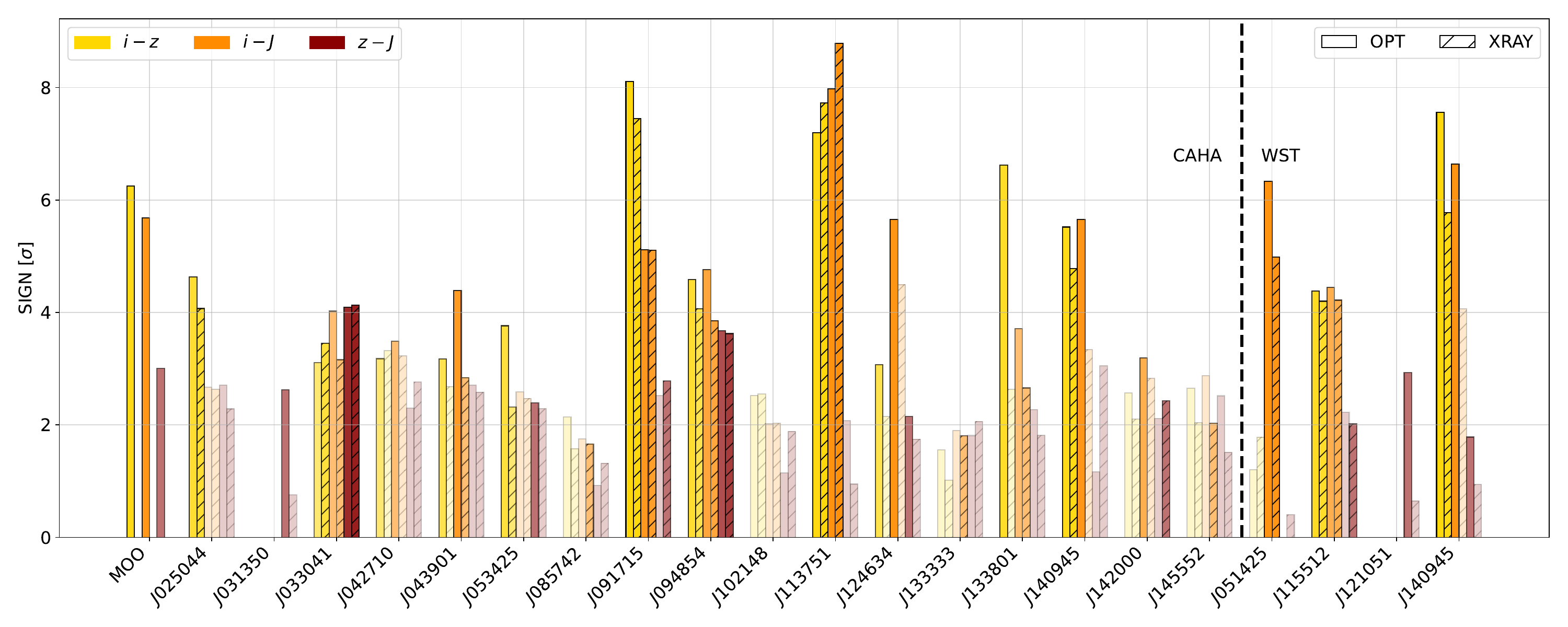}
        \caption{The obtained significances (as defined in Sect. \ref{Sec:RadColDist}) for each target when analysed with respect to the optical (no hatch), and the X-ray (hatch) cluster centre, as well as when using only the $i-z$ (yellow), the $i-J$ (orange), and the $z-J$ (red) colour of the sources. Transparency encodes the number of cluster member galaxies for which we find $p_{\mathrm{clm}} > 0.8$: the more transparent the fewer highly probable cluster member galaxies are found.}
        \label{Fig:SIGNs}
    \end{figure*}

    \begin{table*}
        \caption{
                    Results for finding a radial colour over-density for the $i - J$ colour index around the optical centre.
                }
        \label{Tab:iJOPTResults}
        \centering
        \renewcommand{\arraystretch}{1.2}
            \resizebox{\linewidth}{!}{
                \begin{tabular}{l>{\raggedleft\arraybackslash}p{1.5cm}>{\raggedleft\arraybackslash}p{1.5cm}>{\raggedleft\arraybackslash}p{1.5cm}>{\raggedleft\arraybackslash}p{1.5cm}>{\raggedleft\arraybackslash}p{1.5cm}>{\raggedleft\arraybackslash}p{0.5cm}>{\raggedleft\arraybackslash}p{2.5cm}}
                    \hline
                    \noalign{\smallskip}
                    CLUSTER ID & $\mu$ & $\sigma$ & $A$ & SIGN$_\mathrm{MCMC}$ & $z_{\mathrm{phot}, i-J}$ & $n_{\mathrm{clm}}^{0.8}$ & BEST\_Z \\
                    & [mag$_{\mathrm{AB}}$] &  &  &  &  &  &  \\
                    \hline
                    \multicolumn{8}{c}{CAHA/Omega2000}\\
                    \hline
                    \noalign{\smallskip}
                    1eRASS $J$025044.2$-$044309 & $<2.81$\tablefootmark{(a)} & $<0.17$ & $<0.12$ & 2.67 & - & 1 & $1.02\pm0.02$ \\
                    1eRASS $J$033041.6$-$053608 & $1.88_{-0.05}^{+0.05}$ & $<0.18$ & $0.10_{-0.06}^{+0.17}$ & 4.02 & - & 2 & $1.00\pm0.02$ \\
                    1eRASS $J$042710.8$-$155324 & $1.81_{-0.13}^{+0.13}$ & $<0.18$ & $0.16_{-0.13}^{+0.71}$ & 3.49 & $1.14_{-0.08}^{+0.01}$\tablefootmark{(b)} & 2 & $1.18\pm0.03$ \\
                    1eRASS $J$043901.1$-$022852 & $<2.36$ & $>0.02$ & $>0.00$ & 4.39 & - & 7 & $1.07\pm0.03$ \\
                    1eRASS $J$053425.5$-$183444 & $<2.81$ & $<0.18$ & $<0.42$ & 2.59 & - & 1 & $1.00\pm0.03$ \\
                    1eRASS $J$085742.5$-$054517 & $<2.89$ & $<0.17$ & $<0.29$ & 1.75 & - & 0 & $1.17\pm0.02$ \\
                    1eRASS $J$091715.4$-$102343 & $1.33_{-0.06}^{+0.06}$ & $>0.12$ & $2.77_{-1.26}^{+2.31}$ & 5.11 & $0.92_{-0.03}^{+0.03}$ & 9 & $0.96\pm0.02$ \\
                    1eRASS $J$094854.3$+$133740 & $1.73_{-0.02}^{+0.02}$ & $0.03_{-0.01}^{+0.02}$ & $0.72_{-0.33}^{+0.62}$ & 4.77 & $1.10_{-0.01}^{+0.01}$ & 4 & $1.08\pm0.03$ \\
                    1eRASS $J$102148.4$+$225133 & $<2.88$ & $<0.17$ & $<0.29$ & 2.02 & - & 1 & $1.12\pm0.03$ \\
                    1eRASS $J$113751.5$+$072839 & $1.47_{-0.03}^{+0.03}$ & $0.13_{-0.03}^{+0.03}$ & $2.47_{-0.63}^{+0.85}$ & 7.98 & $0.97_{-0.01}^{+0.01}$ & 15 & $0.94\pm0.02$ \\
                    1eRASS $J$124634.6$+$252236 & $1.15_{-0.04}^{+0.04}$ & $0.15_{-0.03}^{+0.03}$ & $5.45_{-1.75}^{+2.57}$ & 5.66 & $0.84_{-0.02}^{+0.02}$ & 12 & $1.07\pm0.02$ \\
                    1eRASS $J$133333.8$+$062920 & $<2.88$ & $<0.17$ & $<0.26$ & 1.90 & - & 1 & $1.14\pm0.02$ \\
                    1eRASS $J$133801.7$+$175957 & $1.57_{-0.12}^{+0.12}$ & $<0.16$ & $0.36_{-0.33}^{+2.03}$ & 3.71 & $1.02_{-0.05}^{+0.06}$ & 2 & $1.21\pm0.03$ \\
                    1eRASS $J$140945.2$-$130101 & $1.26_{-0.03}^{+0.03}$ & $0.09_{-0.02}^{+0.03}$ & $2.07_{-0.70}^{+1.06}$ & 5.66 & $0.89_{-0.01}^{+0.01}$ & 11 & $0.99\pm0.02$ \\
                    1eRASS $J$142000.6$+$095651 & $1.75_{-0.57}^{+0.57}$ & $<0.17$ & $0.28_{-0.27}^{+3.32}$ & 3.19 & $1.11_{-0.25}^{+0.04}$ & 3 & $1.19\pm0.02$ \\
                    1eRASS $J$145552.2$-$030618 & $<2.77$ & $<0.18$ & $<0.70$ & 2.88 & - & 1 & $1.19\pm0.05$ \\
                    \noalign{\smallskip}
                    \hline
                    \multicolumn{8}{c}{WST/3KK}\\
                    \hline
                    \noalign{\smallskip}
                    1eRASS $J$051425.6$-$093653 & $<2.66$ & $>0.02$ & $>0.00$ & 6.33 & - & 14 & $0.72\pm0.02$ \\
                    1eRASS $J$115512.7$+$125759 & $1.55_{-0.03}^{+0.03}$ & $<0.12$ & $0.41_{-0.24}^{+0.57}$ & 4.44 & $1.00_{-0.01}^{+0.01}$ & 3 & $1.02\pm0.02$ \\
                    1eRASS $J$140945.2$-$130101 & $1.35_{-0.04}^{+0.04}$ & $0.14_{-0.03}^{+0.03}$ & $2.50_{-0.74}^{+1.05}$ & 6.64 & $0.92_{-0.01}^{+0.01}$ & 14 & $0.99\pm0.02$ \\
                    \noalign{\smallskip}            
                    \hline
                    \noalign{\smallskip}
                \end{tabular}
            }
            \tablefoottext{a}{We report upper/lower limits for the parameters if $\mathrm{SIGN}<3.0\sigma$, or if we run into one of our prior boundaries (see Sect. \ref{Sec:RadColDist}) for more details.}\\
            \tablefoottext{b}{We compute a photometric redshift (error) estimate only within the sensitive red-sequence-model-regime for each colour index, respectively (compare with Sect. \ref{Sec:RedSeqModel} and Fig. \ref{Fig:RedSeqModel}).}
    \end{table*}

    In this section, we present the results of the method presented in Sect. \ref{Sec:RadColDist} and the photometric redshift estimation described in Sect. \ref{Sec:RedSeqModel}.
    Additionally, we provide comments on the visual inspection of the targets analysed in this work.
    We show a summary of the significances found using the different colour indicies in Fig. \ref{Fig:SIGNs}, using both the X-ray centres and the optical centres provided by \citet{Kluge2024}.
    In addition, Table \ref{Tab:iJOPTResults} provides detailed results for the optical centres and $i-J$ colours.
    The findings for the other combinations of cluster centre and colour are provided in Appendix \ref{app:SuppD}.
    Listed in the tables are the maximum posterior values or the upper/lower limit values ($95\%$ / $5\%$-percentile of the posterior distribution) of each of the parameters of our model, the significance of our detection as described in Sect. \ref{Sec:RadColDist}, a photometric redshift estimate obtained in this work ($z_{\mathrm{phot}, i-J}$, $i-J$ indicating the used colour index), the number of sources found to have a cluster member probability higher than $80\%$ ($n_{\mathrm{clm}}^{0.8}$), as well as the best redshift estimate computed by \citet{Kluge2024}.

    \subsection{Radial colour over-density}
    
\begin{figure}
            \centering
            \includegraphics[width=0.5\textwidth]{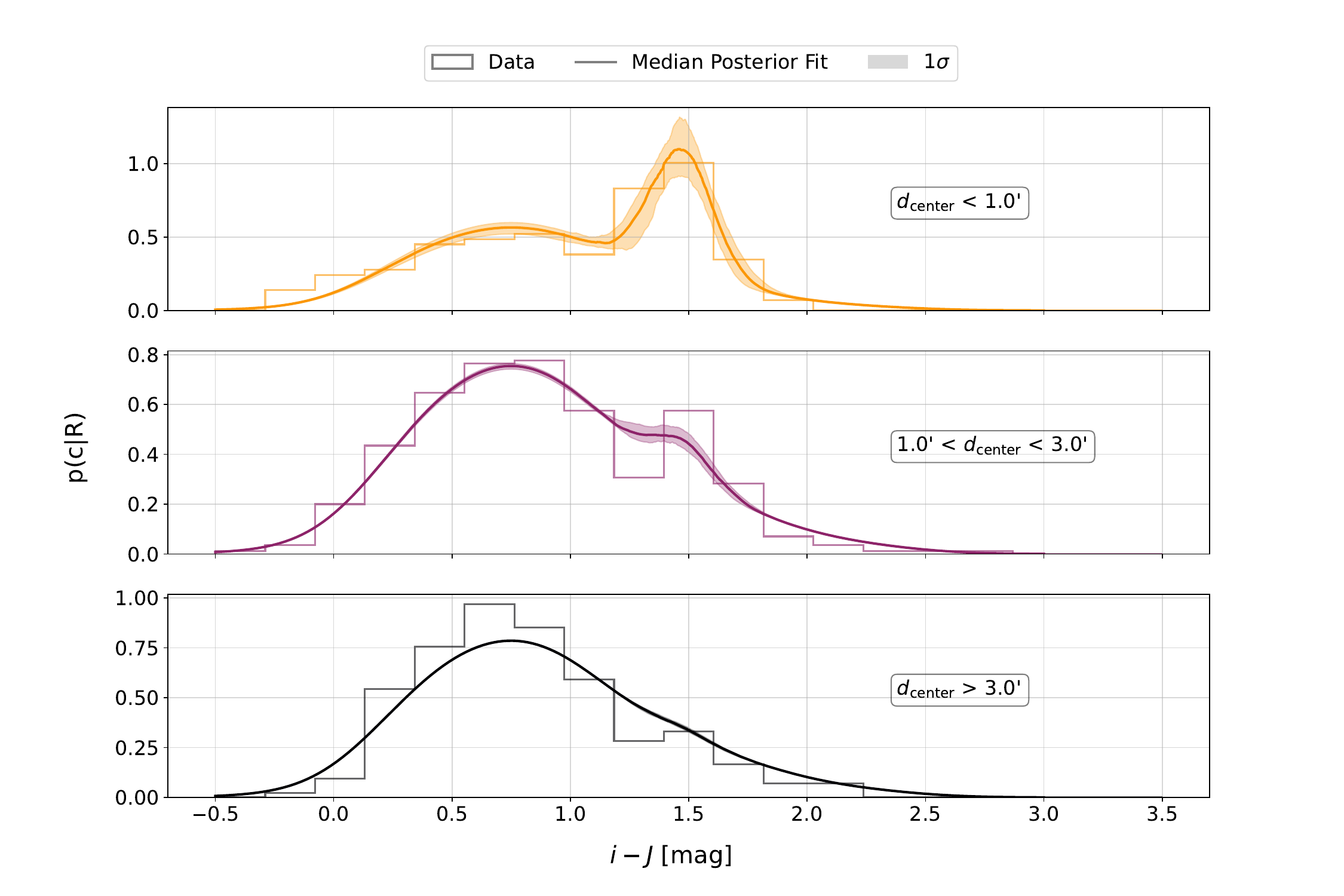}
            \caption{Radial colour distribution of the extracted sources in the field of the cluster 1eRASS $J113751.5+072839$, split into three radial bins with respect to the optical cluster centre: $d_{\mathrm{centre}}<1^\prime$ (orange), $1^\prime < d_{\mathrm{centre}}<3^\prime$ (purple), and $d_{\mathrm{centre}} > 3^\prime$ (black). The histograms show the colour distribution of the data, the curves depict the median fit of the obtained posterior distribution, while the shaded regions show its $1\sigma$ uncertainty. The best fitting parameter values are shown in Table \ref{Tab:iJOPTResults}.}
            \label{Fig:ModelFit_e174}
        \end{figure}

\begin{figure}
            \centering
            \includegraphics[width=0.5\textwidth]{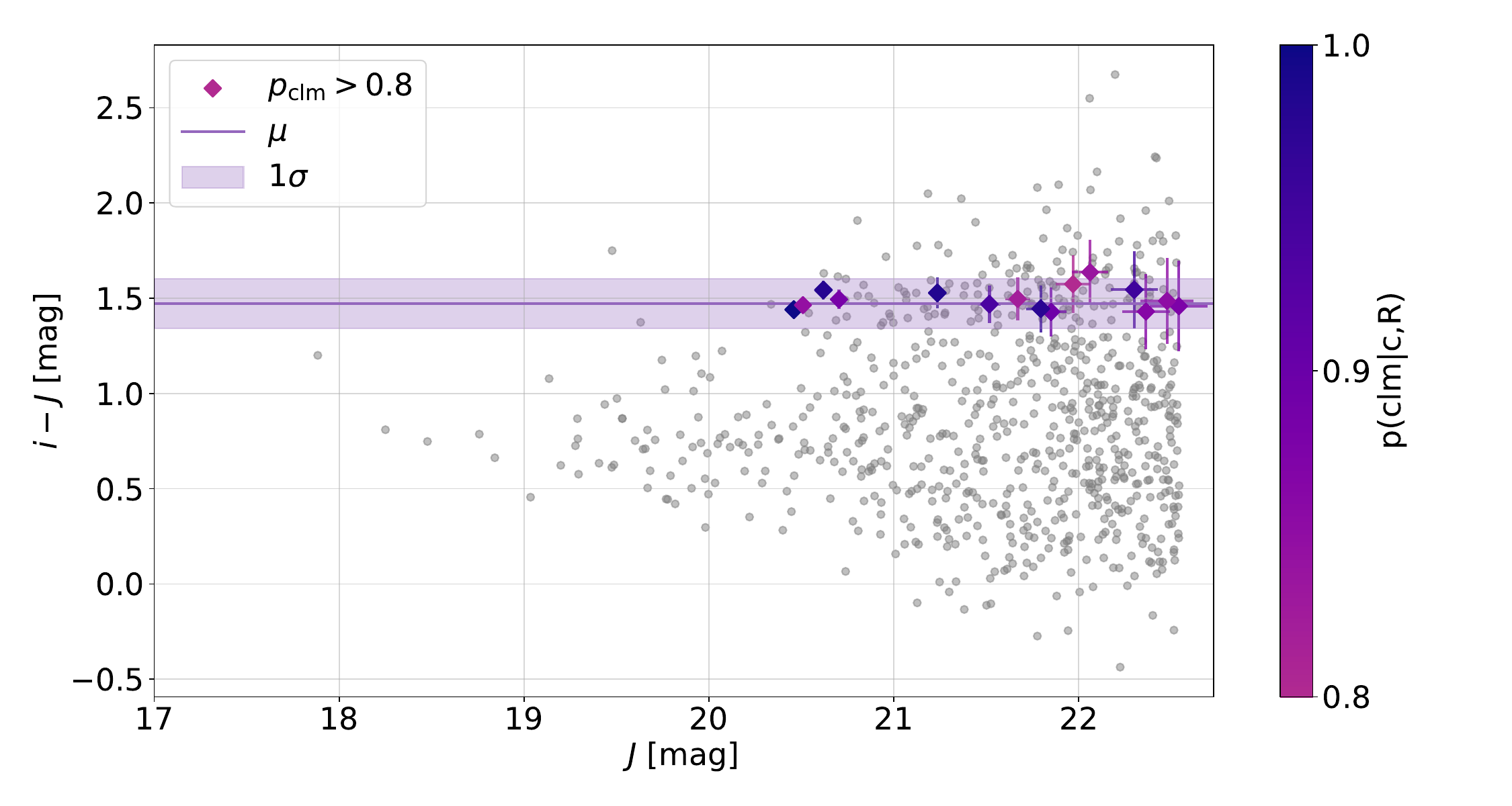}
            \caption{Colour-Magnitude diagram of J113751.5+072839 for the $i-J$ colour index. Shown are all retrieved sources in the field of view (grey). Colour coded are only sources for which we find a cluster member probability larger than $80 \%$. For these we also show the photometric errors estimated by \texttt{SExtractor}.}
            \label{Fig:ColMag_e174}
        \end{figure}

        As described above, we model the pdf of a source at a given angular separation $R_i$ from a supposed cluster centre to have a certain colour $c_i$.
        Thus, it is important which cluster centre and which colour are used.
        As for the latter, it is decisive in what redshift regime one is able to pick up cluster red-sequences. 
        Since eROSITA is expected to pick up galaxy clusters up to redshift \mbox{$z\gtrsim 1$}, the motivation for this work has been to detect and confirm sources at the high-redshift end of the eRASS1 catalogue.
        This would make the $z-J$ colour the most promising, as the corresponding red-sequence model is more sensitive for $z>1$.
        However, since 1eRASS $J$033041.6$-$053608 and 1eRASS $J$094854.3$+$133740 (observed with CAHA/Omega2000, see Fig. \ref{Fig:SIGNs}) are the only significant detections in $z-J$, we focus on the $i-J$ colour in this section.
        For cross-checking, we also analyse the radial colour distribution in the $i-z$ colour.
        As cluster centres, we use both the X-ray centres (we use RA and DEC coming directly out of the detection chain), as well as the optical centres found by \citet{Kluge2024}.
        In principle, our algorithm could simultaneously fit for a new cluster centre as the mean position of all sources in the field of view, weighted by their respective cluster member probability.
        However, since we barely have the signal to fit for the amplitude of the cluster profile, the width and the mean colour of the red-sequence, we decided to fix the location of the centre at either the X-ray, or the optical centre.\\
        \noindent
        If we look at the summarised findings in Table \ref{Tab:iJOPTResults}, we find 12 out of 18 targets that show significant, i.e. $\mathrm{SIGN}>3\sigma$, colour over-densities in $i-J$.
        If we combine the colour indices $i - z$ and $i - J$, we can confirm 14 out of 18 targets which have $i$-band data and an extended likelihood parameter of $\mathrm{EXT\_LIKE}>5$.
        This would correspond to a $22.2\%$ contaminant fraction at the high-redshift end of the eRASS1 galaxy cluster catalogue.\\
        For the most significant detection (1eRASS $J113751.5+072839$) we display the fit of our model to the data in three different radial bins in Fig. \ref{Fig:ModelFit_e174}.    
        We manage to accurately find the location of the second peak in the radial colour distribution.
        The amplitude of the second peak is well matched in the first and the second bin, but over-estimates the over-density in the outer bin.
        Furthermore we show the colour-magnitude-diagram of all analysed sources in the observed field around 1eRASS $J113751.5+072839$ in Fig. \ref{Fig:ColMag_e174}, where we colour-code the sources for which we find $p_{\mathrm{clm}}^{i}>0.8$ as described in Sect. \ref{Sec:RadColDist}, also plotting their photometric errors.
        \begin{figure}
            \centering
            \includegraphics[width=0.5\textwidth]{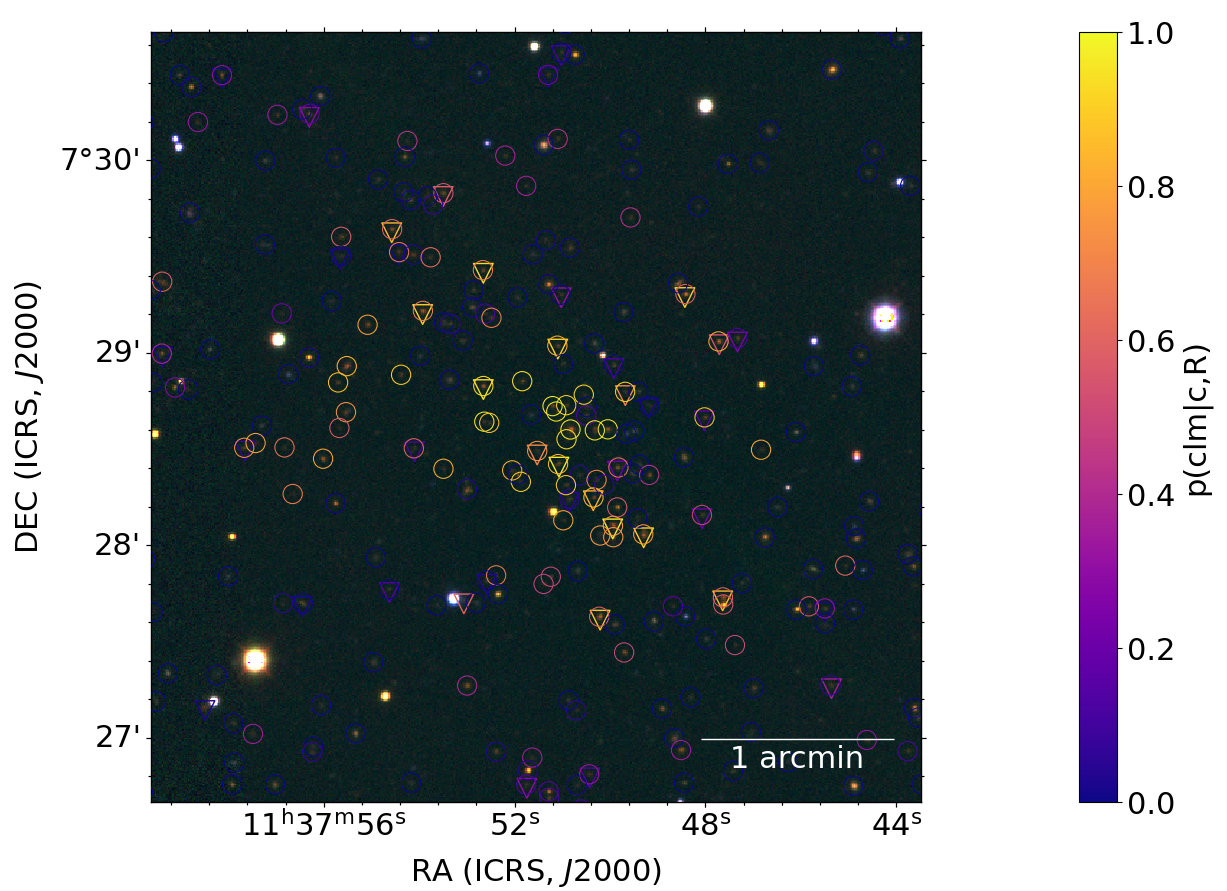}
            \caption{Comparison of sources and their assigned cluster member probability found in this work (colour-coded circles) to the ones reported in \citet{Kluge2024} (colour-coded triangles). For better readability, we change the colour-code of cluster probability compared to Fig. \ref{Fig:ColMag_e174}.}
            \label{Fig:CompareCLMs}
        \end{figure}
        We compare the cluster member galaxies of 1eRASS $J113751.5+072839$ found in this work with the ones found by \citet{Kluge2024} in Fig. \ref{Fig:CompareCLMs}, colour coding them according to cluster member probability.
        We recover all of the probable cluster members from \citet{Kluge2024}, while we also find additional high-probability cluster members.\\
        As has been noted above, 1eRASS $J140945.2-130101$ has been observed with both CAHA3.5/Omega2000 and WST/3KK, allowing us to compare the findings from the two different telescopes.
        The observational strategies and properties have been discussed in Sect. \ref{chap:Data}, but it is worth to point out the difference in exposure time needed to compensate for the smaller mirror size of the WST/3KK.
        The observing conditions were worse for the CAHA run compared to WST, which resulted in a poorer image quality after PSF homogenisation (see Sect. \ref{subsec:PreCal}) also in the $i$- and $z$-band (see Table \ref{Tab:TargetListIQ}).
        \begin{figure}
            \centering
            \includegraphics[width=0.5\textwidth]{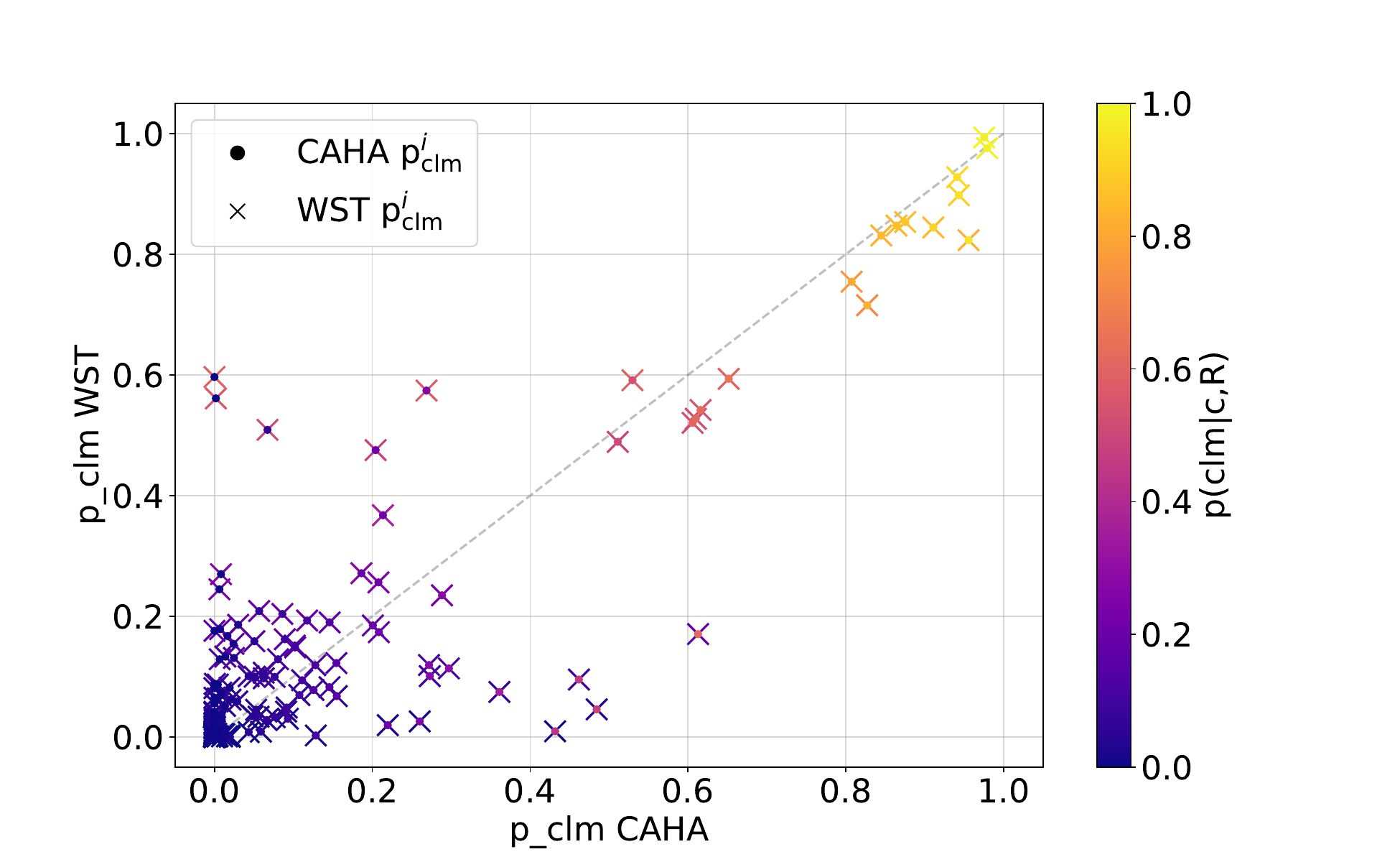}
            \caption{Comparison of found cluster member probabilities from the CAHA (dots) and the WST (crosses) data.}
            \label{Fig:CAHAvsWST}
        \end{figure}
        We compare the estimated cluster member probabilities from the CAHA data and the the WST data in Fig. \ref{Fig:CAHAvsWST}.
        The most probable cluster members (upper right corner of Fig. \ref{Fig:CAHAvsWST}) are identified with both of the data sets, whereas the estimated cluster member probability of less probable cluster member galaxies scatters symmetrically around the one-to-one relation.
        For an image comparison, please see Fig \ref{Fig:CAHAvsWST_JBand} in Appendix \ref{app:CAHAvsWST}.\\
        \noindent
        We do not fit simultaneously for more than one red-sequence colour.
        Instead, we compare the mean colour of the over-densities for all targets confirmed as significant detections in both $i - z$ and $i - J$ to the two re-calibrated (see Sect. \ref{subsec:PhotZEst}) red-sequence models plotted against each other in Fig. \ref{Fig:CompareRedSeqs}.
        \begin{figure}
            \centering
            \includegraphics[width=0.49\textwidth]{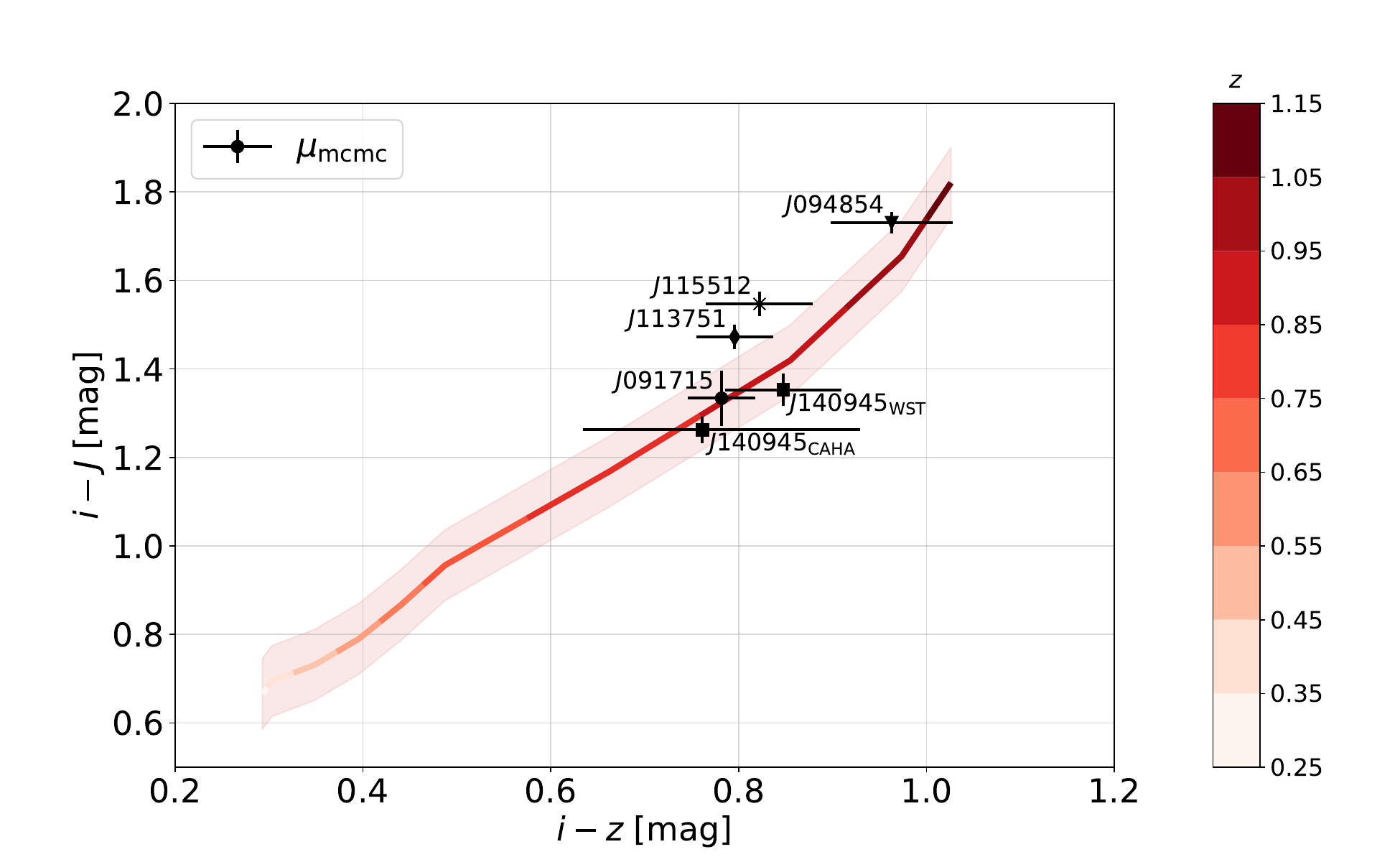}
            \caption{Comparison of re-calibrated (see Sect \ref{subsec:PhotZEst}) red-sequence models for the $i-z$ and the $i-J$ colour index (red line), together with the $1\sigma$ uncertainty on the $i - J$ model estimated from the calibration using MOO $J$0319$-$0025. Over-plotted are the measured mean values for the colour over-densities in $i-z$ and $i-J$.}
            \label{Fig:CompareRedSeqs}
        \end{figure}
        We find that the modelled and the measured red-sequence colour combinations are broadly consistent given their respective uncertainties.
        Note the improvement regarding the error of the mean colour when including $J$-band imaging.

    \subsection{Photometric redshift calibration}\label{subsec:PhotZEst}

        \begin{figure}
            \centering
            \includegraphics[width=0.5\textwidth]{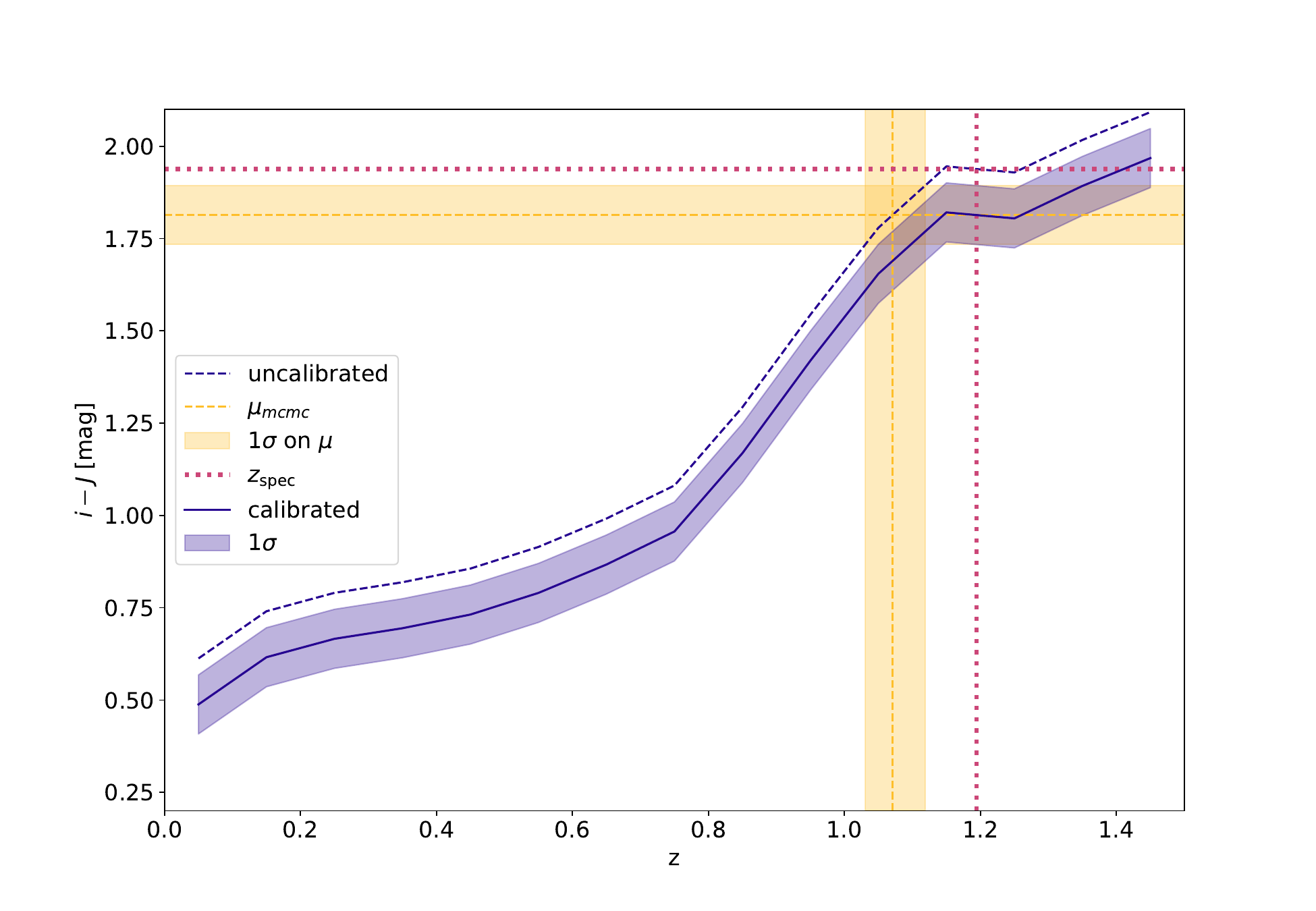}
            \caption{The uncalibrated (dashed blue), as well as the calibrated (continuous blue) red-sequence model for the colour index $i - J$. The blue shaded region shows the offset uncertainty due to the uncertainty on the retrieved mean colour value $\mu$ (yellow dashed line/shaded region) for MOO $J$0319$-$0025. The dotted magenta line shows the spectroscopic redshift from \citet{Gonzales2019} (vertical), and the corresponding $i - J$ colour (horizontal).}
            \label{Fig:RedSeqModelCal}
        \end{figure}
        To be able to empirically check and recalibrate the theoretical red-sequence models, we also observed a galaxy cluster with a known spectroscopic redshift, MOO $J$0319$-$0025, an optically selected galaxy cluster with a spectroscopic redshift of $z = 1.194$ presented in Table 5 of \citet{Gonzales2019}.
        We used this target to estimate a $0^{\mathrm{th}}$-order re-calibration of the red-sequence model including $J$-band imaging data introduced in Sec \ref{Sec:RedSeqModel} (see Fig. \ref{Fig:RedSeqModelCal}).
        However, since we only detected a significant colour over-density for this target for the colour index $i - J$, the re-calibration is only possible for the photometric redshift estimations for this colour index.\\
        \noindent
        We compare the mean colours of significant radial colour over-densities with the respective red-sequence models by interpolating in between the nodes of the red-sequence model, in order to obtain the photometric redshift estimate reported in Table \ref{Tab:iJOPTResults}.
        The errors on $z_\mathrm{phot}$ shown here are solely statistical errors emerging from the MCMC run.
        We do not show any additional statistical uncertainties from the photometric calibration, or systematic uncertainties emerging from our red-sequence-calibration.
        We estimate the uncertainty on the zero-point calibration to be $\SI{0.04}{mag}$, which would translate to photometric redshift error of $\delta z_\mathrm{zp} \sim 0.03$.
        We estimate the systematic uncertainty emerging from our red-sequence calibration to be $\delta z_\mathrm{rs\_cal} \sim 0.08$.
        Both depend on the colour and the redshift regime of the used red-sequence model.

    \subsection{Visual inspection}\label{subsec:VisualInspection}
    
    For illustration we provide image cutouts around the respective centres (see Appendix \ref{app:SuppD}), and comment on each cluster candidate, as well as our respective findings.
    We order the targets in three different classes defined by a visual assessment and their statistical significance emerging from our analysis:\\
    
    \noindent\textbf{Robust Confirmation}
    \begin{itemize}
        \item \underline{1eRASS $J$033041.6$-$053608}: A very bright star lies in the vicinity (left of the image cutout), so we centred the FoV not on the X-ray centre, but off-setted to the right.
        The cutout shown in Fig. \ref{subfig:J0330} is centred on the optical centre found by \citet{Kluge2024}.
        The cluster lies at the left border of the FoV, therefore the rgb image does not fill the whole cutout radius of $\SI{3}{\arcmin}$.
        A visually clear over-density of red galaxies can be spotted at the supposed cluster centre.
        We find a statistical significant detection in $z - J$.
        \item \underline{1eRASS $J$091715.4$-$102343}: An over-density of similarly coloured galaxies close to the cluster centre is clearly visible.
        A possible candidate for a BCG can be identified.
        The posterior distribution shows a clear preference for a value inside the prior range, and the colour distribution changes significantly throughout radial bins with respect to the cluster centre.
        Compare with Fig. \ref{subfig:J0917}.
        \item \underline{1eRASS $J$094854.3$+$133740}: Possibly a high-redshift detection.
        The rgb image shows a group of three similar coloured galaxies close to the cluster centre proxy, as well as several fainter sources around them.
        Some readout issue left its imprint in the top right corner of the image, this region is masked for the analysis.
        Our algorithm picks up a clear cluster component in the colour distributions of all analysed colour indices and around both cluster centre proxies (see Fig. \ref{Fig:SIGNs}).
        The comparison with the red-sequence model gives a photometric redshift estimate at $z_{\mathrm{phot}} \sim 1.1$ (see Table \ref{Tab:iJOPTResults}), indicating a high-redshift object.
        Compare with Fig. \ref{subfig:J0948}.
        \item \underline{1eRASS $J$113751.5$+$072839}: A very clear and rich over-density of red galaxies, BCG candidate is visible, some elongated structure could be attributed.
        This is the most significant detection in our analysis.
        Compare with Fig. \ref{subfig:J1137}.
        \item \underline{1eRASS $J$124634.6$+$252236}: A visible concentration of faint red galaxies around the cluster centre proxy.
        The colour distribution changes significantly in radial bins towards the outskirts, our algorithm finds a significant cluster component in the distribution of the $i - J$ colour around both the optical and the X-ray centre.
        Compare with Fig. \ref{subfig:J1246}.
        \item \underline{1eRASS $J$140945.2$-$130101}: The target which was observed with both CAHA/Omega2000 and WST/3KK.
        A bright red galaxy visible close to the cluster centres, surrounded by several similarly coloured, fainter galaxies.
        The WST/3KK image is slightly deeper (see Table \ref{Tab:TargetListIQ}), there an over-density of red galaxies is clearly visible.
        For both instruments, around both cluster centre proxies, in both the $i - z$ and the $i - J$ colour our algorithm finds a significant detection.
        Compare with Fig. \ref{subfig:J1409} and Fig. \ref{subfig:J1409_WST}.
        \item \underline{1eRASS $J$115512.7$+$125759}: Some foreground structures in the top left corner of the cutout, but an additional over-density of redder galaxies can be seen at slightly higher declination with respect to the two cluster centres.
        A more suitable candidate for the BCG could be identified, closer to the cluster centres.
        The posterior distributions for all three parameters show a clear peak within in the prior ranges in both the $i - z$ and $i - J$ colour.
        Compare with Fig. \ref{subfig:J1155}.
        \item \underline{1eRASS $J$121051.6$+$315520}: A very clear over-density of red galaxies can be seen around the cluster centre proxies.
        No $i$-band available for this target, so we only analysed it in $z - J$, but our algorithm finds a significant detection around the optical centre for this colour index.
        Compare with Fig. \ref{subfig:J1210}.
    \end{itemize}
    \noindent\textbf{Tentative}
        \begin{itemize}
            \item \underline{1eRASS $J$025044.2$-$044309}: An apparent colour over-density close to the centre proxy, but the overall colour distribution in different radial bins only shows an additional, small component in $i - z$.
            In $i - J$ and $z - J$, the colour distribution does not change significantly in different radial bins.
            Compare with Fig. \ref{subfig:J0250}.
            \item \underline{1eRASS $J$031350.5$-$000546}: A group of galaxies close to the cluster centre proxy, with similar colours redder than the surrounding sources.
            No $i$-band data available, so this target was only analysed in $z - J$.
            Here the colour distribution does not show a significant change in radial bins and is not considered to be a significant detection.
            Compare with Fig. \ref{subfig:J0313}.
            \item \underline{1eRASS $J$043901.1$-$022852}: A compact group of three faint, but similarly coloured sources sit just at the cluster centre proxy, but with a rather under-dense region around them.
            Picked up as a significant detection by our algorithm both in $i - z$ and $i - J$.
            Compare with Fig. \ref{subfig:J0439}.
            \item \underline{1eRASS $J$102148.4$+$225133}: A group of faint, similarly red coloured galaxies are visible a bit off-centre in the cutout.
            However, these sources were cut from the analysis because of their large photometric uncertainty.
            This target was not picked up as a significant detection by our algorithm, but could be a potential candidate for deeper observations.
            Compare with Fig. \ref{subfig:J1021}.
            \item \underline{1eRASS $J$133801.7$+$175957}: No obvious over-density of red galaxies close to the cluster centre is visible.
            One bright red galaxy sits in the centre, with some very faint sources surrounding it, which however have a different colour.
            Otherwise it is quite empty around the cluster centres.
            Our algorithm picked up some signal in $i - z$ and $i - J$, but runs into the prior boundaries for both the mean colour $\mu$ and the width of the red-sequence $\sigma$.
            Compare with Fig. \ref{subfig:J1338}.
            \item \underline{1eRASS $J$142000.6$+$095651}: There is one slightly brighter red galaxy visible, close to some fainter red galaxies in the vicinity of the cluster centre proxies.
            Our algorithm finds a significant detection around the optical centre in the $i - J$ colour.
            Compare with Fig. \ref{subfig:J1420}.
            \item \underline{1eRASS $J$051425.6$-$093653}: Many red galaxies located in the central region, a BCG candidate is identifiable.
            The radial colour distribution shows some changes towards outer bins, but our algorithm only picks up some signal in the $i - J$ colour, and runs into the prior boundaries for all three parameters $\mu$, $\sigma$, $A$.
            Especially for the amplitude, the posterior runs into the upper boundary of the prior, indicating either a very rich cluster, or a fit for the field colour distribution.
            Compare with Fig. \ref{subfig:J0514}.
            Due to missing dither steps and the survey border, the $z$-band image (green channel) does not show any data in a horizontal and in a vertical strip, as well as a very noisy and bright strip at the top border of the horizontal strip without data, hence the different colour towards the lower edge of the image.
            The $J$-band image (red channel) has a a noisy vertical strip in centre of the cutout (defect in a read-out channel), it shows here as a red vertical strip in the rgb-image.
        \end{itemize}
    \noindent\textbf{Poor}
        \begin{itemize}
            \item \underline{1eRASS $J$042710.8$-$155324}: There is no over-density close to the cluster centre proxies visible.
            The target was picked up as a significant detection by our algorithm for the colour index $i - J$ around both the optical and the X-ray centre, as well as for the colour index $i - z$ around the optical centre.
            However, the found cluster member candidates are not concentrated around cluster centre proxies.
            Furthermore, the X-ray contours for this target suggest that the X-ray signal stems in fact from an AGN contaminant.
            Compare with Fig. \ref{subfig:J0427}.
            \item \underline{1eRASS $J$053425.5$-$183444}: There is no visible over-density anywhere in the image cutout, a high chance of being a contaminant according to \citet{Kluge2024} ($\mathrm{PCONT} = 0.92$).
            It was detected as a significant (SIGN $= 3.8$) over-density with our algorithm in $i - z$ around the optical cluster centre, but not very concentrated around the centre proxy.
            The rgb image overlaid with the X-ray contours suggests a point source contamination.
            Compare with Fig. \ref{subfig:J0534}.
            \item \underline{1eRASS $J$085742.5$-$054517}: There is no visible over-density anywhere in the image cutout, but it is crowded and has many bright stars in the FoV.
            The colour distribution does not change significantly in radial bins around the cluster centre proxies.
            The posterior distribution covers the whole prior range in all colour indices and around all centres.
            Compare with Fig. \ref{subfig:J0857}.
            \item \underline{1eRASS $J$133333.8$+$062920}: No clear concentration of red galaxies close to the cluster centre proxies, or anywhere in the FoV.
            The posterior distribution fills the whole prior range.
            The blue point source close to the centre of the X-ray contour is likely the cause of the strong X-ray signal.
            Compare with Fig. \ref{subfig:J1333}.
            \item \underline{1eRASS $J$145552.2$-$030618}: No over-density in red galaxies close to the cluster centre proxies, contrastively the central region appears rather empty besides one red source.
            Our algorithm does not pick up any signal.
            Furthermore, there is no obvious candidate for the origin of the X-ray signal visible.
            This could hint towards a very high redshift cluster, or could be due to background fluctuations as described in \citet{Seppi2022}.
            Compare with Fig. \ref{subfig:J1455}.
        \end{itemize}


\section{Discussion}\label{chap:Discussion}

    We highlight some implications and limitations of our results in the following subsections.

    \subsection{Confirmation of only a fraction of the sub-sample}

        As mentioned above, we only find two significant detections in $z - J$ (1eRASS $J$033041.6$-$053608 and 1eRASS $J$094854.3$+$133740).
        For both targets, the posterior distribution of the mean colour $\mu$ runs into the lower prior boundary, indicating an attempted fit for the field colour distribution.
        Thus the photometric redshift estimation as explained above changes from an actual estimate into a lower boundary.
        1eRASS $J$033041.6$-$053608 was not found to be a significant detection neither in $i - z$, nor in $i - J$, while 1eRASS $J$094854.3$+$133740 showed a significant additional component red-wards of the field colour distribution for both colour indices.
        For the latter, both the analysis in $i - z$ and in $i -J$ give a consistent photometric redshift estimation at $z>1$ (see Table \ref{Tab:iJOPTResults} and Table \ref{Tab:izOPTResults}).
        However, it is notable that
        \begin{enumerate}[1)]
            \item we do not find again any significant high-redshift ($z \sim 1.1$) targets in $z - J$.
            \item we can only confirm a fraction of the targets in our selected sub-sample.
        \end{enumerate}
        We did reach the required depth of $J = \SI{21.4}{mag}$ with a S/N of $10 \sigma$, thus we would expect to find cluster member galaxies up until a redshift of $z\sim1.1$.
        Given the fact that we can only confirm a fraction of the galaxy cluster candidates at $z \leq 1.1$, none in $z - J$, leaves three possible implications (ordered from least to most probable from our perspective):
        \begin{enumerate}[i)]
            \item  Some of the selected targets are not real clusters, but are in fact contaminants in the original eROSITA/eRASS1 cluster catalogue.
            \item  We detect up-scattered lower-redshift objects, but not down-scattered high-$z$ objects.
            \item  The assumed required depth which corresponds to $M^* + 0.5$ at $z\sim1.1$ is not deep enough to pick up a significant number of the cluster member galaxies.
        \end{enumerate}
        Regarding i), some of the targets in our sub-sample are highly likely to be contaminants (e.g. 1eRASS $J$042710.8$-$155324, 1eRASS $J$053425.5$-$183444, 1eRASS $J$133801.7$+$175957) as also concluded from visual inspection and overlaying cluster member candidates and X-ray contours\footnote{\href{https://erass-cluster-inspector.com/index.php}{https://erass-cluster-inspector.com/index.php}}.
        We argue that this is at least one of the likely reasons why we confirm only a fraction of our target list.
        Concerning ii), the scatter of photometric redshift estimates for true redshifts \mbox{$z\gtrsim 0.9$} is expected to become larger due to increasing photometric uncertainties.
        Thus, given our findings it seems plausible that we find again up-scattered lower-redshift objects, but not down-scattered higher-redshift objects.
        Whether or not the observed depth mentioned in iii) is not deep enough is difficult to answer as it depends on how much one trusts our target luminosity $L^*$ and if that is sufficient.
        Our choice of $M^* + 0.5$ corresponds to $0.6L^*$, where $L^*$ is the characteristic luminosity defining the transition between the exponential part of the  modified Schechter luminosity function (see e.g. \citealt{Schechter1976, To2020, Thai2023}) (at bright luminosity) and the power law part (at faint luminosity).
        For comparison, \citet{Kluge2024} chose a threshold of $0.2L^*$ which would correspond to $M^* + 1.7$.
        This is based on \citet{Rykoff2012} choosing this threshold to minimise their richness scatter and increases the richness estimate by $65\%$ compared to a limit of $0.4L^*$ ($M^* + 1$).
        Thus, an aspired depth of $M^* + 0.5$ might just not be enough to pick up enough cluster member galaxies to find a significant over-density with our algorithm.
        We think this is a valid concern, requiring deeper NIR data to be investigated further.\\
        The contamination in the eRASS1 cluster catalogue is quantified with the contamination estimator $\mathrm{PCONT}$, as defined in \cite{Ghirardini2024} and \cite{Kluge2020}.
        According to this estimator, in our subsample, one would expect a mean contamination of $\overline{\mathrm{PCONT}}(\mathrm{BEST\_Z}>0.9, {\rm EXT\_LIKE>5}) = 16.4\%$.
        Our analysis shows a contamination fraction of $22.2\%$ in our subsample, which can be understood as an upper limit on the true contamination, keeping in mind scenarios i) and ii) described above.
        Although these numbers can technically be considered consistent, we suspect that $\mathrm{PCONT}$ underestimates the true contamination at the highest redshifts.
        That is, because the flag IN\_ZVLIM \citep[${\rm S/N}>10$ for a 0.4\,L$^*$ galaxy,][]{Kluge2020} was applied to the comparison samples of randoms and AGNs during the computation of $\mathrm{PCONT}$, but not to the cluster sample.
        This artificially decreases the ratio of comparison samples over identified clusters, hence reducing $\mathrm{PCONT}$.
        We note that we did not pre-clean our subsample using the flag IN\_ZVLIM, which would have eliminated 80\% of our targets, because we specifically aimed at the highest-$z$ clusters.
        According to \citet{Seppi2022}, we would expect $\sim 80\%$ real clusters (here: $75\%$), $\sim 20\%$ point source contamination (here: $15\%$), and $\sim 1\%$ background fluctuations (here: $5\%$) with an $\mathrm{EXT\_LIKE}>5$.
        
    \subsection{Photometric redshifts}
        We approximate a calibration of the red-sequence model including $J$-band data against a single, optically selected, galaxy cluster with a confirmed spectroscopic redshift \citep{Gonzales2019}.
        To zeroth order, the effect of this calibration is a constant offset \citep{Kluge2024}, and thus this approach constitutes a reasonable calibration attempt, given the lack of additional clusters in our sample with spectroscopic reference redshifts.
        However, calibrating only against one single spectroscopic redshift carries the risk of introducing bias.
        Furthermore, the calibration of the red-sequence may have a redshift dependence.
        The literature value for the spectroscopic redshift of MOO $J$0319$-$0025 lies at $z = 1.194$, so at the upper end of the redshift regime explored in this work.\\
        We do not observe a good agreement when comparing the photometric redshifts obtained by analysing the $i - z$ colour index in this work and the ones obtained in \citet{Kluge2024}.
        However, when looking at Fig. C.1 of \citet{Kluge2024}, the spectroscopic source sample to calibrate the red-sequence model becomes quite scarce at $z>0.9$.
        Additionally, the photometric uncertainty when including $w1$ of NEOWISE \citep{NEOWISE2014} becomes comparatively large.
        Thus, it can be expected to have a larger scatter in the photometric redshift estimation at $z>0.9$ (compare also to Fig. 17 of \citealt{Kluge2024}).\\
        Nonetheless, it is worth noting the improvement of the photometric errors when including $J$-band imaging for sources at this redshift (compare to Fig. \ref{Fig:CompareRedSeqs}), highlighting the necessity of wide and deep $J$-band (or other, redder NIR band) observations to improve cluster photometric redshift estimation and confirmation at redshifts beyond $z>1$.
        We quantify the improvement of the photometric errors by comparing the median of the relative errors on the measured colours of probable cluster members found in both $i-z$ and $i-J$.
        We assert an improvement in colour measurement precision from $8\%$ to $4\%$ when including the $J$-band data, showing a more precise measurement for $i - J$ than for $i - z$.
        This could help in future analyses where better spectroscopic-redshift-calibration-samples are available.

    \subsection{Cluster centre proxies}
        We test the eRASS1 X-ray centres as reported in \citet{Bulbul2024} (see Appendix \ref{app:SuppD}), as well as optical centres found in \citet{Kluge2024} (see Table \ref{Tab:iJOPTResults} and Appendix \ref{app:SuppD}).
        In principle, the optical centre is a more suitable centre proxy for our purposes because most of the times it is located at the position of a high probability, bright cluster member galaxy found in earlier works \citep{Kluge2024}.
        The colour of the central galaxy has an influence of the extracted mean colour, due to the designed radial weighting of the sources' colours with the weighting function $f(A,R_i)$ in Eq. (\ref{Eq:p_cGr}).
        Thus, if the cluster centre is placed on an actual cluster member galaxy, the measured colour is weighted accurately by $f(A,R_i)$ to find the correct red-sequence colour.
        If the cluster centre is wrongly placed on a random field galaxy, the measured colour does not correspond to the correct red-sequence colour and is mistakenly weighted strongly by $f(A,R_i)$.
        This effect is amplified if the region around the optical centre is scarcely populated after $5\sigma$ S/N and colour cuts and suppressed if there are many cluster member candidates (i.e. sources of similar colour).
        When looking at Fig. \ref{Fig:SIGNs}, it is evident that this effect does not play a role for most of our significant detections.
        Similarly, this can also happen when using the X-ray centre proxy, if the X-ray centre is misaligned with the optical cluster centre, i.e. the sources closest to the X-ray centre are not necessarily cluster member galaxies.
        This can be observed when comparing the results using optical vs X-ray centre proxy for 1eRASS $J$140945.2$-$130101 in Fig. \ref{Fig:SIGNs} where the two centre proxies are shifted by $\SI{1.2}{arcmin}$ in RA (compare with Table \ref{Tab:TargetList}).\\
    \subsection{Implications for cluster cosmology}
        The cosmological constraints derived from the weak lensing calibrated number counts of eROSITA-selected clusters \citep{Ghirardini2024, Artis2024, Artis2025} used a sample limited to $z_\mathrm{BEST}<0.8$ and $\mathrm{EXT\_LIKE} > 6$ (see also \citealt{Bulbul2024} for the definition of the "cosmology sample").
        This is clearly below the redshift range explored here.
        As shown by \citet{Kluge2024}, at $z<0.8$ the $i-z$ ($r-i$) colours still provide a reliable photometric redshift estimate and confirmation tool.
        Thus, our findings do not impact the cosmological constraints derived from eROSITA selected clusters, but help further validate the upper redshift limit imposed in the selection of the cosmology sample.
        They also underscore clearly the necessity for deep and wide near-infrared data to reliably confirm clusters at $z\gtrsim0.9$, as well as to measure their redshifts.

\section{Conclusion}\label{chap:Conclusion}
    In this work we conducted targeted NIR follow-up observations of 17  high-redshift galaxy cluster candidate from eRASS1 and the optically selected high-redshift cluster  MOO $J$0319$-$0025 at the Calar Alto Observatory.
    We supplemented the target list with additional data of 4 more targets taken with the Wendelstein Wide Field Imager, of which we also observed one at Calar Alto (see Appendix \ref{app:CAHAvsWST}), totalling to 20 different targets analysed in this work.
    We performed the prevalent reduction steps using the end-to-end pipeline \texttt{THELI} for the Calar Alto observations and the pipeline by \citet{Obermeier2020} for the Wendelstein observations, and calibrated the $J$-band magnitudes against 2MASS.
    We complemented the acquired NIR data with the \emph{griz}-bands of the LSDR10, performed a forced-photometry analysis and detected radial colour over-densities around both X-ray and optical cluster centres in $i - J$, $z - J$, and $i - z$ for a subsample of our selected targets.
    For these targets we were able to estimate a photometric redshift.
    For $2$ of the $20$ different targets no $i$-band data is available from the Legacy Survey at this point in time, thus they were solely analysed in $z - J$.
    We only obtain two significant detections in $z - J$, therefore our main analysis is based on the $i - J$ colour.
    Out of the $18$ cluster candidates with $i$-band data, we can confirm $12$ targets.
    If we combine the colour indices $i - z$ and $i - J$, we can confirm 14 out of 18 targets which have an extended likelihood parameter of $\mathrm{EXT\_LIKE}>5$.
    This would correspond to a $22.2\%$ contaminant fraction at the high-redshift end of the eRASS1 galaxy cluster catalogue.
    If we combine the visual inspection classes good and tentative to a confirmed fraction, we would obtain a lenient estimate for the contamination fraction of $25\%$.
    This would give support to our choices in the used cuts in detections significance ($\mathrm{SIGN}>3\sigma$).\\
    Aside from this apparent contamination fraction, we deduce that we are affected by two relevant effects:
    Firstly, the target depth of $M^* + 0.5$ limits our ability to detect high-$z$ targets with sufficient S/N to extract a significant colour over-density.
    Secondly, the zero point uncertainty when converting 2MASS Vega to AB magnitudes, as well as the uncertainty of the applied re-calibration of the red-sequence model including the $J$-band observations are likely to introduce additional systematic uncertainties in the photometric-redshift estimates, which we currently cannot quantify given the lack of additional systems with spectroscopic redshifts.\\
    In conclusion, when comparing to earlier works, we would argue that $M^* + 0.5$ ($J=\SI{21.4}{mag}$) is not deep enough to pick up enough cluster member galaxies to detect a significant over-density of galaxies of similar colour.
    Both issues will soon be overcome by ESA's new space mission {\it Euclid}.
    Its Wide Survey will push significantly deeper, with a limiting magnitude of $J=24.4$ (for $5\sigma$ point sources, \citealt{Schirmer2022}), allowing for the detection of lower-luminosity cluster members.
    In addition, thanks to its large sky coverage it will overlap with existing spectroscopic follow-up programmes of high-$z$ clusters (e.g. \citealt{Balogh2017}), improving empirical calibrations of red-sequence models.
    The capabilities of {\it Euclid} in the follow-up and confirmation of high-redshift galaxy clusters is already glanced at in \citet{Klein2025}, using data from the {\it Euclid} Quick Release 1 (Q1).


\bibliographystyle{aa} 
\bibliography{biblio} 



\begin{appendix}

    \section{Acknowledgements}

    We would like to thank the whole staff of the Calar Alto Observatory for their support while operating the telescope and making this experience possible.
    Special thanks to Ana Guijarro, Gilles Bergond, and Alba Fernández-Martín for the quick and informative exchanges we had also after our visits at the site.\\
    We thank Dr.~Lindsey Bleem 
    for her contributions to the preparation of the observing proposal.\\
    VG acknowledges the financial contribution from the contracts Prin-MUR 2022 supported by Next Generation EU (M4.C2.1.1, n.20227RNLY3 {\it The concordance cosmological model: stress-tests with galaxy clusters}).\\
    The Innsbruck authors acknowledge support provided by the Austrian Research Promotion Agency (FFG) and the Federal Ministry of the Republic of  Austria for Climate Action, Environment, Mobility, Innovation and Technology (BMK) via the Austrian Space Applications Programme with grant numbers 899537, 900565, and 911971.
    This project has received funding from the European Union's Horizon 2020 research and innovation programme under grant agreement No 101004719. This material reflects only the authors views and the Commission is not liable for any use that may be made of the information contained therein.
    This research has made use of the Spanish Virtual Observatory (http://svo.cab.inta-csic.es) supported by the MINECO/FEDER through grant AyA2017-84089.7
    The Wendelstein 2m telescope project was funded by the Bavarian government and by the German Federal government through a common funding process. Part of the 2m instrumentation including some of the upgrades for the infrastructure were funded by the Cluster of Excellence ``Origin of the Universe" of the German Science foundation DFG.
    This work is based on data from eROSITA, the soft X-ray instrument aboard SRG, a joint Russian-German science mission supported by the Russian Space Agency (Roskosmos), in the interests of the Russian Academy of Sciences represented by its Space Research Institute (IKI), and the Deutsches Zentrum f{\"{u}}r Luft und Raumfahrt (DLR). The SRG spacecraft was built by Lavochkin Association (NPOL) and its subcontractors and is operated by NPOL with support from the Max Planck Institute for Extraterrestrial Physics (MPE).
    The development and construction of the eROSITA X-ray instrument was led by MPE, with contributions from the Dr. Karl Remeis Observatory Bamberg \& ECAP (FAU Erlangen-Nuernberg), the University of Hamburg Observatory, the Leibniz Institute for Astrophysics Potsdam (AIP), and the Institute for Astronomy and Astrophysics of the University of T{\"{u}}bingen, with the support of DLR and the Max Planck Society. The Argelander Institute for Astronomy of the University of Bonn and the Ludwig Maximilians Universit{\"{a}}t Munich also participated in the science preparation for eROSITA.
    The eROSITA data shown here were processed using the eSASS software system developed by the German eROSITA consortium.
    E. Bulbul, A. Liu, and X. Zhang acknowledge financial support from the European Research Council (ERC) Consolidator Grant under the European Union’s Horizon 2020 research and innovation program (grant agreement CoG DarkQuest No 101002585). VG acknowledges the financial contribution from the contracts Prin-MUR 2022 supported by Next Generation EU (M4.C2.1.1, n.20227RNLY3 {\it The concordance cosmological model: stress-tests with galaxy clusters}).
    The Legacy Surveys consist of three individual and complementary projects: the Dark Energy Camera Legacy Survey (DECaLS; Proposal ID \#2014B-0404; PIs: David Schlegel and Arjun Dey), the Beijing-Arizona Sky Survey (BASS; NOAO Prop. ID \#2015A-0801; PIs: Zhou Xu and Xiaohui Fan), and the Mayall z-band Legacy Survey (MzLS; Prop. ID \#2016A-0453; PI: Arjun Dey). DECaLS, BASS and MzLS together include data obtained, respectively, at the Blanco telescope, Cerro Tololo Inter-American Observatory, NSF’s NOIRLab; the Bok telescope, Steward Observatory, University of Arizona; and the Mayall telescope, Kitt Peak National Observatory, NOIRLab. Pipeline processing and analyses of the data were supported by NOIRLab and the Lawrence Berkeley National Laboratory (LBNL). The Legacy Surveys project is honored to be permitted to conduct astronomical research on Iolkam Du’ag (Kitt Peak), a mountain with particular significance to the Tohono O’odham Nation.
    NOIRLab is operated by the Association of Universities for Research in Astronomy (AURA) under a cooperative agreement with the National Science Foundation. LBNL is managed by the Regents of the University of California under contract to the U.S. Department of Energy.
    This project used data obtained with the Dark Energy Camera (DECam), which was constructed by the Dark Energy Survey (DES) collaboration. Funding for the DES Projects has been provided by the U.S. Department of Energy, the U.S. National Science Foundation, the Ministry of Science and Education of Spain, the Science and Technology Facilities Council of the United Kingdom, the Higher Education Funding Council for England, the National Center for Supercomputing Applications at the University of Illinois at Urbana-Champaign, the Kavli Institute of Cosmological Physics at the University of Chicago, Center for Cosmology and Astro-Particle Physics at the Ohio State University, the Mitchell Institute for Fundamental Physics and Astronomy at Texas A\&M University, Financiadora de Estudos e Projetos, Fundacao Carlos Chagas Filho de Amparo, Financiadora de Estudos e Projetos, Fundacao Carlos Chagas Filho de Amparo a Pesquisa do Estado do Rio de Janeiro, Conselho Nacional de Desenvolvimento Cientifico e Tecnologico and the Ministerio da Ciencia, Tecnologia e Inovacao, the Deutsche Forschungsgemeinschaft and the Collaborating Institutions in the Dark Energy Survey. The Collaborating Institutions are Argonne National Laboratory, the University of California at Santa Cruz, the University of Cambridge, Centro de Investigaciones Energeticas, Medioambientales y Tecnologicas-Madrid, the University of Chicago, University College London, the DES-Brazil Consortium, the University of Edinburgh, the Eidgenossische Technische Hochschule (ETH) Zurich, Fermi National Accelerator Laboratory, the University of Illinois at Urbana-Champaign, the Institut de Ciencies de l’Espai (IEEC/CSIC), the Institut de Fisica d’Altes Energies, Lawrence Berkeley National Laboratory, the Ludwig Maximilians Universitat Munchen and the associated Excellence Cluster Universe, the University of Michigan, NSF’s NOIRLab, the University of Nottingham, the Ohio State University, the University of Pennsylvania, the University of Portsmouth, SLAC National Accelerator Laboratory, Stanford University, the University of Sussex, and Texas A\&M University.
    BASS is a key project of the Telescope Access Program (TAP), which has been funded by the Nationaal Astronomical Observatories of China, the Chinese Academy of Sciences (the Strategic Priority Research Program “The Emergence of Cosmological Structures” Grant \# XDB09000000), and the Special Fund for Astronomy from the Ministry of Finance. The BASS is also supported by the External Cooperation Program of Chinese Academy of Sciences (Grant \# 114A11KYSB20160057), and Chinese National Natural Science Foundation (Grant \# 12120101003, \# 11433005).
    The Legacy Survey team makes use of data products from the Near-Earth Object Wide-field Infrared Survey Explorer (NEOWISE), which is a project of the Jet Propulsion Laboratory/California Institute of Technology. NEOWISE is funded by the National Aeronautics and Space Administration.
    The Legacy Surveys imaging of the DESI footprint is supported by the Director, Office of Science, Office of High Energy Physics of the U.S. Department of Energy under Contract No. DE-AC02-05CH1123, by the National Energy Research Scientific Computing Center, a DOE Office of Science User Facility under the same contract; and by the U.S. National Science Foundation, Division of Astronomical Sciences under Contract No. AST-0950945 to NOAO.

    \section{Mock Validation}\label{app:MockVal}

        \begin{figure}[H]\label{Fig:MockVal}
            \centering
            \includegraphics[width=\linewidth]{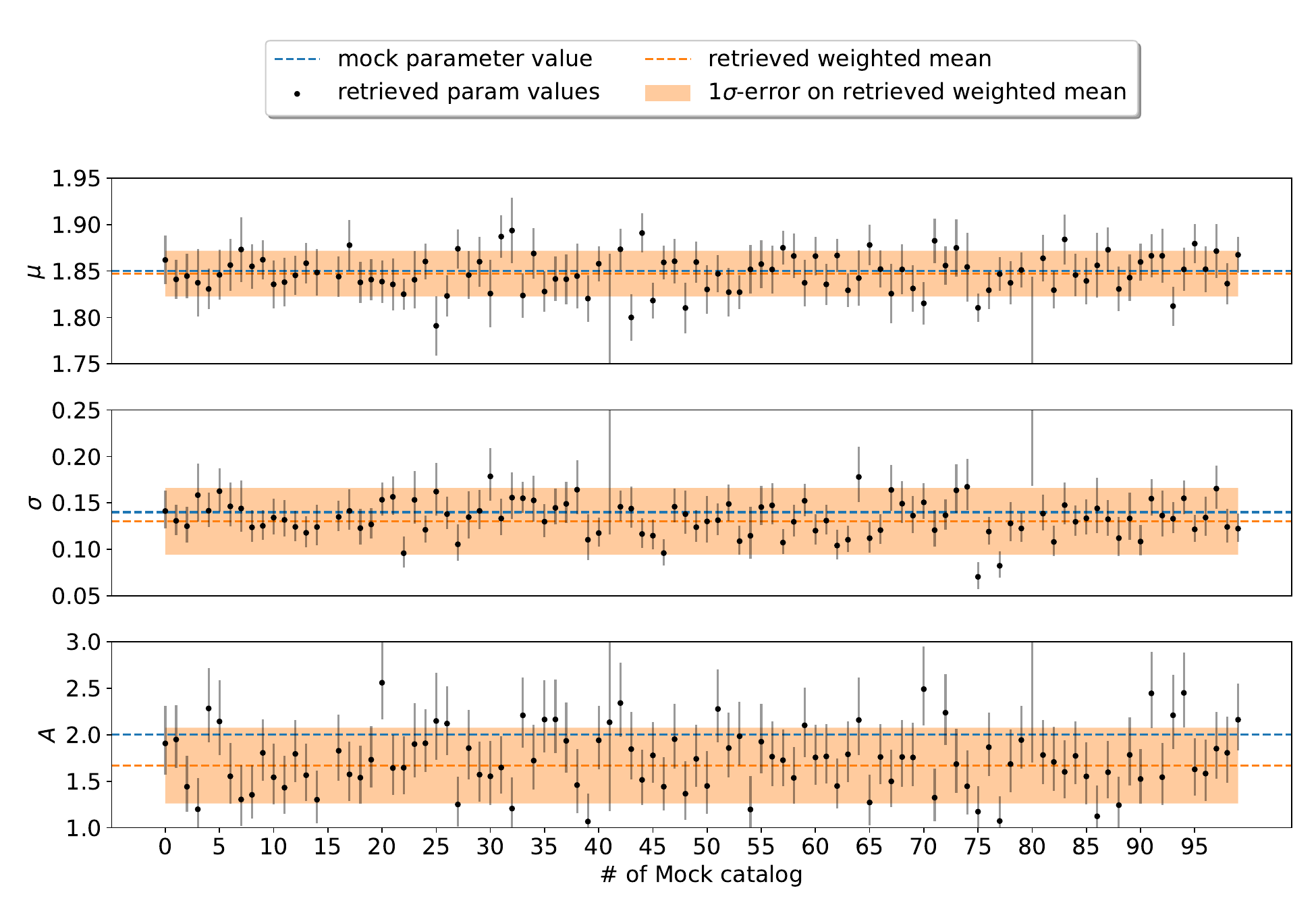}
            \caption{Retrieved parameter values for mean excess colour (top), its standard deviation (middle), and amplitude (bottom), as defined in Sect.\ref{Sec:RadColDist} over $100$ mock realizations. The blue dashed line shows the respective input mock parameter values. The black dots show the individual retrieved parameter values and their uncertainty ($68\%$ interval around the maximum likelihood value). The orange dashed line shows the mean parameter values of all $100$ mock realizations, weighted by their respective uncertainty. The orange shaded region shows the $1\sigma$ error on the weighted mean parameter values as computed in Eq. (\ref{Eq:MeanWeErr})}
            \label{Fig:MockVal}
        \end{figure}

        We validate the inference framework presented and used throughout this work on a series of mock catalogues.
        For this we define a set of mock parameter values \{$A$, $\mu$, $\sigma$\}$_{\mathrm{mock}}$ = \{$2.0$, $1.85$, $0.14$\}, for which we initialise both an instance of our weight function $f(A,R_i)$ (see Eq. \ref{Eq:Frac}) for a set of real angular separations $R_i$, as well as a field colour distribution $p_{\mathrm{field}}$ (see Sect.\ref{Sec:RadColDist}) learned from a set of real galaxy colours.
        These real data sets are the data vectors from 1eRASS $J113751.5$$+$072839.
        For every value $f_i = f(A,R_i)$ we draw randomly from a uniform distribution $u \sim \mathcal{U}(0,1)$.
        If $u < f_i$, the mock galaxy is classified as a cluster member galaxy and its colour is drawn from a normal distribution $\mathcal{N}(\mu_{\mathrm{mock}}, \sigma_{\mathrm{mock}})$.
        Contrarily, if $u \geq f_i$, the source is classified as random field galaxy and its colour is drawn from $p_{\mathrm{field}}$.
        This provides us with a data set of real angular separations, but artificial colours on which we can then test our algorithm.
        We create $100$ different mock catalogues to obtain a statistically more robust estimate of the accuracy and uncertainty of our algorithm (see Fig. \ref{Fig:MockVal}).
        We compute the mean value of the retrieved parameter value by inverse variance weighting and compute the error on the retrieved mean value as
        \begin{eqnarray}
            \Bar{x}_{\mathrm{wtd}} &=& \frac{\sum_i w_i x_i}{\sum_i w_i}\\
            \mathrm{Var}(x)_{\mathrm{wtd}} &=& \frac{\sum_i w_i \left( x_i - \Bar{x}_{\mathrm{wtd}} \right)^2}{\sum_i w_i},\label{Eq:MeanWeErr}
        \end{eqnarray}
        where $w_i = (1/\sigma_{x_i})^2$ with $\sigma_{x_i}$ being the error of $x_i$.
        The retrieved weighted mean values for the mean colour excess ($\mu$) and standard deviation ($\sigma$) agree very well with the input mock values.
        We thus conclude that we can retrieve unbiased estimates for both these parameter values.
        While the amplitude $A$ of the NFW-profile seems to be retrieved reasonably well within the error bars, it is noticeably more noisy and less accurate than the other two parameters, showing a bias towards lower amplitudes.

    \section{CAHA vs WST}\label{app:CAHAvsWST}

\begin{figure*}[ht]
            \centering
            \includegraphics[width=0.99\textwidth]{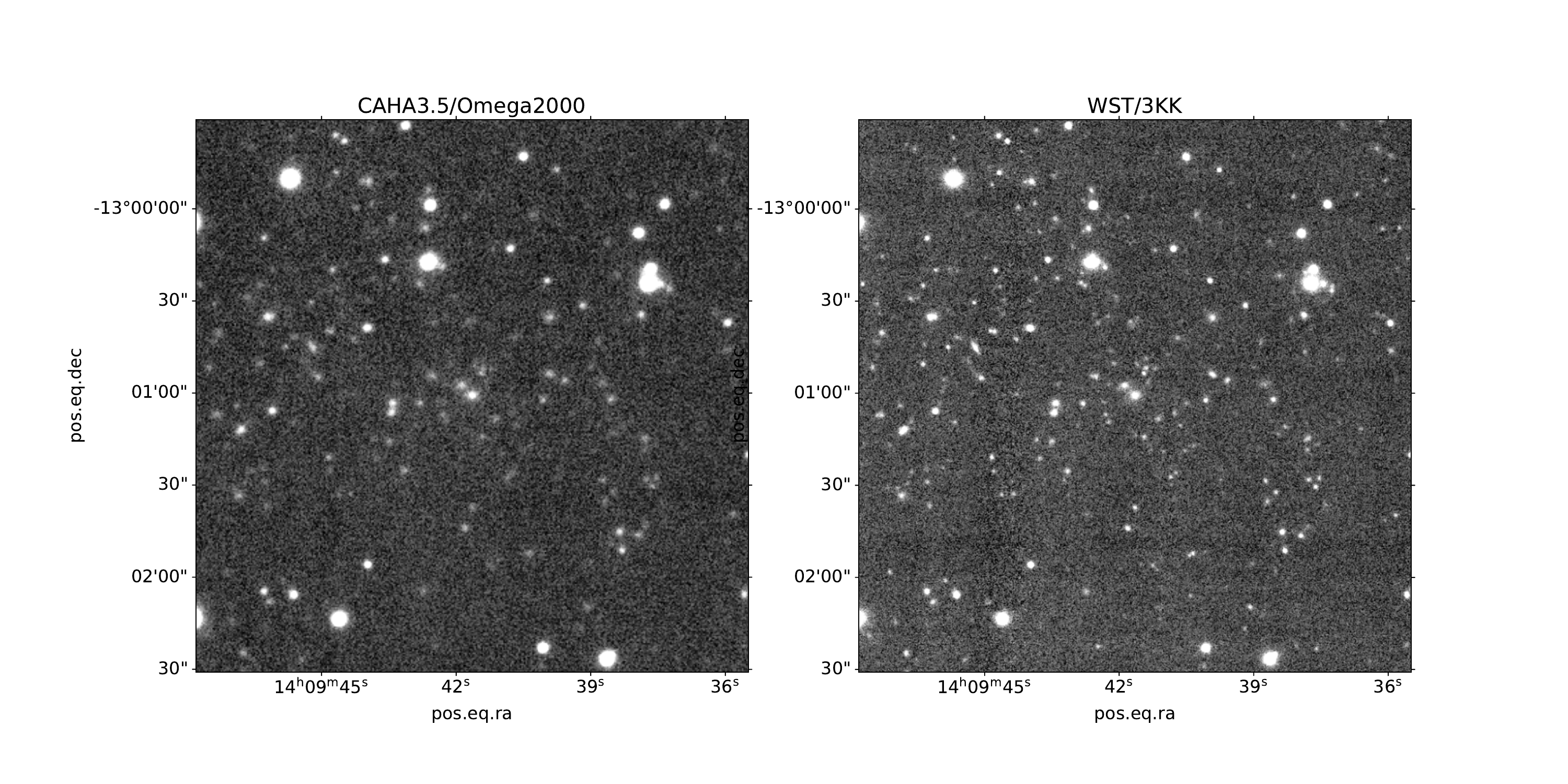}
            \caption{Direct comparison of the $J$-band imaging data obtained using the CAHA3.5/Omega2000 and the WST/3KK instruments. Note the clearer signal of the faint sources close to the centre, as well as the better seeing when comparing the stellar sources. For a quantitative comparison of the imaging data, please see Table \ref{Tab:TargetListIQ}. One should consider the fact that this target had the worst image quality in all the CAHA run, and the best in the WST run.}
            \label{Fig:CAHAvsWST_JBand}
        \end{figure*}

        In Fig. \ref{Fig:CAHAvsWST_JBand} we show the direct comparison of the imaging data between CAHA3.5/Omega2000 and WST/3KK to showcase the difference in depth and seeing obtained with the two different instruments.
        WST/3KK provides us with a better seeing and slightly deeper source detection, at the price of 6 times the integration time of CAHA3.5/Omega2000, accounting for the smaller mirror size at the Wendelstein observatory.
        Additionally, this target had the worst image quality in the CAHA run, and the best in the WST run.
        
    \section{Supplementary Data Section}\label{app:SuppD}

\begin{figure*}[htp] 
        \centering
        
        \vspace{.1cm} 
        \begin{subfigure}{0.4\textwidth}
            \includegraphics[width=\linewidth]{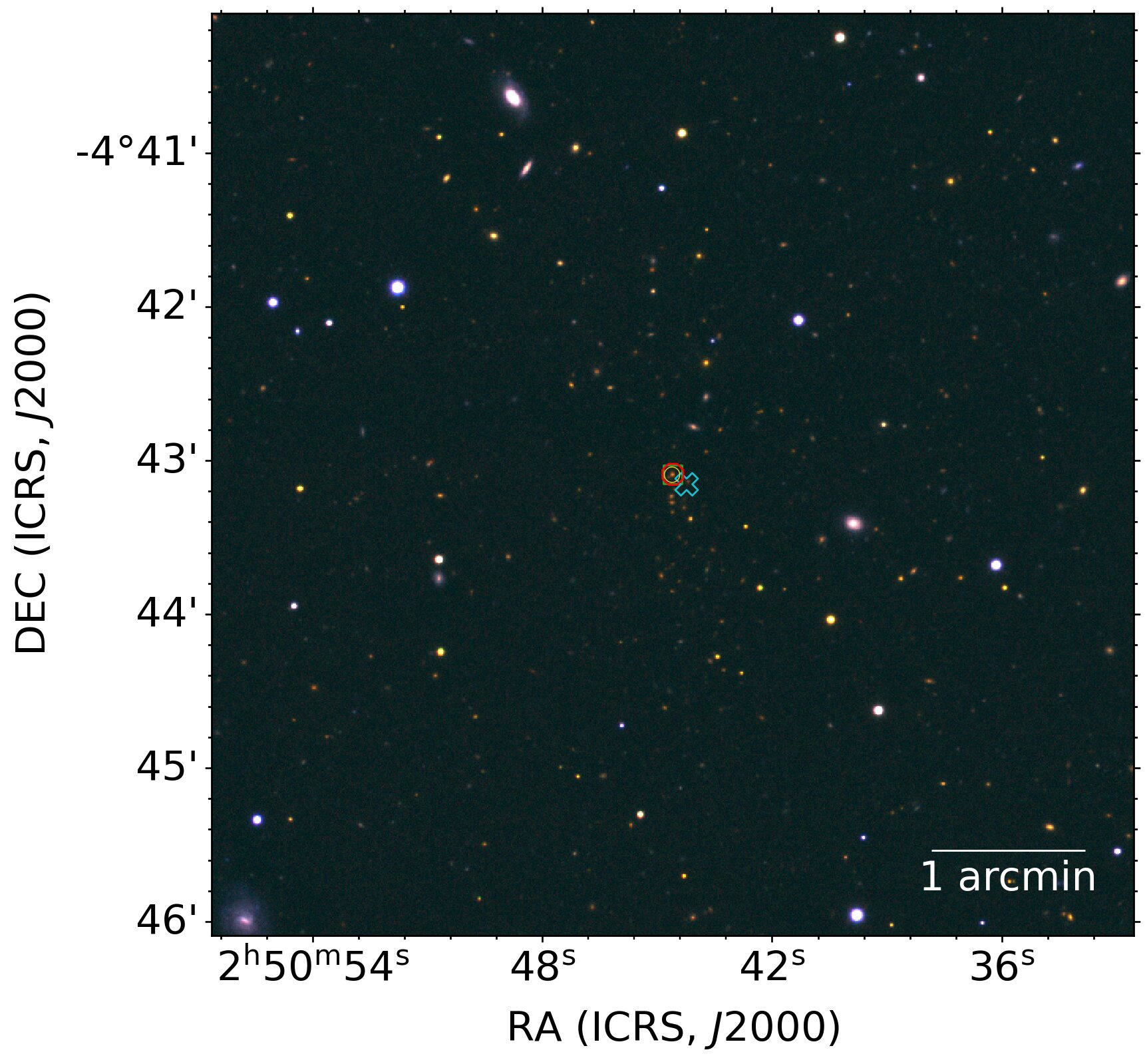}
            \caption{1eRASS J025044.2-044309, tentative}
            \label{subfig:J0250}
        \end{subfigure}
        \begin{subfigure}{0.4\textwidth}
            \includegraphics[width=\linewidth]{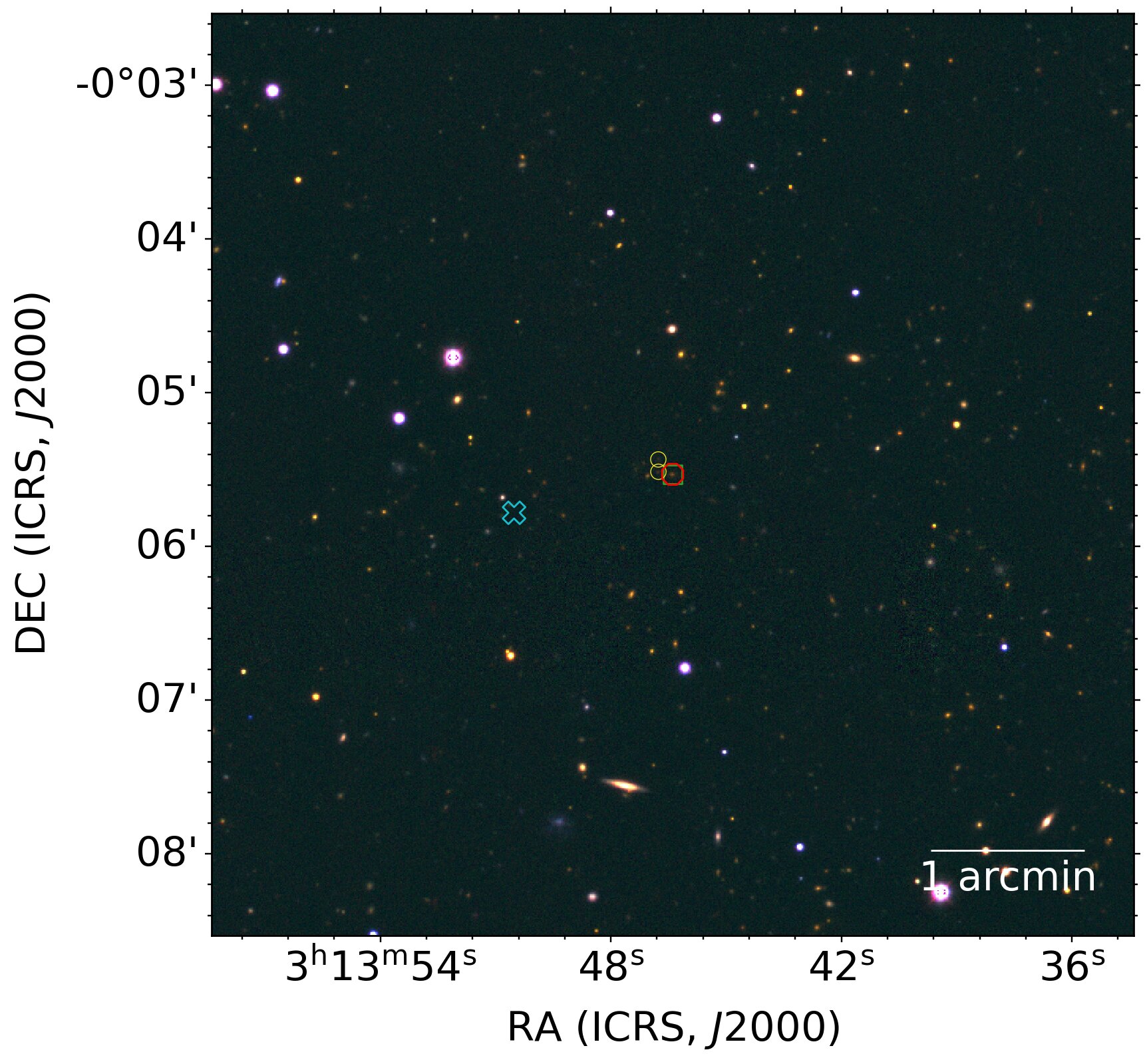}
            \caption{1eRASS J031350.5-000546, tentative}
            \label{subfig:J0313}
        \end{subfigure}
        
        \vspace{.1cm} 
        \begin{subfigure}{0.4\textwidth}
            \includegraphics[width=\linewidth]{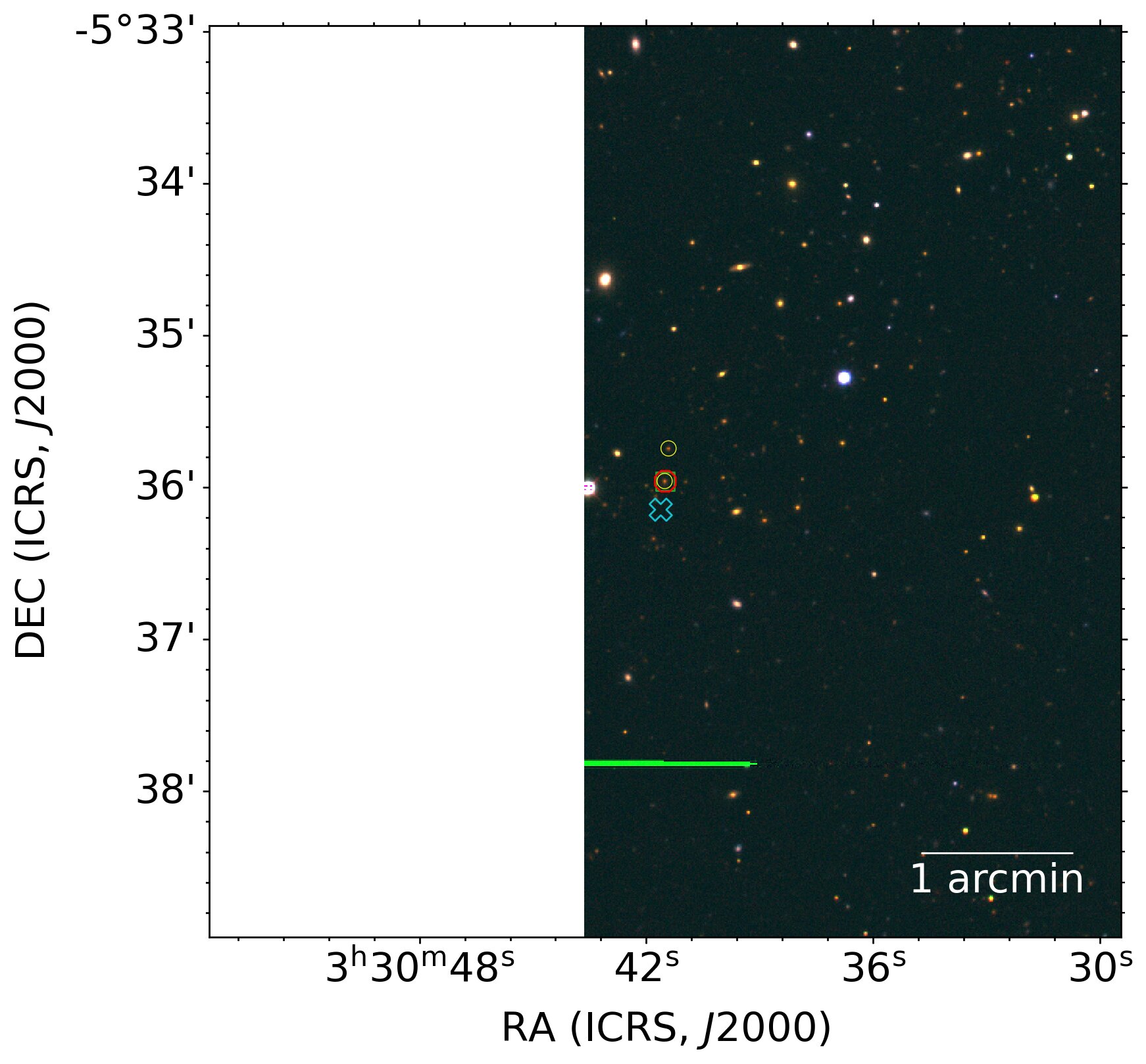}
            \caption{1eRASS J033041.6-053608, robust confirmation}
            \label{subfig:J0330}
        \end{subfigure}
        \begin{subfigure}{0.4\textwidth}
            \includegraphics[width=\linewidth]{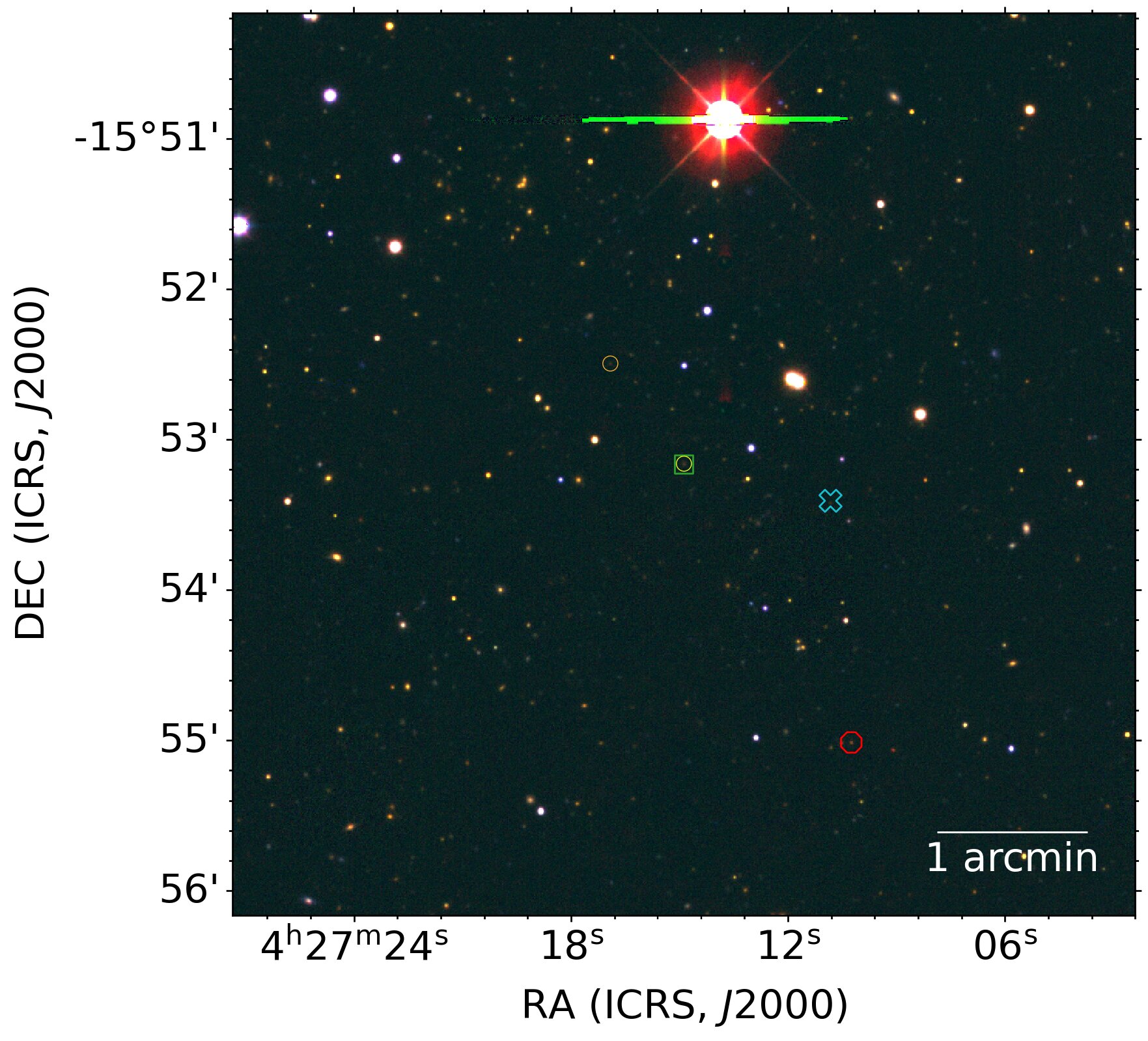}
            \caption{1eRASS J042710.8-155324, poor}
            \label{subfig:J0427}
        \end{subfigure}

        \vspace{.1cm} 
        \begin{subfigure}{0.4\textwidth}
            \includegraphics[width=\linewidth]{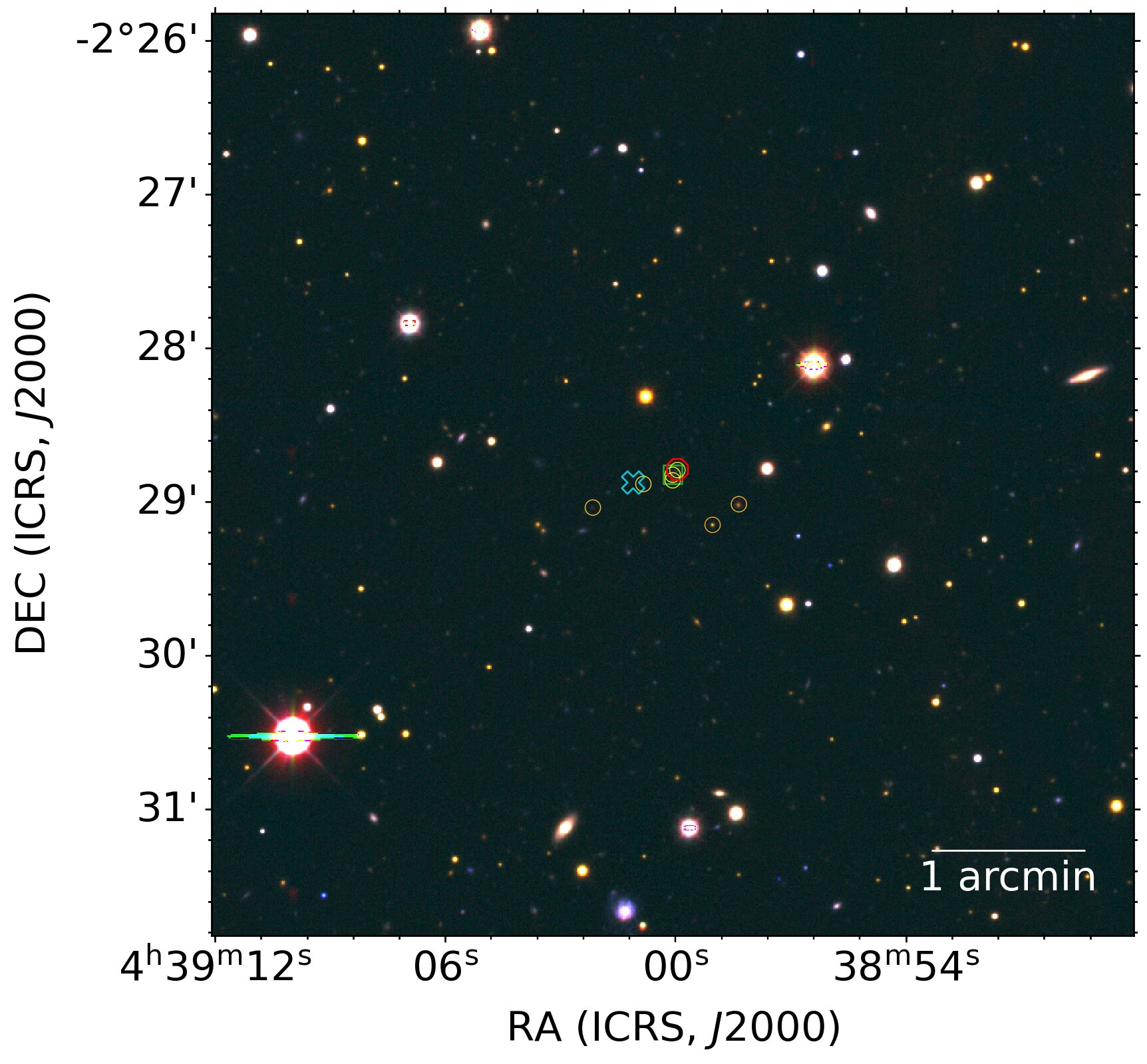}
            \caption{1eRASS J043901.1-022852, tentative}
            \label{subfig:J0439}
        \end{subfigure}
        \begin{subfigure}{0.4\textwidth}
            \includegraphics[width=\linewidth]{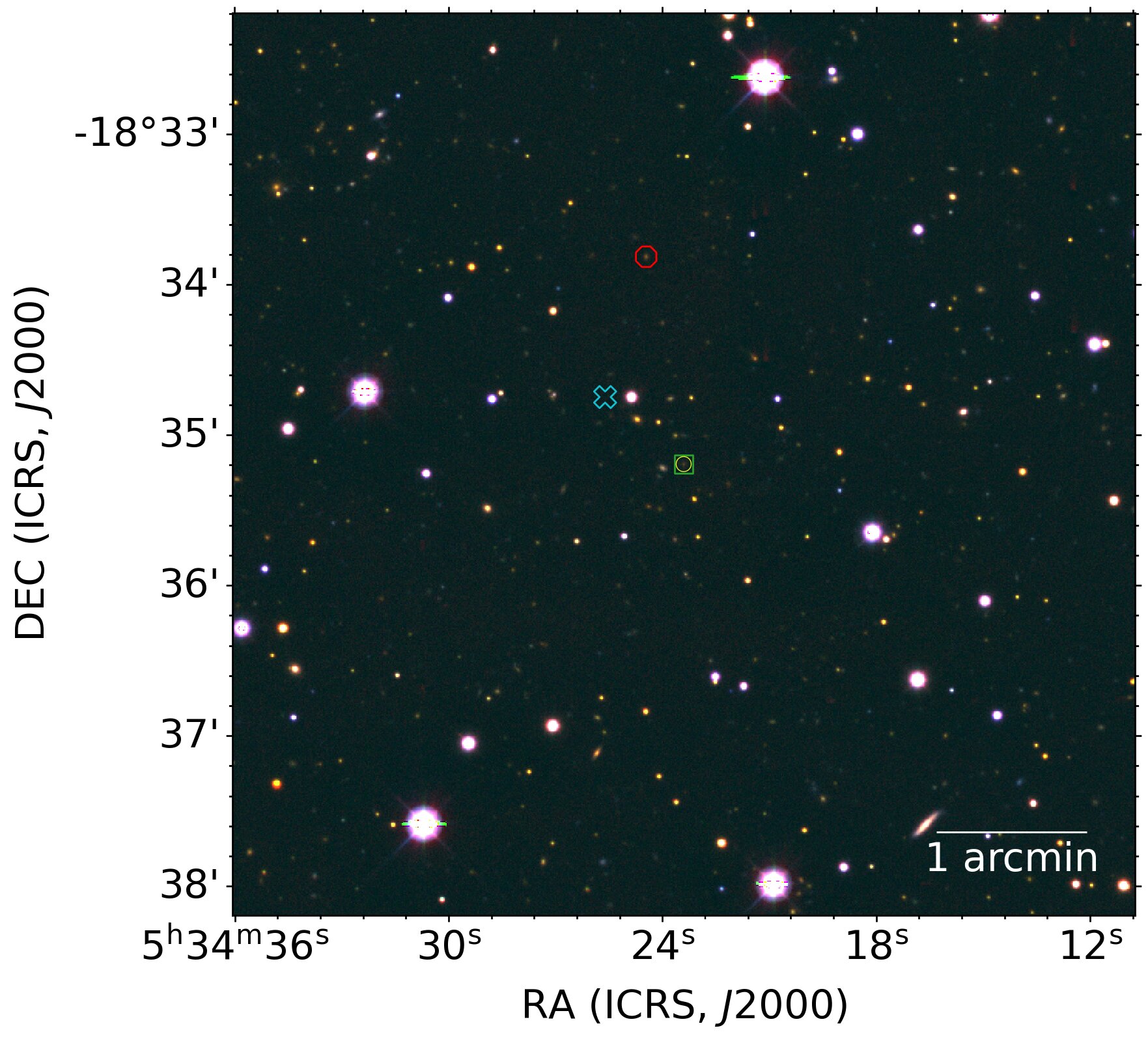}
            \caption{1eRASS J053425.5-183444, poor}
            \label{subfig:J0534}
        \end{subfigure}

    \caption{Fake rgb images (channels: r:$J$, g:$z$, b:$g$; $J$-band from CAHA/Omega2000, $z$- and $g$-band from the Legacy Survey) of the selected galaxy cluster candidates. We show the optical centre retrieved by \citet{Kluge2024} (green square), the X-ray centre (cyan cross), and a BCG candidate found in \citet{Kluge2024} (red octagon). We also include the classification of the candidate introduced in Sect. \ref{subsec:VisualInspection} in the caption, as well as the sources found to have a cluster member probability higher than $80\%$ (yellow circles).}
    \label{fig:RGBCutouts1}
    \end{figure*}

\begin{figure*}[htp] 
        \centering
        
        \vspace{.1cm} 
        
        \begin{subfigure}{0.4\textwidth}
            \includegraphics[width=\linewidth]{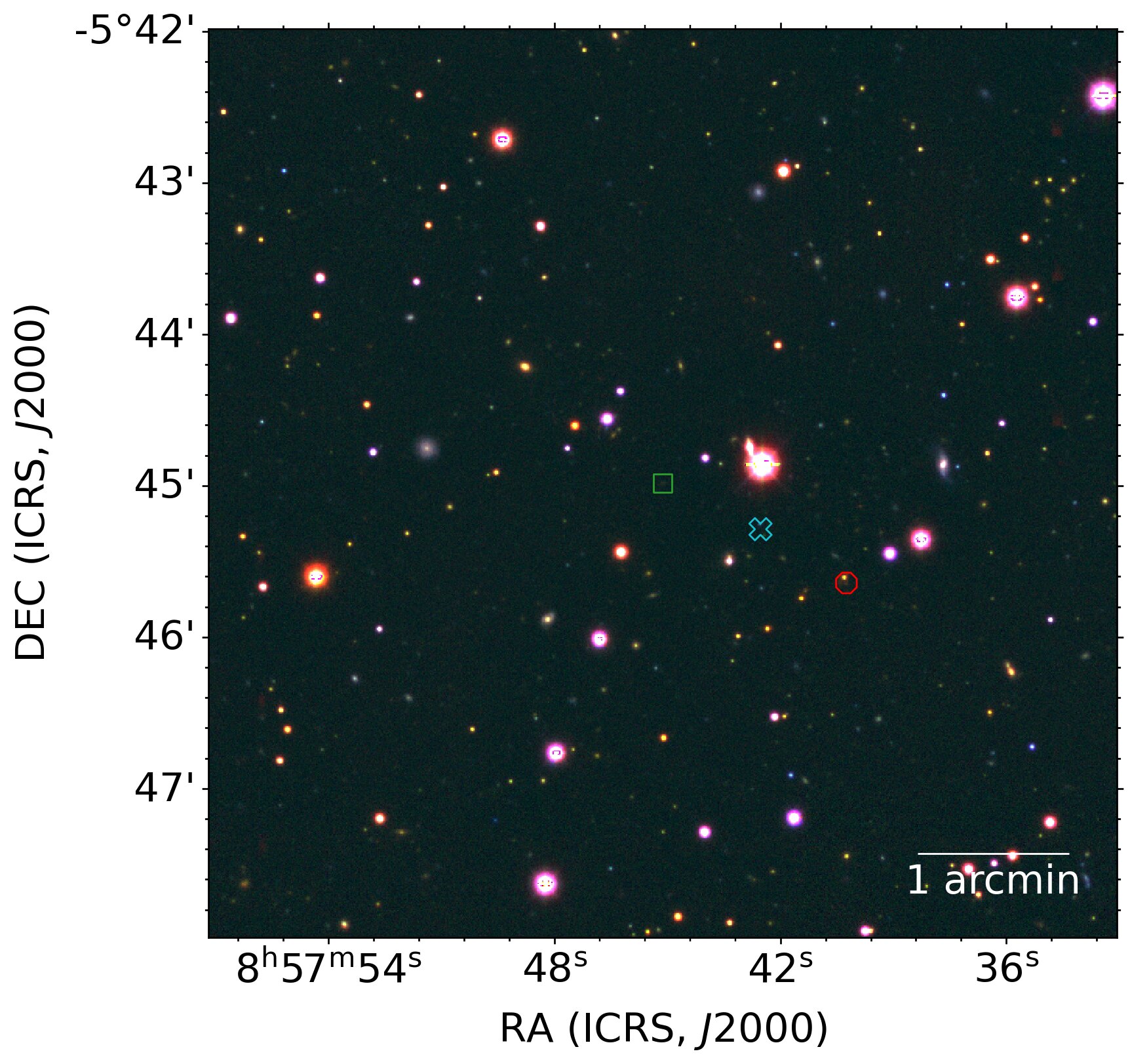}
            \caption{1eRASS J085742.5-054517, poor}
            \label{subfig:J0857}
        \end{subfigure}
        \begin{subfigure}{0.4\textwidth}
            \includegraphics[width=\linewidth]{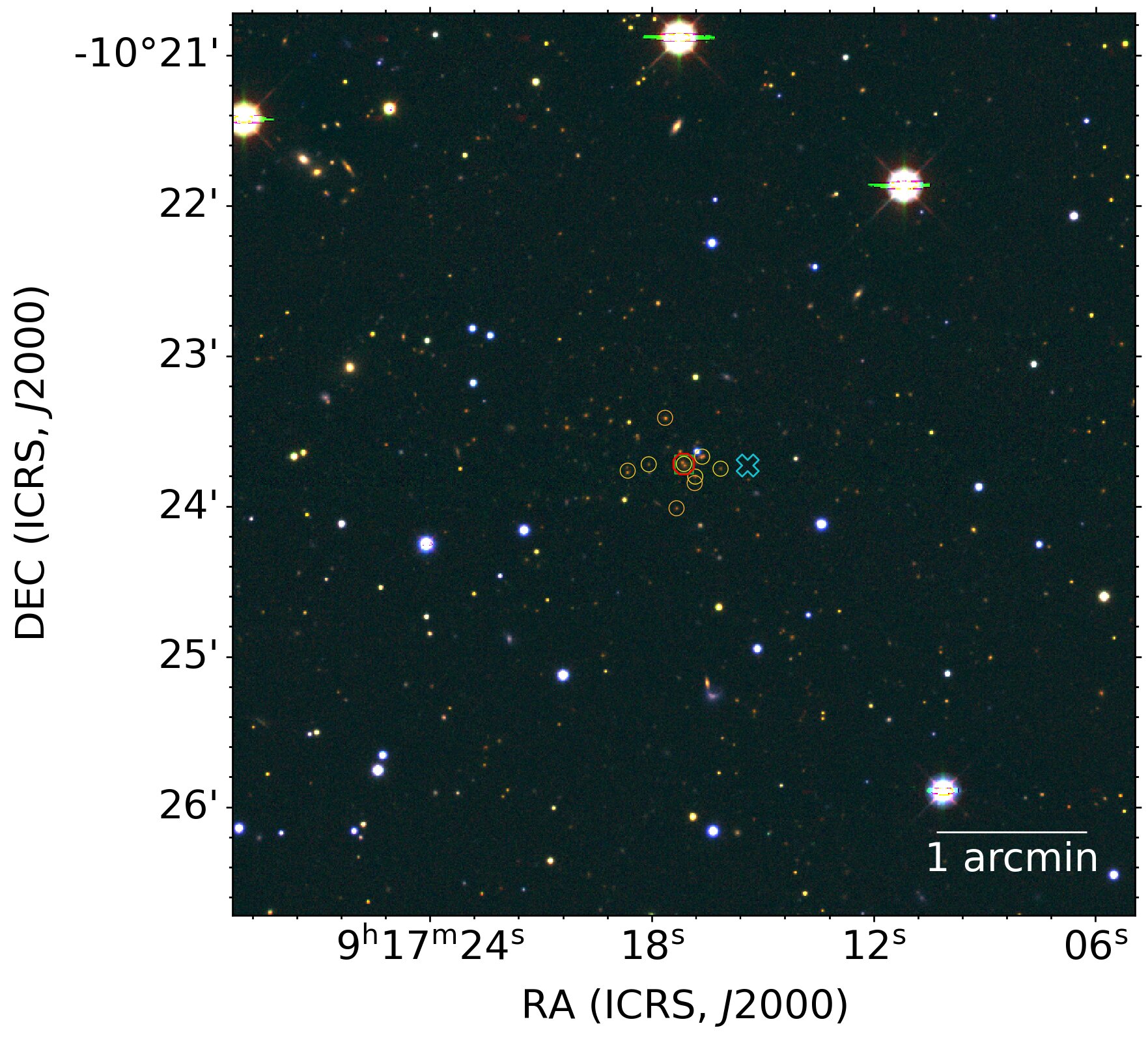}
            \caption{1eRASS J091715.4-102343, robust confirmation}
            \label{subfig:J0917}
        \end{subfigure}

        \vspace{.1cm} 
        \begin{subfigure}{0.4\textwidth}
            \includegraphics[width=\linewidth]{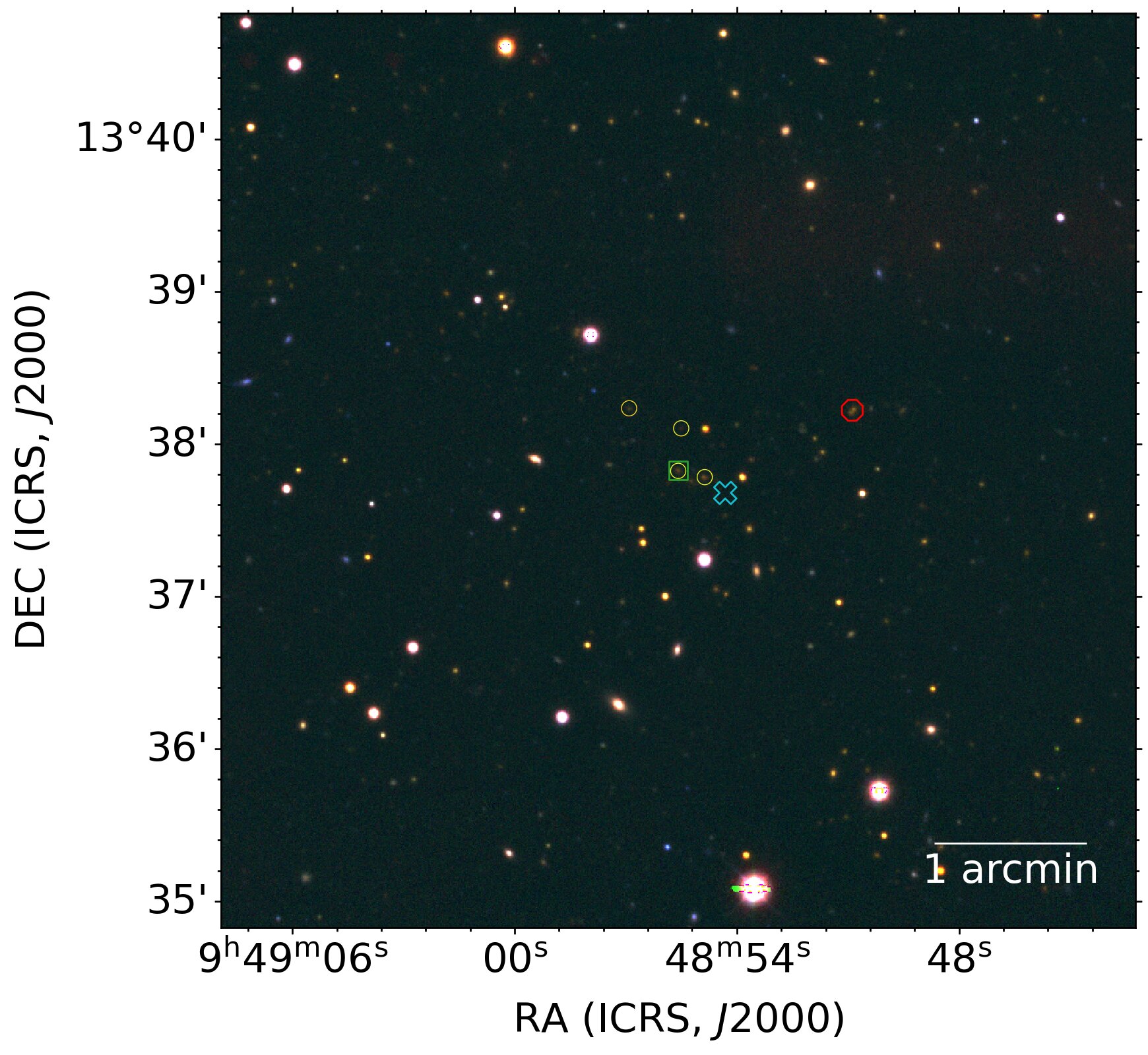}
            \caption{1eRASS J094854.3+133740, robust confirmation}
            \label{subfig:J0948}
        \end{subfigure}
        \begin{subfigure}{0.4\textwidth}
            \includegraphics[width=\linewidth]{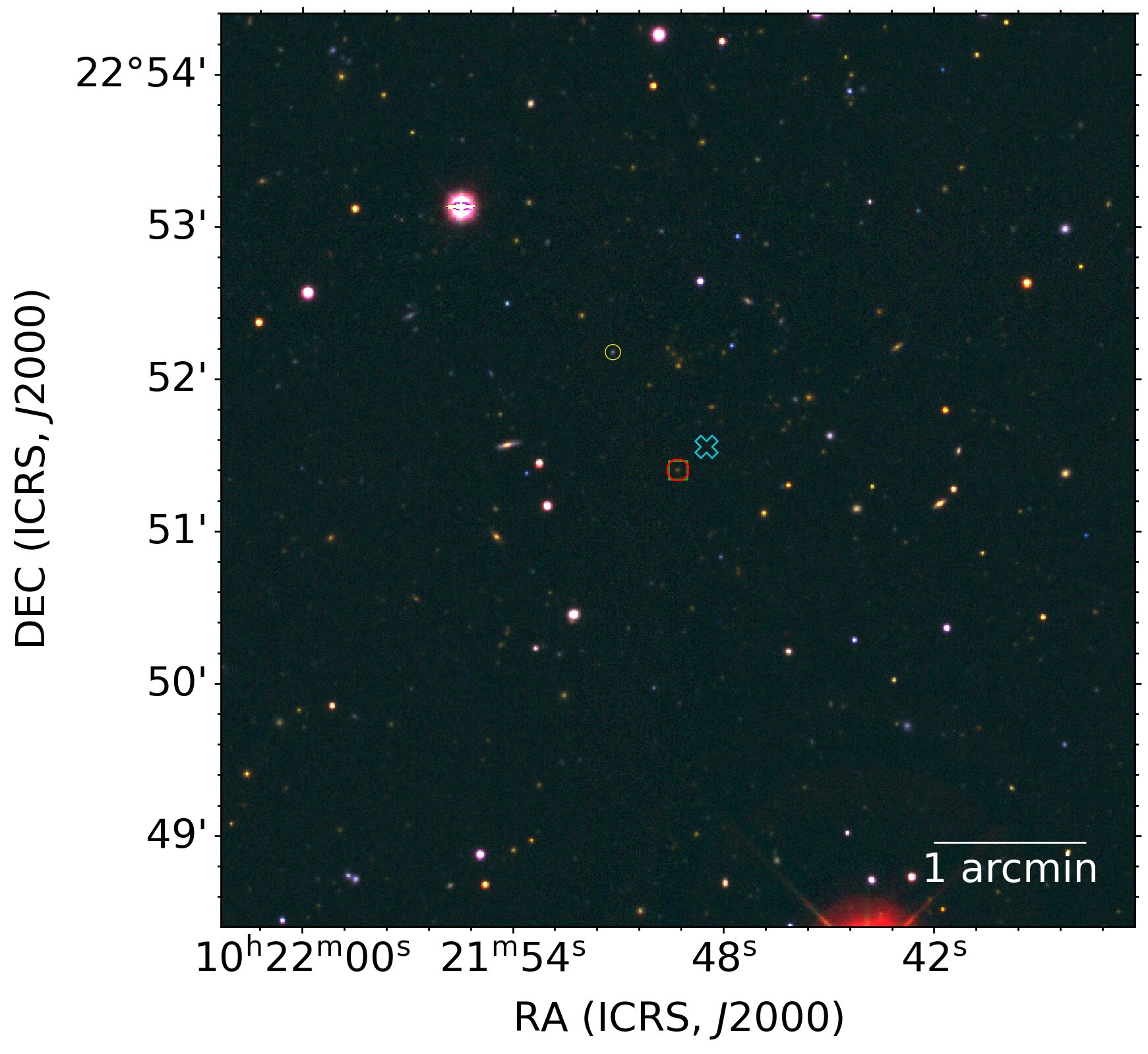}
            \caption{1eRASS J102148.4+225133, tentative}
            \label{subfig:J1021}
        \end{subfigure}
        
        \vspace{.1cm} 
        \begin{subfigure}{0.4\textwidth}
            \includegraphics[width=\linewidth]{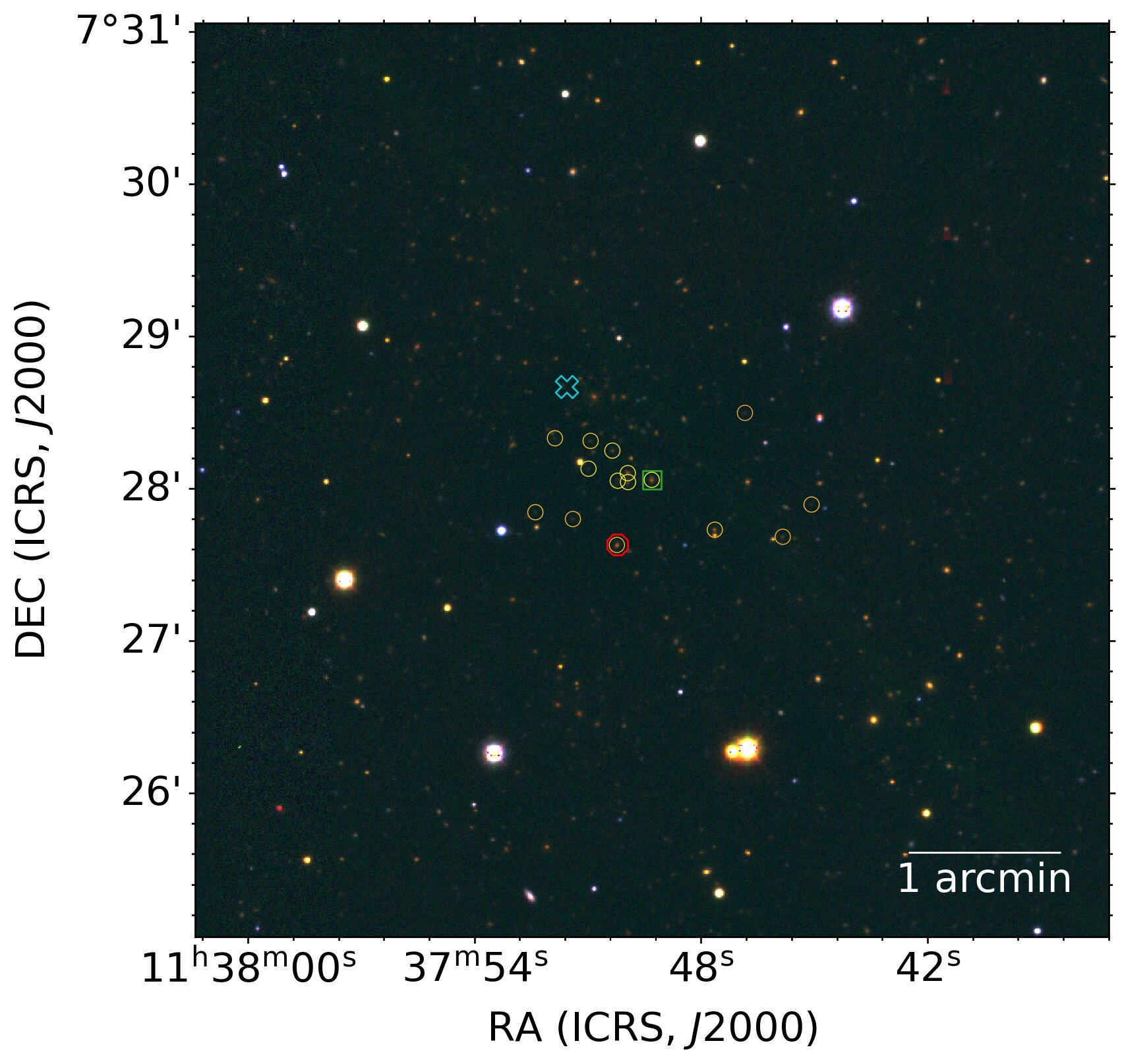}
            \caption{1eRASS J113751.5+072839, robust confirmation}
            \label{subfig:J1137}
        \end{subfigure}
        \begin{subfigure}{0.4\textwidth}
            \includegraphics[width=\linewidth]{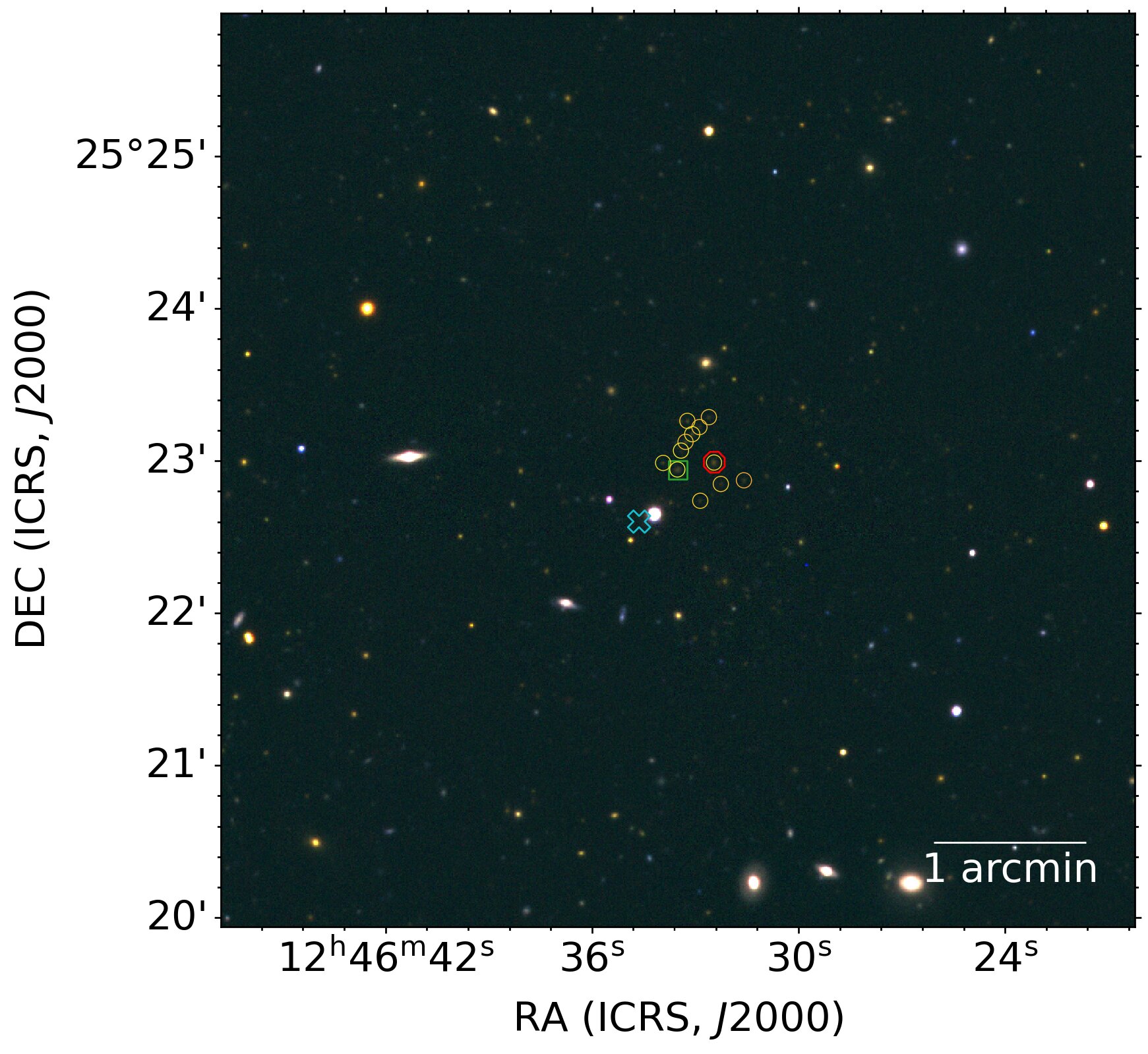}
            \caption{1eRASS J124634.6+252236, robust confirmation}
            \label{subfig:J1246}
        \end{subfigure}

    \caption{Fake rgb images (channels: r:$J$, g:$z$, b:$g$; $J$-band from CAHA/Omega2000, $z$- and $g$-band from the Legacy Survey) of the selected galaxy cluster candidates. We show the optical centre retrieved by \citet{Kluge2024} (green square), the X-ray centre (cyan cross), and a BCG candidate found in \citet{Kluge2024} (red octagon). We also include the classification of the candidate introduced in Sect. \ref{subsec:VisualInspection} in the caption, as well as the sources found to have a cluster member probability higher than $80\%$ (yellow circles).}
    \label{fig:RGBCutouts2}
    \end{figure*}

\begin{figure*}[htp] 
        \centering
        
        \vspace{.1cm} 
        \begin{subfigure}{0.4\textwidth}
            \includegraphics[width=\linewidth]{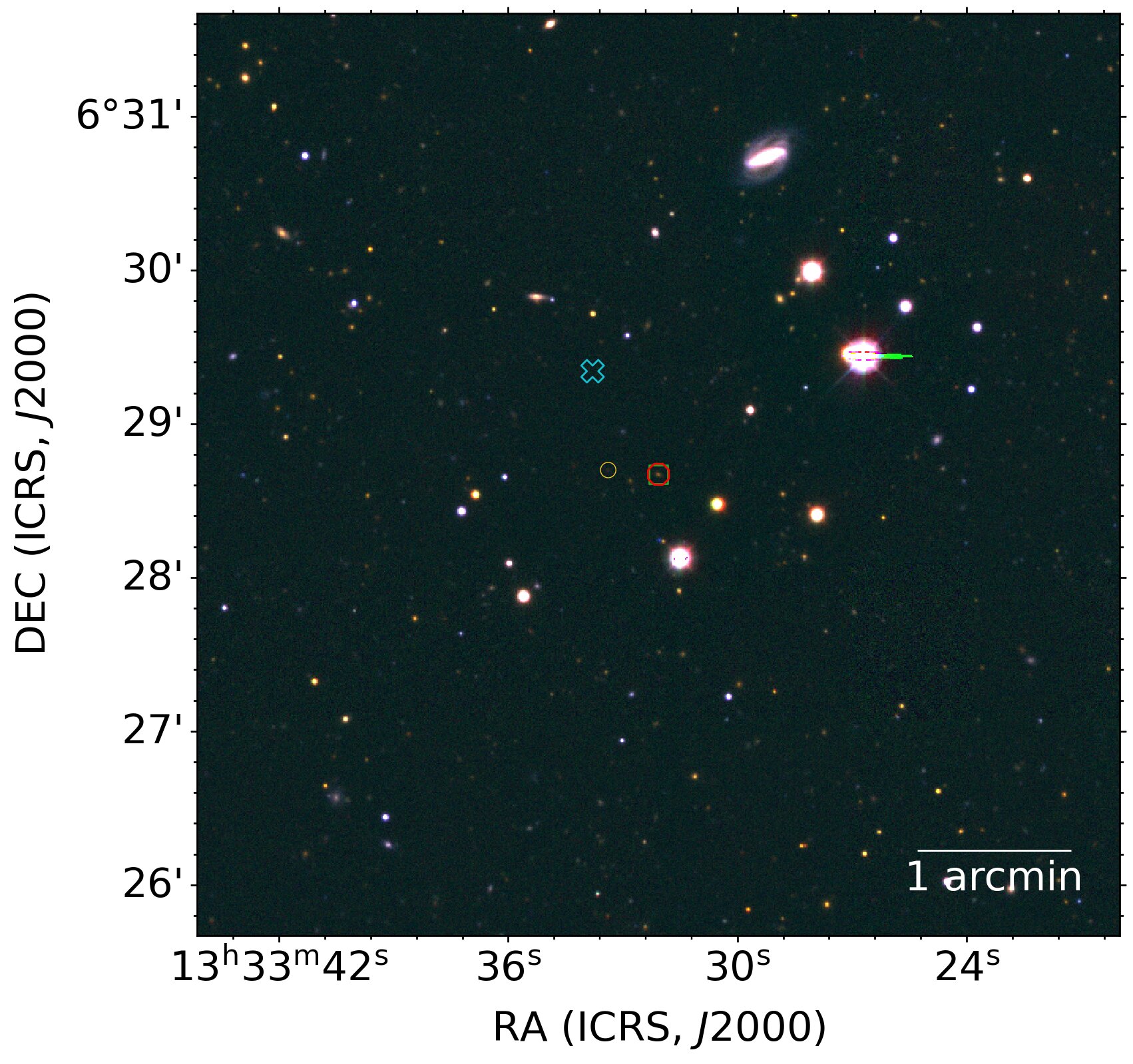}
            \caption{1eRASS J133333.8+062920, poor}
            \label{subfig:J1333}
        \end{subfigure}
        \begin{subfigure}{0.4\textwidth}
            \includegraphics[width=\linewidth]{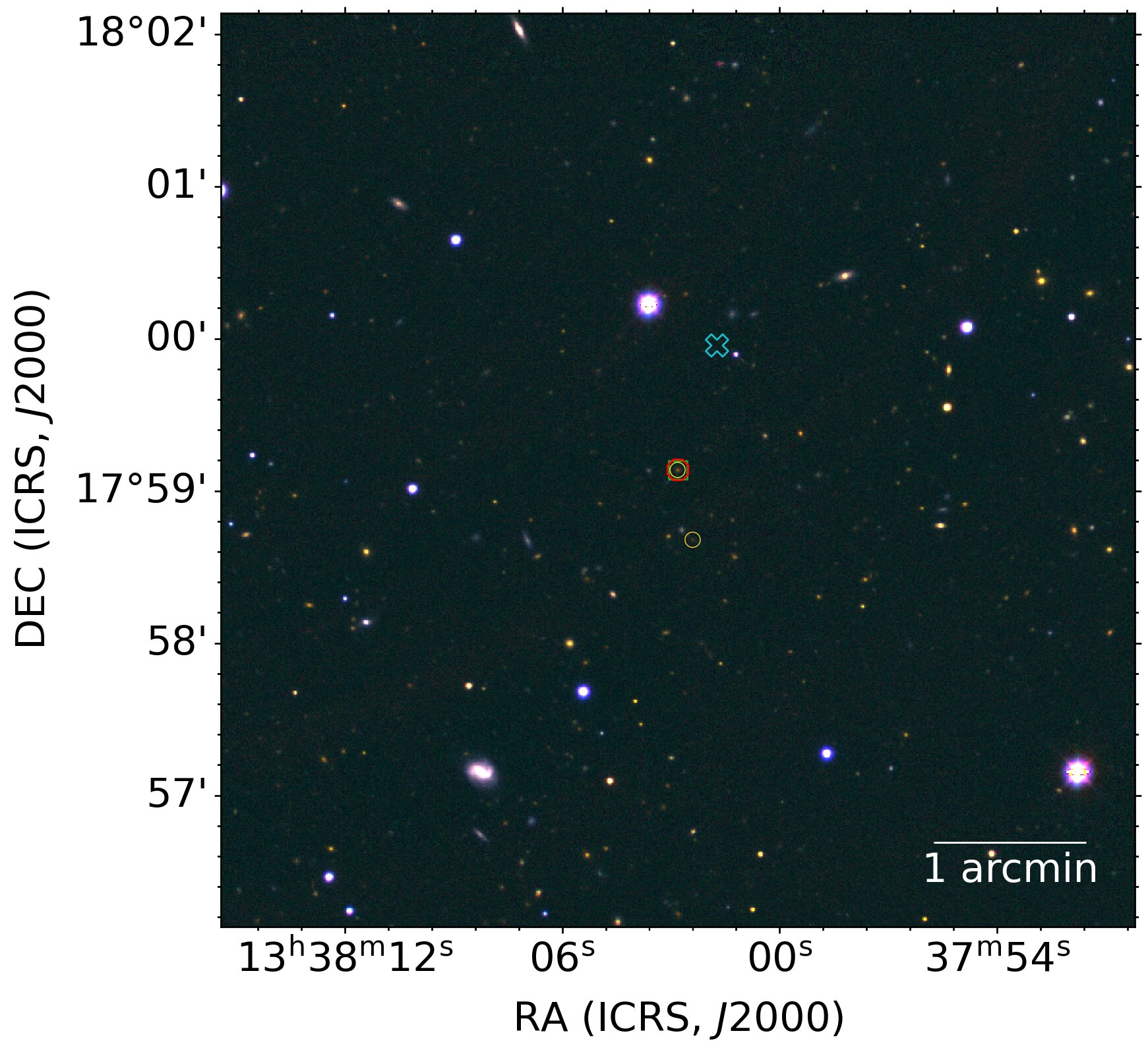}
            \caption{1eRASS J133801.7+175957, tentative}
            \label{subfig:J1338}
        \end{subfigure}
        
        \vspace{.1cm} 
        \begin{subfigure}{0.4\textwidth}
            \includegraphics[width=\linewidth]{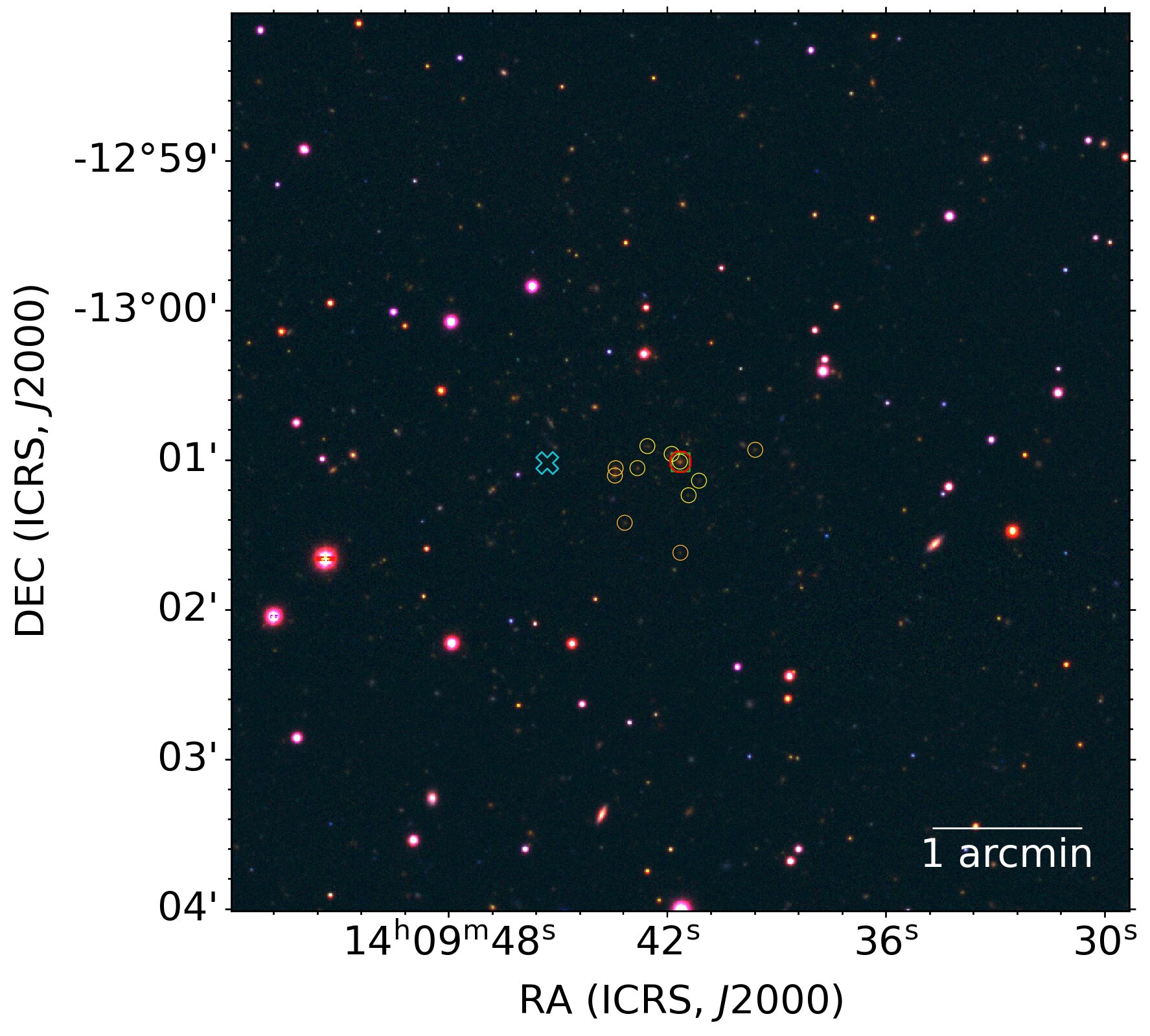}
            \caption{1eRASS J140945.2-130101, robust confirmation}
            \label{subfig:J1409}
        \end{subfigure}
        \begin{subfigure}{0.4\textwidth}
            \includegraphics[width=\linewidth]{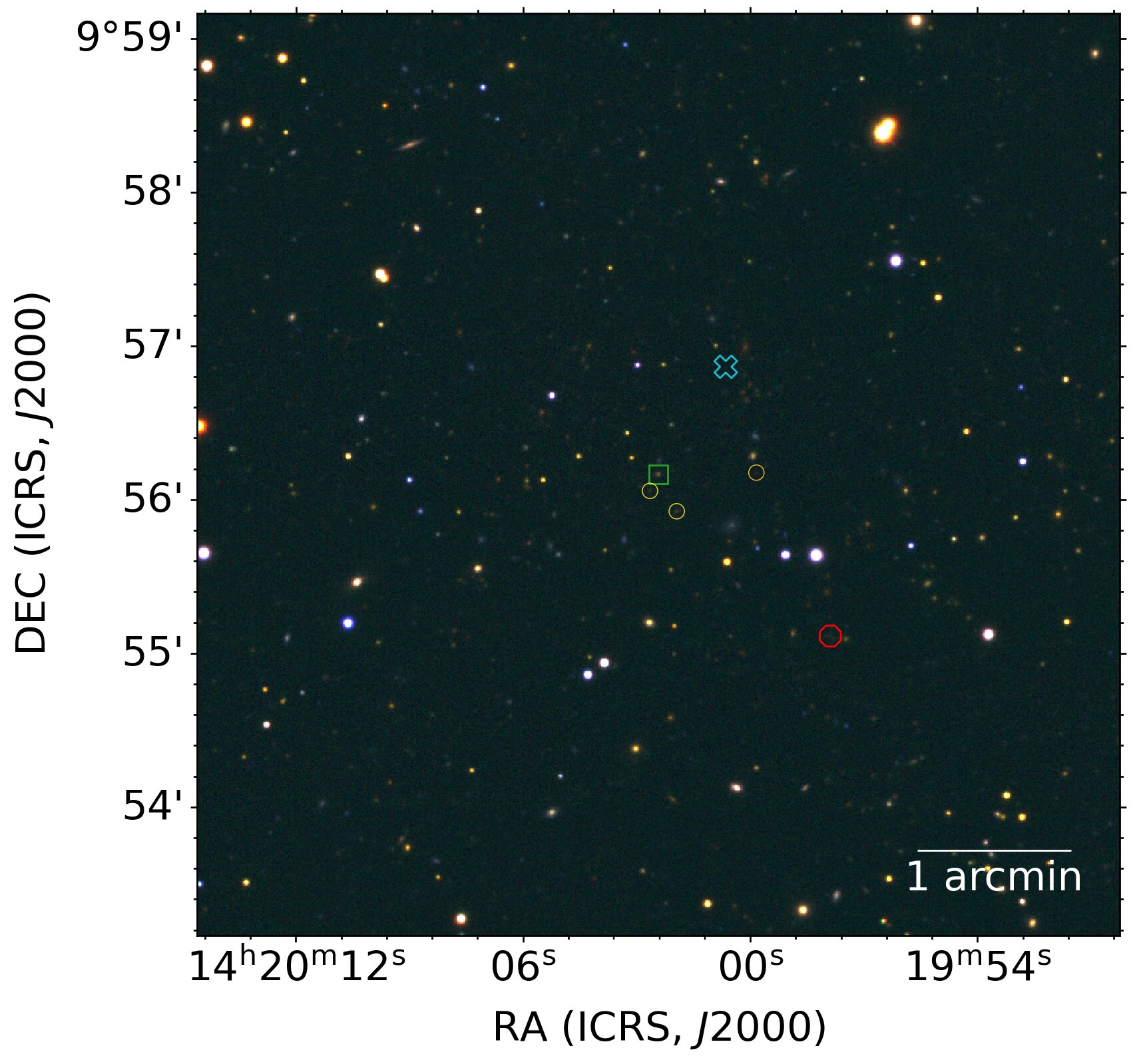}
            \caption{1eRASS J142000.6+095651, tentative}
            \label{subfig:J1420}
        \end{subfigure}

        \vspace{.1cm} 
        \begin{subfigure}{0.4\textwidth}
            \includegraphics[width=\linewidth]{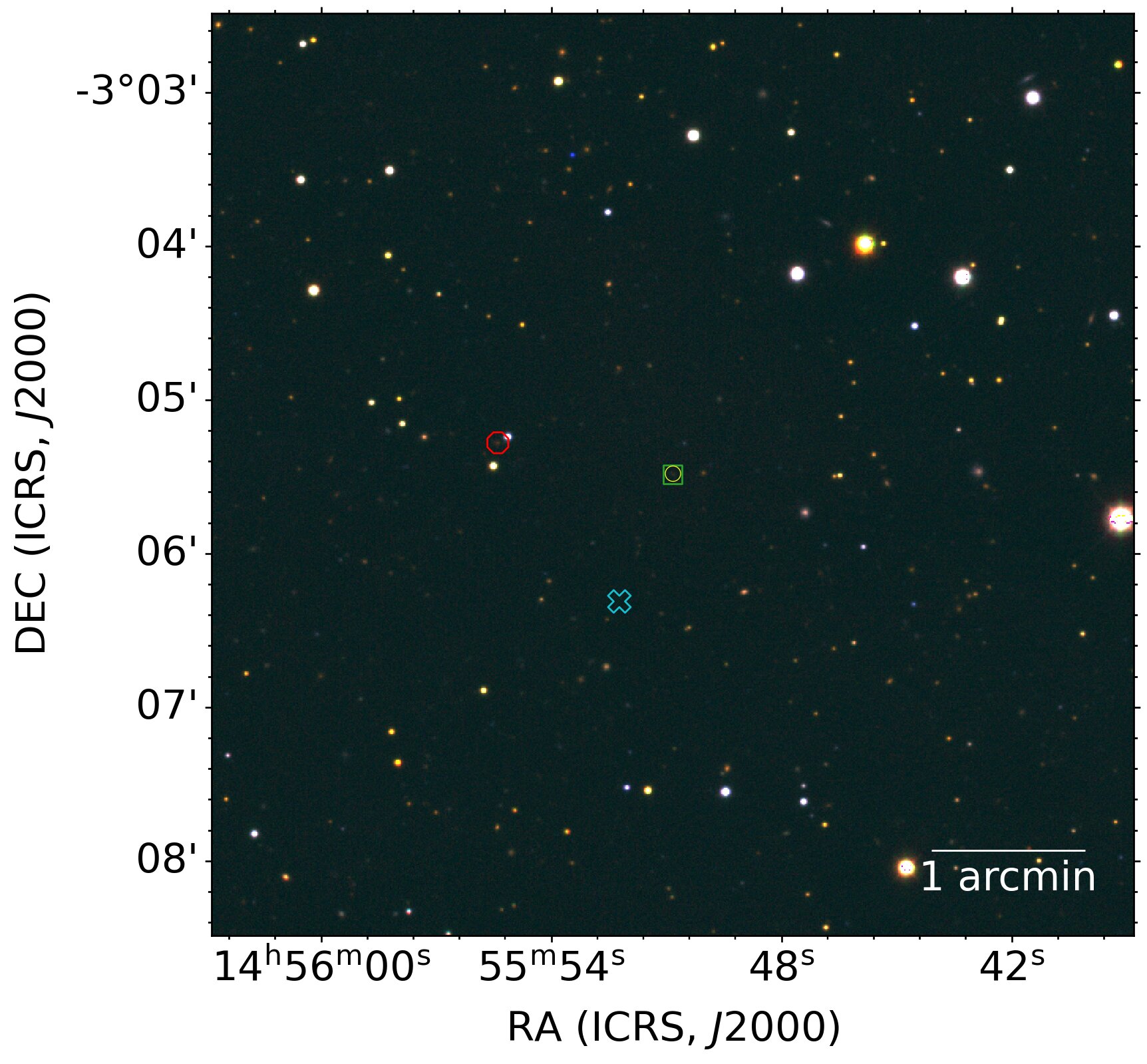}
            \caption{1eRASS J145552.2-030618, poor}
            \label{subfig:J1455}
        \end{subfigure}
        \begin{subfigure}{0.4\textwidth}
            \includegraphics[width=\linewidth]{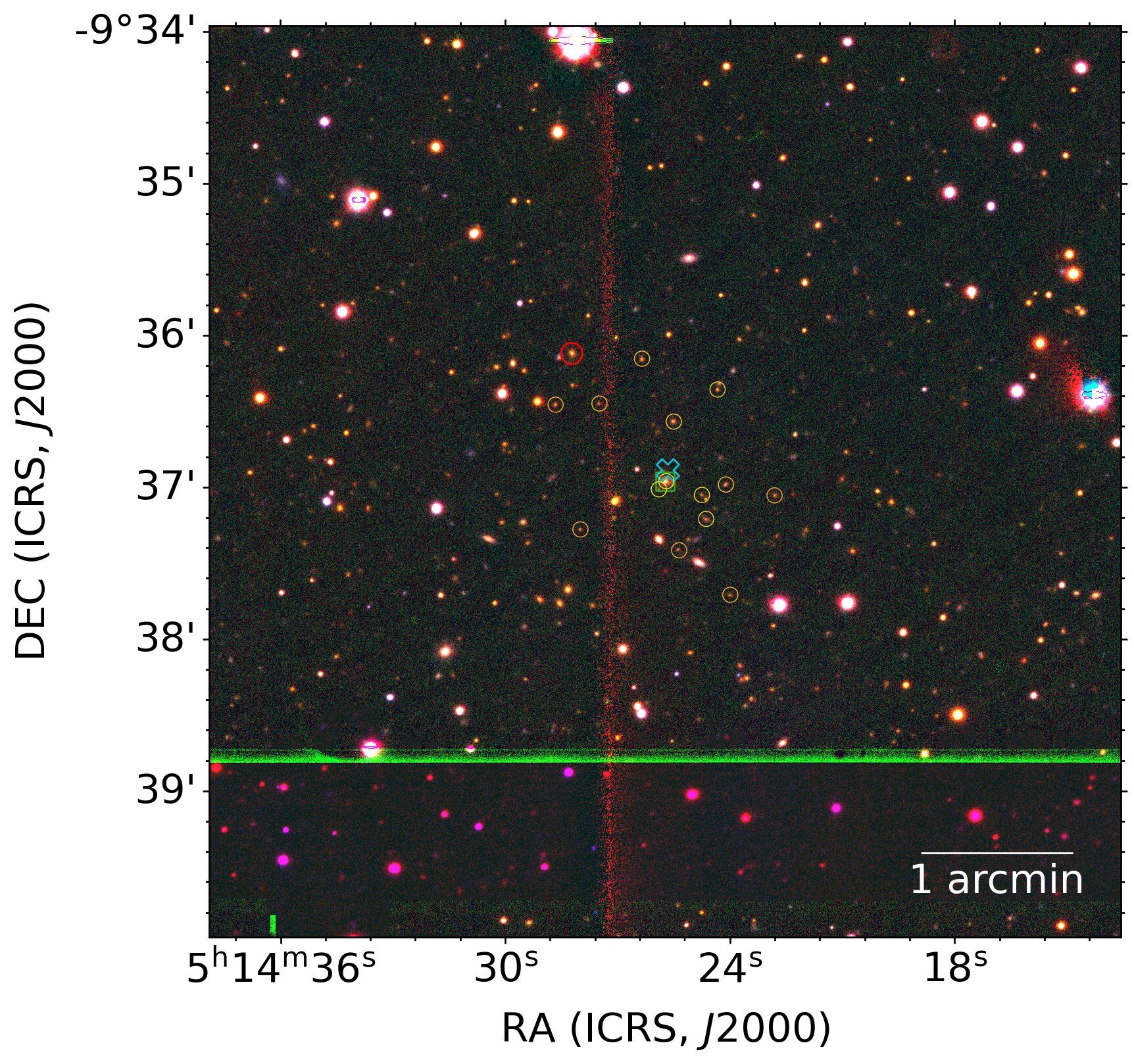}
            \caption{1eRASS J051425.6-093653, tentative.}
            \label{subfig:J0514}
        \end{subfigure}

    \caption{Fake rgb images (channels: r:$J$, g:$z$, b:$g$; $J$-band from WST/3KK (bottom right) and from CAHA/Omega2000 (rest), $z$- and $g$-band from the Legacy Survey) of the selected galaxy cluster candidates. We show the optical centre retrieved by \citet{Kluge2024} (green square), the X-ray centre (cyan cross), and a BCG candidate found in \citet{Kluge2024} (red octagon). We also include the classification of the candidate introduced in Sect. \ref{subsec:VisualInspection} in the caption, as well as the sources found to have a cluster member probability higher than $80\%$ (yellow circles).}
    \label{fig:RGBCutouts3}
    \end{figure*}

\begin{figure*}[htp] 
        \centering
        
        \vspace{.1cm} 
        \begin{subfigure}{0.4\textwidth}
            \includegraphics[width=\linewidth]{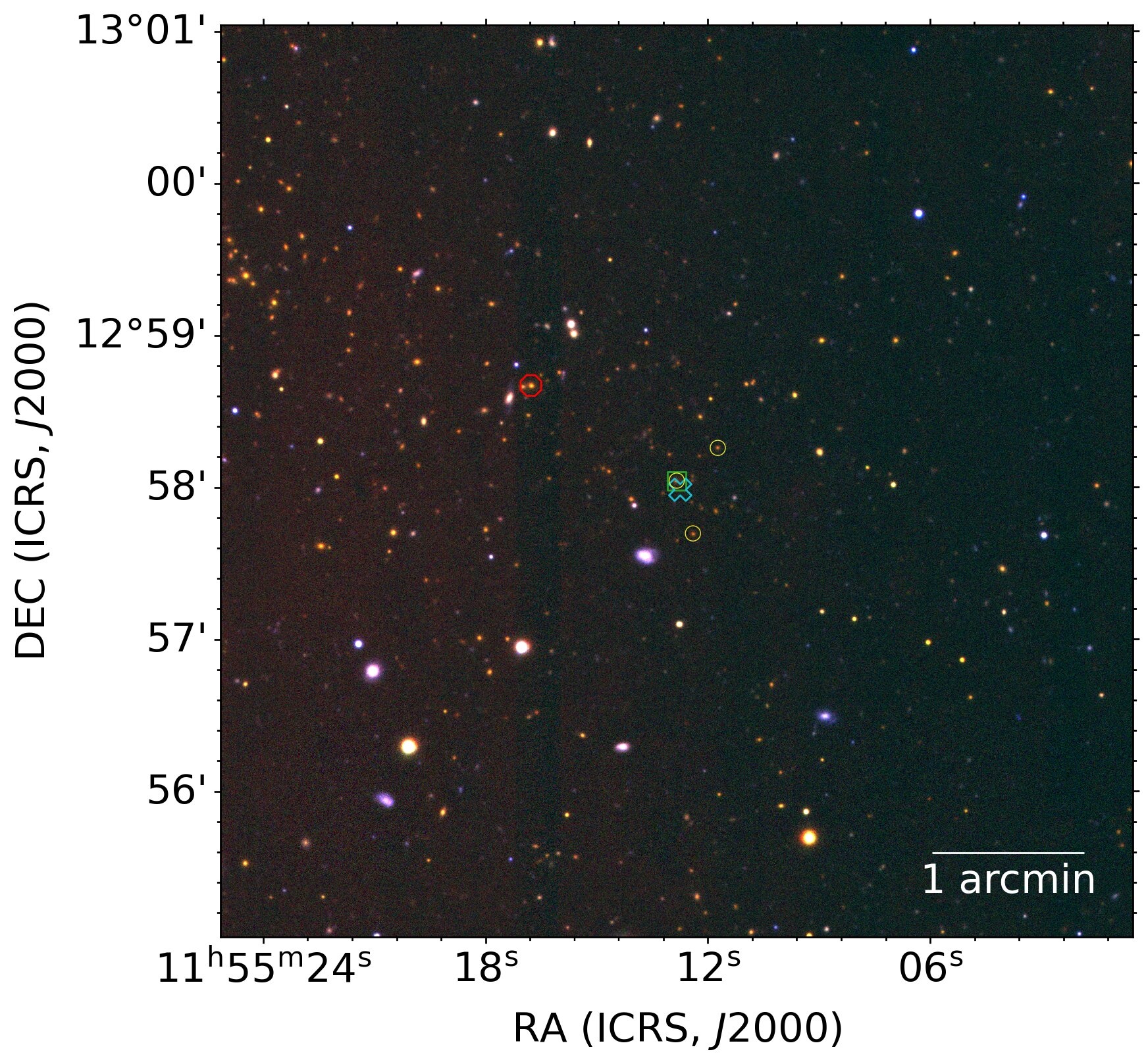}
            \caption{1eRASS J115512.7+125759, robust confirmation}
            \label{subfig:J1155}
        \end{subfigure}
        \begin{subfigure}{0.4\textwidth}
            \includegraphics[width=\linewidth]{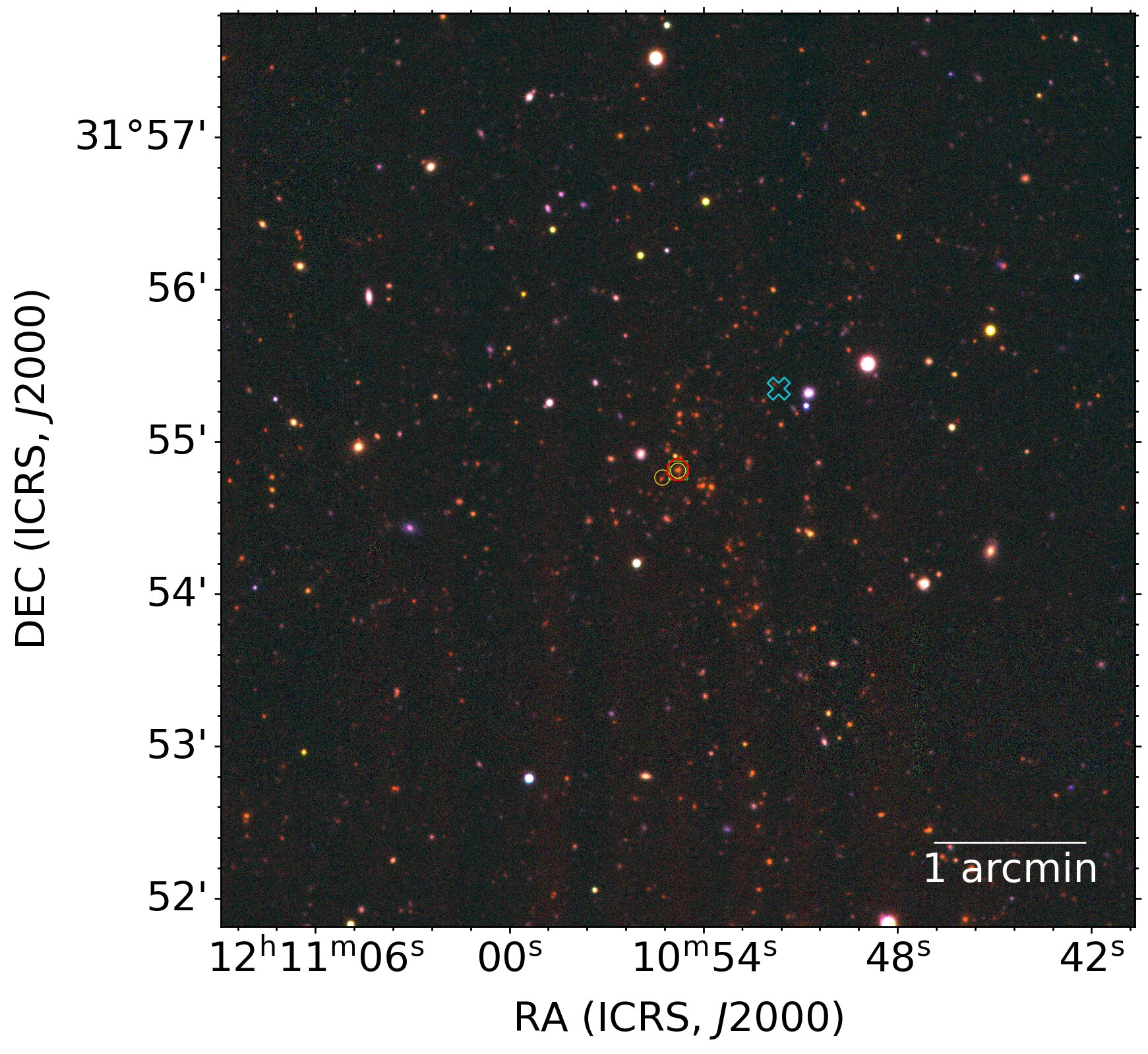}
            \caption{1eRASS J121051.6+315520, robust confirmation}
            \label{subfig:J1210}
        \end{subfigure}

        \vspace{.1cm} 
        \begin{subfigure}{0.4\textwidth}
            \includegraphics[width=\linewidth]{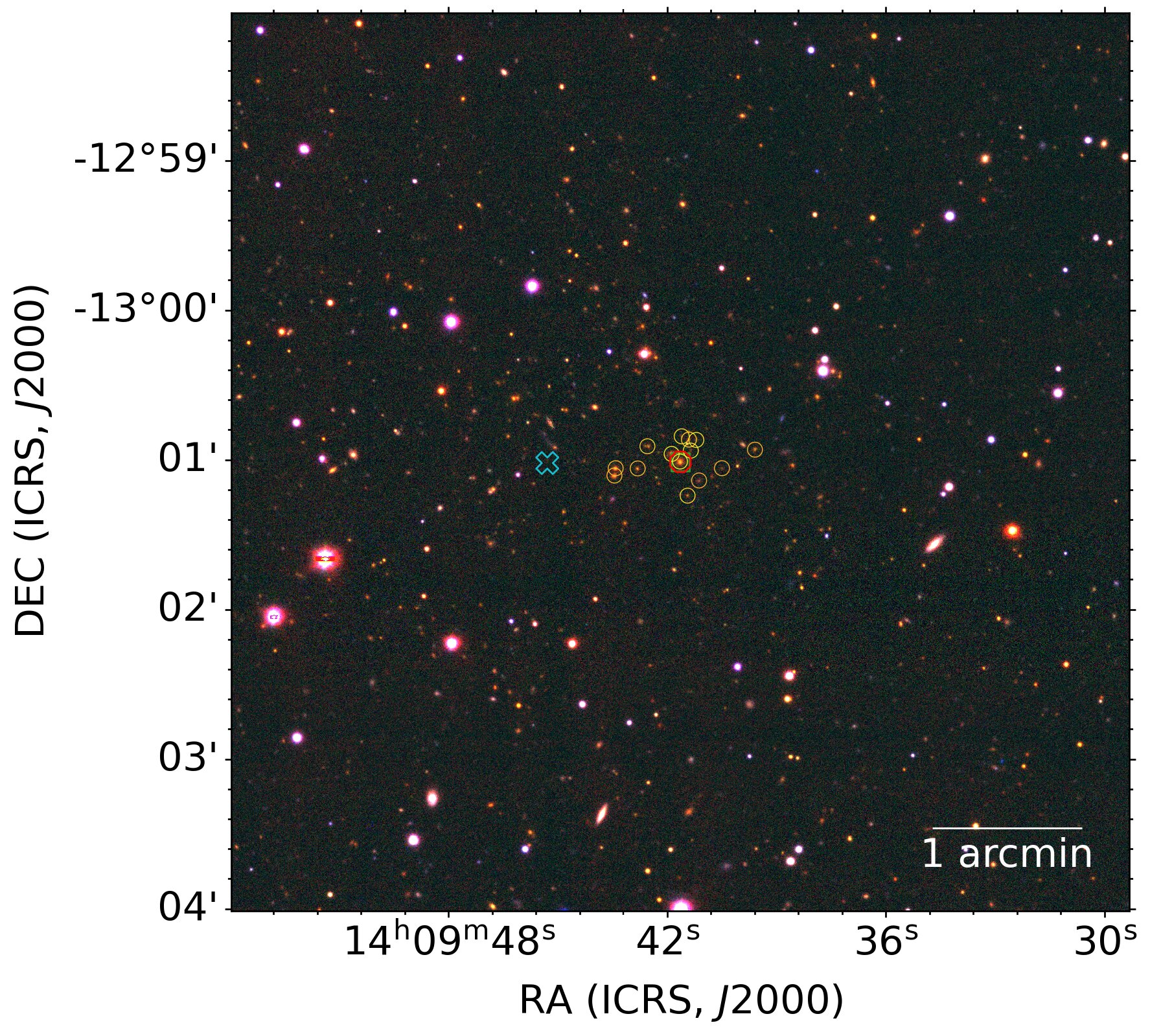}
            \caption{1eRASS J140945.2-130101, robust confirmation}
            \label{subfig:J1409_WST}
        \end{subfigure}

    \caption{Fake rgb images (channels: r:$J$, g:$z$, b:$g$; $J$-band from WST/3KK, $z$- and $g$-band from the Legacy Survey) of the selected galaxy cluster candidates. We show the optical centre retrieved by \citet{Kluge2024} (green square), the X-ray centre (cyan cross), and a BCG candidate found in \citet{Kluge2024} (red octagon). We also include the classification of the candidate introduced in Sect. \ref{subsec:VisualInspection} in the caption, as well as the sources found to have a cluster member probability higher than $80\%$ (yellow circles).}
    \label{fig:RGBCutouts4}
    \end{figure*}

        Here we summarise the results we obtain when analysing the targets with respect to the X-ray centre and $i-J$ colour (Table \ref{Tab:iJXRAYResults}, the optical and X-ray centre for $i-z$ colour (Table \ref{Tab:izOPTResults} and \ref{Tab:izXRAYResults}), and the optical and X-ray centre for the $z-J$ colour (Table \ref{Tab:zJOPTResults} and \ref{Tab:zJXRAYResults}).
        Furthermore, we show false-colour image cutouts of the selected cluster candidates around the optical centres in Fig. \ref{fig:RGBCutouts1}, \ref{fig:RGBCutouts2}, \ref{fig:RGBCutouts3}, and \ref{fig:RGBCutouts4}.
        We include the class of the respective target as has been assigned in the visual inspection in Sect.\ref{subsec:VisualInspection} in the caption of each individual image.

    \begin{table*}
        \caption{
                    Results for finding a radial colour over-density for the $i - J$ colour index around the X-ray centre.
                }
        \label{Tab:iJXRAYResults}
        \centering
        \renewcommand{\arraystretch}{1.2}
            \resizebox{0.9\linewidth}{!}{
                \begin{tabular}{l>{\raggedleft\arraybackslash}p{1.5cm}>{\raggedleft\arraybackslash}p{1.5cm}>{\raggedleft\arraybackslash}p{1.5cm}>{\raggedleft\arraybackslash}p{1.5cm}>{\raggedleft\arraybackslash}p{1.5cm}>{\raggedleft\arraybackslash}p{0.5cm}>{\raggedleft\arraybackslash}p{2.5cm}}
                    \hline
                    \noalign{\smallskip}
                    CLUSTER ID & $\mu$ & $\sigma$ & $A$ & SIGN$_\mathrm{MCMC}$ & $z_{\mathrm{phot}, i-J}$ & $n_{\mathrm{clm}}^{0.8}$ & BEST\_Z \\
                    & [mag$_{\mathrm{AB}}$] &  &  &  &  &  &  \\
                    \hline
                    \multicolumn{8}{c}{CAHA/Omega2000}\\
                    \hline
                    \noalign{\smallskip}
                    1eRASS $J$025044.2$-$044309 & $<2.83$ & $<0.17$ & $<0.11$ & 2.63 & - & 1 & $1.02\pm0.02$ \\
                    1eRASS $J$033041.6$-$053608 & $1.88_{-0.53}^{+0.53}$ & $<0.18$ & $0.10_{-0.09}^{+0.76}$ & 3.16 & - & 2 & $1.00\pm0.02$ \\
                    1eRASS $J$042710.8$-$155324 & $1.72_{-0.24}^{+0.24}$ & $<0.18$ & $0.32_{-0.27}^{+2.00}$ & 3.23 & $1.09_{-0.11}^{+0.06}$ & 1 & $1.18\pm0.03$ \\
                    1eRASS $J$043901.1$-$022852 & $<2.84$ & $<0.17$ & $<0.70$ & 2.84 & - & 2 & $1.07\pm0.03$ \\
                    1eRASS $J$053425.5$-$183444 & $<2.86$ & $<0.17$ & $<0.32$ & 2.47 & - & 1 & $1.00\pm0.03$ \\
                    1eRASS $J$085742.5$-$054517 & $<2.89$ & $<0.17$ & $<0.16$ & 1.66 & - & 2 & $1.17\pm0.02$ \\
                    1eRASS $J$091715.4$-$102343 & $1.37_{-0.07}^{+0.07}$ & $>0.10$ & $2.84_{-1.25}^{+2.22}$ & 5.11 & $0.93_{-0.03}^{+0.03}$ & 6 & $0.96\pm0.02$ \\
                    1eRASS $J$094854.3$+$133740 & $1.73_{-0.05}^{+0.05}$ & $0.03_{-0.02}^{+0.05}$ & $0.68_{-0.41}^{+1.07}$ & 3.85 & $1.10_{-0.03}^{+0.03}$ & 4 & $1.08\pm0.03$ \\
                    1eRASS $J$102148.4$+$225133 & $<2.88$ & $<0.17$ & $<0.29$ & 2.03 & - & 1 & $1.12\pm0.03$ \\
                    1eRASS $J$113751.5$+$072839 & $1.45_{-0.03}^{+0.03}$ & $0.15_{-0.02}^{+0.02}$ & $2.76_{-0.59}^{+0.74}$ & 8.78 & $0.96_{-0.01}^{+0.01}$ & 26 & $0.94\pm0.02$ \\
                    1eRASS $J$124634.6$+$252236 & $1.16_{-0.05}^{+0.05}$ & $0.13_{-0.03}^{+0.04}$ & $4.96_{-1.87}^{+2.99}$ & 4.49 & $0.84_{-0.02}^{+0.02}$ & 0 & $1.07\pm0.02$ \\
                    1eRASS $J$133333.8$+$062920 & $<2.87$ & $<0.17$ & $<0.24$ & 1.81 & - & 2 & $1.14\pm0.02$ \\
                    1eRASS $J$133801.7$+$175957 & $<2.89$ & $<0.17$ & $<0.25$ & 2.66 & - & 2 & $1.21\pm0.03$ \\
                    1eRASS $J$140945.2$-$130101 & $1.20_{-0.18}^{+0.18}$ & $>0.01$ & $>0.00$ & 3.34 & $0.86_{-0.08}^{+0.07}$ & 0 & $0.99\pm0.02$ \\
                    1eRASS $J$142000.6$+$095651 & $<2.84$ & $<0.18$ & $<0.46$ & 2.83 & - & 1 & $1.19\pm0.02$ \\
                    1eRASS $J$145552.2$-$030618 & $<2.87$ & $<0.17$ & $<0.30$ & 2.03 & - & 2 & $1.19\pm0.05$ \\
                    \noalign{\smallskip}
                    \hline
                    &   &   & WST/3KK &   &   &   &   \\
                    \hline
                    \noalign{\smallskip}
                    1eRASS $J$051425.6$-$093653 & $<2.67$ & $>0.02$ & $>0.00$ & 4.98 & - & 11 & $0.72\pm0.02$ \\
                    1eRASS $J$115512.7$+$125759 & $1.55_{-0.03}^{+0.03}$ & $<0.12$ & $0.42_{-0.24}^{+0.58}$ & 4.22 & $1.00_{-0.01}^{+0.01}$ & 3 & $1.02\pm0.02$ \\
                    1eRASS $J$140945.2$-$130101 & $1.33_{-0.05}^{+0.05}$ & $0.11_{-0.04}^{+0.07}$ & $1.66_{-0.78}^{+1.46}$ & 4.06 & $0.92_{-0.02}^{+0.02}$ & 1 & $0.99\pm0.02$ \\
                    \noalign{\smallskip}            
                    \hline
                    \noalign{\smallskip}
                \end{tabular}
            }
    \end{table*}

    \begin{table*}
        \caption{
                    Results for finding a radial colour over-density for the $i - z$ colour index around the optical centre.
                }
        \label{Tab:izOPTResults}
        \centering
        \renewcommand{\arraystretch}{1.2}
            \resizebox{0.9\linewidth}{!}{
                \begin{tabular}{l>{\raggedleft\arraybackslash}p{1.5cm}>{\raggedleft\arraybackslash}p{1.5cm}>{\raggedleft\arraybackslash}p{1.5cm}>{\raggedleft\arraybackslash}p{1.5cm}>{\raggedleft\arraybackslash}p{1.5cm}>{\raggedleft\arraybackslash}p{0.5cm}>{\raggedleft\arraybackslash}p{2.5cm}}
                    \hline
                    \noalign{\smallskip}
                    CLUSTER ID & $\mu$ & $\sigma$ & $A$ & SIGN$_\mathrm{MCMC}$ & $z_{\mathrm{phot}, i-J}$ & $n_{\mathrm{clm}}^{0.8}$ & BEST\_Z \\
                    & [mag$_{\mathrm{AB}}$] &  &  &  &  &  &  \\
                    \hline
                    \multicolumn{8}{c}{CAHA/Omega2000}\\
                    \hline
                    \noalign{\smallskip}
                    1eRASS $J$025044.2$-$044309 & $0.92_{-0.01}^{+0.01}$ & $<0.11$ & $0.06_{-0.03}^{+0.06}$ & 4.63 & $1.01_{-0.01}^{+0.01}$ & 5 & $1.02\pm0.02$ \\
                    1eRASS $J$033041.6$-$053608 & $1.19_{-0.16}^{+0.16}$ & $<0.19$ & $0.08_{-0.07}^{+0.33}$ & 3.11 & - & 2 & $1.00\pm0.02$ \\
                    1eRASS $J$042710.8$-$155324 & $0.92_{-0.19}^{+0.19}$ & $0.08_{-0.04}^{+0.10}$ & $>0.00$ & 3.18 & $1.01_{-0.12}^{+0.14}$ & 2 & $1.18\pm0.03$ \\
                    1eRASS $J$043901.1$-$022852 & $0.77_{-0.11}^{+0.11}$ & $0.05_{-0.03}^{+0.07}$ & $0.39_{-0.36}^{+2.70}$ & 3.17 & $0.91_{-0.06}^{+0.07}$ & 3 & $1.07\pm0.03$ \\
                    1eRASS $J$053425.5$-$183444 & $0.82_{-0.09}^{+0.09}$ & $>0.02$ & $1.20_{-0.80}^{+2.42}$ & 3.76 & $0.93_{-0.05}^{+0.06}$ & 4 & $1.00\pm0.03$ \\
                    1eRASS $J$085742.5$-$054517 & $<2.86$ & $<0.17$ & $<0.15$ & 2.15 & - & 1 & $1.17\pm0.02$ \\
                    1eRASS $J$091715.4$-$102343 & $0.78_{-0.04}^{+0.04}$ & $>0.16$ & $4.56_{-1.14}^{+1.52}$ & 8.11 & $0.91_{-0.02}^{+0.02}$ & 23 & $0.96\pm0.02$ \\
                    1eRASS $J$094854.3$+$133740 & $0.96_{-0.06}^{+0.06}$ & $>0.08$ & $1.40_{-0.64}^{+1.17}$ & 4.59 & $1.04_{-0.05}^{+0.11}$ & 5 & $1.08\pm0.03$ \\
                    1eRASS $J$102148.4$+$225133 & $<2.85$ & $<0.17$ & $<0.25$ & 2.52 & - & 1 & $1.12\pm0.03$ \\
                    1eRASS $J$113751.5$+$072839 & $0.80_{-0.04}^{+0.04}$ & $>0.12$ & $2.71_{-0.77}^{+1.08}$ & 7.20 & $0.92_{-0.02}^{+0.02}$ & 15 & $0.94\pm0.02$ \\
                    1eRASS $J$124634.6$+$252236 & $<2.78$ & $>0.01$ & $>0.00$ & 3.07 & - & 3 & $1.07\pm0.02$ \\
                    1eRASS $J$133333.8$+$062920 & $<2.87$ & $<0.18$ & $<0.12$ & 1.56 & - & 1 & $1.14\pm0.02$ \\
                    1eRASS $J$133801.7$+$175957 & $<0.73$ & $>0.19$ & $3.82_{-1.44}^{+2.31}$ & 6.62 & <0.88 & 6 & $1.21\pm0.03$ \\
                    1eRASS $J$140945.2$-$130101 & $<0.96$ & $>0.02$ & $>0.40$ & 5.52 & <1.04 & 10 & $0.99\pm0.02$ \\
                    1eRASS $J$142000.6$+$095651 & $<2.76$ & $<0.18$ & $<0.64$ & 2.57 & - & 1 & $1.19\pm0.02$ \\
                    1eRASS $J$145552.2$-$030618 & $<2.83$ & $<0.17$ & $<0.29$ & 2.66 & - & 1 & $1.19\pm0.05$ \\
                    \noalign{\smallskip}
                    \hline
                    &   &   & WST/3KK &   &   &   &   \\
                    \hline
                    \noalign{\smallskip}
                    1eRASS $J$051425.6$-$093653 & $<2.90$ & $<0.17$ & $<0.06$ & 1.21 & - & 1 & $0.72\pm0.02$ \\
                    1eRASS $J$115512.7$+$125759 & $0.82_{-0.06}^{+0.06}$ & $0.14_{-0.04}^{+0.05}$ & $1.03_{-0.46}^{+0.83}$ & 4.38 & $0.93_{-0.03}^{+0.04}$ & 7 & $1.02\pm0.02$ \\
                    1eRASS $J$140945.2$-$130101 & $0.85_{-0.06}^{+0.06}$ & $>0.15$ & $3.61_{-1.29}^{+2.00}$ & 7.56 & $0.95_{-0.03}^{+0.05}$ & 21 & $0.99\pm0.02$ \\
                    \noalign{\smallskip}            
                    \hline
                    \noalign{\smallskip}
                \end{tabular}
            }
    \end{table*}
        
    \begin{table*}
        \caption{
                    Results for finding a radial colour over-density for the $i - z$ colour index around the X-ray centre.
                }
        \label{Tab:izXRAYResults}
        \centering
        \renewcommand{\arraystretch}{1.2}
            \resizebox{0.9\linewidth}{!}{
                \begin{tabular}{l>{\raggedleft\arraybackslash}p{1.5cm}>{\raggedleft\arraybackslash}p{1.5cm}>{\raggedleft\arraybackslash}p{1.5cm}>{\raggedleft\arraybackslash}p{1.5cm}>{\raggedleft\arraybackslash}p{1.5cm}>{\raggedleft\arraybackslash}p{0.5cm}>{\raggedleft\arraybackslash}p{2.5cm}}
                    \hline
                    \noalign{\smallskip}
                    CLUSTER ID & $\mu$ & $\sigma$ & $A$ & SIGN$_\mathrm{MCMC}$ & $z_{\mathrm{phot}, i-J}$ & $n_{\mathrm{clm}}^{0.8}$ & BEST\_Z \\
                    & [mag$_{\mathrm{AB}}$] &  &  &  &  &  &  \\
                    \hline
                    \multicolumn{8}{c}{CAHA/Omega2000}\\
                    \hline
                    \noalign{\smallskip}
                    1eRASS $J$025044.2$-$044309 & $0.93_{-0.08}^{+0.08}$ & $<0.16$ & $0.06_{-0.04}^{+0.10}$ & 4.07 & $1.01_{-0.06}^{+0.10}$ & 5 & $1.02\pm0.02$ \\
                    1eRASS $J$033041.6$-$053608 & $<2.49$ & $>0.02$ & $0.77_{-0.69}^{+2.73}$ & 3.45 & - & 4 & $1.00\pm0.02$ \\
                    1eRASS $J$042710.8$-$155324 & $<2.78$ & $<0.17$ & $0.27_{-0.26}^{+1.98}$ & 3.32 & - & 1 & $1.18\pm0.03$ \\
                    1eRASS $J$043901.1$-$022852 & $<2.83$ & $<0.17$ & $<0.43$ & 2.68 & - & 1 & $1.07\pm0.03$ \\
                    1eRASS $J$053425.5$-$183444 & $<2.85$ & $<0.17$ & $<0.59$ & 2.32 & - & 2 & $1.00\pm0.03$ \\
                    1eRASS $J$085742.5$-$054517 & $<2.87$ & $<0.17$ & $<0.11$ & 1.58 & - & 0 & $1.17\pm0.02$ \\
                    1eRASS $J$091715.4$-$102343 & $0.79_{-0.04}^{+0.04}$ & $>0.15$ & $4.56_{-1.31}^{+1.84}$ & 7.45 & $0.92_{-0.02}^{+0.02}$ & 18 & $0.96\pm0.02$ \\
                    1eRASS $J$094854.3$+$133740 & $0.96_{-0.07}^{+0.07}$ & $>0.05$ & $1.30_{-0.68}^{+1.42}$ & 4.07 & $1.04_{-0.06}^{+0.11}$ & 3 & $1.08\pm0.03$ \\
                    1eRASS $J$102148.4$+$225133 & $<2.84$ & $<0.17$ & $<0.23$ & 2.55 & - & 1 & $1.12\pm0.03$ \\
                    1eRASS $J$113751.5$+$072839 & $0.83_{-0.03}^{+0.03}$ & $0.14_{-0.02}^{+0.03}$ & $1.97_{-0.46}^{+0.60}$ & 7.73 & $0.94_{-0.01}^{+0.01}$ & 21 & $0.94\pm0.02$ \\
                    1eRASS $J$124634.6$+$252236 & $<2.82$ & $<0.17$ & $<0.80$ & 2.15 & - & 0 & $1.07\pm0.02$ \\
                    1eRASS $J$133333.8$+$062920 & $<2.87$ & $<0.17$ & $<0.13$ & 1.02 & - & 0 & $1.14\pm0.02$ \\
                    1eRASS $J$133801.7$+$175957 & $<2.83$ & $<0.17$ & $<0.28$ & 2.63 & - & 1 & $1.21\pm0.03$ \\
                    1eRASS $J$140945.2$-$130101 & $<0.96$ & $>0.05$ & $>0.08$ & 4.78 & <1.04 & 12 & $0.99\pm0.02$ \\
                    1eRASS $J$142000.6$+$095651 & $<2.85$ & $<0.18$ & $<0.19$ & 2.10 & - & 0 & $1.19\pm0.02$ \\
                    1eRASS $J$145552.2$-$030618 & $<2.86$ & $<0.17$ & $<0.26$ & 2.04 & - & 1 & $1.19\pm0.05$ \\
                    \noalign{\smallskip}
                    \hline
                    &   &   & WST/3KK &   &   &   &   \\
                    \hline
                    \noalign{\smallskip}
                    1eRASS $J$051425.6$-$093653 & $<2.92$ & $<0.17$ & $<0.08$ & 1.78 & - & 1 & $0.72\pm0.02$ \\
                    1eRASS $J$115512.7$+$125759 & $0.82_{-0.06}^{+0.06}$ & $0.13_{-0.04}^{+0.05}$ & $0.92_{-0.40}^{+0.72}$ & 4.20 & $0.93_{-0.03}^{+0.04}$ & 6 & $1.02\pm0.02$ \\
                    1eRASS $J$140945.2$-$130101 & $0.74_{-0.06}^{+0.06}$ & $>0.17$ & $5.09_{-1.68}^{+2.52}$ & 5.78 & $0.89_{-0.03}^{+0.03}$ & 12 & $0.99\pm0.02$ \\
                    \noalign{\smallskip}            
                    \hline
                    \noalign{\smallskip}
                \end{tabular}
            }
    \end{table*}
    
    \begin{table*}
        \caption{
                    Results for finding a radial colour over-density for the $z - J$ colour index around the optical centre.
                }
        \label{Tab:zJOPTResults}
        \centering
        \renewcommand{\arraystretch}{1.2}
            \resizebox{0.9\linewidth}{!}{
                \begin{tabular}{l>{\raggedleft\arraybackslash}p{1.5cm}>{\raggedleft\arraybackslash}p{1.5cm}>{\raggedleft\arraybackslash}p{1.5cm}>{\raggedleft\arraybackslash}p{1.5cm}>{\raggedleft\arraybackslash}p{1.5cm}>{\raggedleft\arraybackslash}p{0.5cm}>{\raggedleft\arraybackslash}p{2.5cm}}
                    \hline
                    \noalign{\smallskip}
                    CLUSTER ID & $\mu$ & $\sigma$ & $A$ & SIGN$_\mathrm{MCMC}$ & $z_{\mathrm{phot}, i-J}$ & $n_{\mathrm{clm}}^{0.8}$ & BEST\_Z \\
                    & [mag$_{\mathrm{AB}}$] &  &  &  &  &  &  \\
                    \hline
                    \multicolumn{8}{c}{CAHA/Omega2000}\\
                    \hline
                    \noalign{\smallskip}
                    1eRASS $J$025044.2$-$044309 & $<2.79$ & $<0.17$ & $<0.03$ & 2.71 & - & 1 & $1.02\pm0.02$ \\
                    1eRASS $J$031350.5$-$000546 & $<2.84$ & $<0.18$ & $<0.22$ & 2.63 & - & 2 & $1.15\pm0.02$ \\
                    1eRASS $J$033041.6$-$053608 & $<2.28$ & $>0.02$ & $2.52_{-2.16}^{+6.90}$ & 4.10 & - & 6 & $1.00\pm0.02$ \\
                    1eRASS $J$042710.8$-$155324 & $<2.84$ & $<0.17$ & $<0.20$ & 2.30 & - & 1 & $1.18\pm0.03$ \\
                    1eRASS $J$043901.1$-$022852 & $<2.83$ & $<0.17$ & $<0.11$ & 2.71 & - & 1 & $1.07\pm0.03$ \\
                    1eRASS $J$053425.5$-$183444 & $<2.85$ & $<0.17$ & $<0.14$ & 2.39 & - & 2 & $1.00\pm0.03$ \\
                    1eRASS $J$085742.5$-$054517 & $<2.87$ & $<0.17$ & $<0.20$ & 0.92 & - & 0 & $1.17\pm0.02$ \\
                    1eRASS $J$091715.4$-$102343 & $<2.87$ & $<0.17$ & $<0.09$ & 2.53 & - & 1 & $0.96\pm0.02$ \\
                    1eRASS $J$094854.3$+$133740 & $<2.27$ & $<0.17$ & $0.47_{-0.42}^{+2.42}$ & 3.68 & - & 3 & $1.08\pm0.03$ \\
                    1eRASS $J$102148.4$+$225133 & $<2.90$ & $<0.17$ & $<0.07$ & 1.15 & - & 1 & $1.12\pm0.03$ \\
                    1eRASS $J$113751.5$+$072839 & $<2.84$ & $<0.17$ & $<0.09$ & 2.08 & - & 1 & $0.94\pm0.02$ \\
                    1eRASS $J$124634.6$+$252236 & $<2.88$ & $<0.17$ & $<0.13$ & 2.15 & - & 2 & $1.07\pm0.02$ \\
                    1eRASS $J$133333.8$+$062920 & $<2.87$ & $<0.17$ & $<0.09$ & 1.81 & - & 1 & $1.14\pm0.02$ \\
                    1eRASS $J$133801.7$+$175957 & $<2.86$ & $<0.17$ & $<0.14$ & 2.28 & - & 1 & $1.21\pm0.03$ \\
                    1eRASS $J$140945.2$-$130101 & $<2.88$ & $<0.17$ & $<0.07$ & 1.17 & - & 0 & $0.99\pm0.02$ \\
                    1eRASS $J$142000.6$+$095651 & $<2.87$ & $<0.17$ & $<0.11$ & 2.11 & - & 1 & $1.19\pm0.02$ \\
                    1eRASS $J$145552.2$-$030618 & $<2.86$ & $<0.17$ & $<0.28$ & 2.52 & - & 1 & $1.19\pm0.05$ \\
                    \noalign{\smallskip}
                    \hline
                    &   &   & WST/3KK &   &   &   &   \\
                    \hline
                    \noalign{\smallskip}
                    1eRASS $J$051425.6$-$093653 & $<2.88$ & $<0.17$ & $<0.09$ & 0.00 & - & 0 & $0.72\pm0.02$ \\
                    1eRASS $J$115512.7$+$125759 & $<2.87$ & $<0.18$ & $<0.19$ & 2.22 & - & 1 & $1.02\pm0.02$ \\
                    1eRASS $J$121051.6$+$315520 & $<2.85$ & $<0.17$ & $<0.62$ & 2.93 & - & 2 & $0.97\pm0.01$ \\
                    1eRASS $J$140945.2$-$130101 & $<2.87$ & $<0.18$ & $<0.08$ & 1.78 & - & 2 & $0.99\pm0.02$ \\
                    \noalign{\smallskip}            
                    \hline
                    \noalign{\smallskip}
                \end{tabular}
            }
    \end{table*}

    \begin{table*}
        \caption{
                    Results for finding a radial colour over-density for the $z - J$ colour index around the X-ray centre.
                }
        \label{Tab:zJXRAYResults}
        \centering
        \renewcommand{\arraystretch}{1.2}
            \resizebox{0.9\linewidth}{!}{
                \begin{tabular}{l>{\raggedleft\arraybackslash}p{1.5cm}>{\raggedleft\arraybackslash}p{1.5cm}>{\raggedleft\arraybackslash}p{1.5cm}>{\raggedleft\arraybackslash}p{1.5cm}>{\raggedleft\arraybackslash}p{1.5cm}>{\raggedleft\arraybackslash}p{0.5cm}>{\raggedleft\arraybackslash}p{2.5cm}}
                    \hline
                    \noalign{\smallskip}
                    CLUSTER ID & $\mu$ & $\sigma$ & $A$ & SIGN$_\mathrm{MCMC}$ & $z_{\mathrm{phot}, i-J}$ & $n_{\mathrm{clm}}^{0.8}$ & BEST\_Z \\
                    & [mag$_{\mathrm{AB}}$] &  &  &  &  &  &  \\
                    \hline
                    \multicolumn{8}{c}{CAHA/Omega2000}\\
                    \hline
                    \noalign{\smallskip}
                    1eRASS $J$025044.2$-$044309 & $<2.83$ & $<0.18$ & $<0.03$ & 2.29 & - & 1 & $1.02\pm0.02$ \\
                    1eRASS $J$031350.5$-$000546 & $<2.87$ & $<0.18$ & $<0.08$ & 0.76 & - & 0 & $1.15\pm0.02$ \\
                    1eRASS $J$033041.6$-$053608 & $<2.54$ & $>0.02$ & $>0.00$ & 4.13 & - & 8 & $1.00\pm0.02$ \\
                    1eRASS $J$042710.8$-$155324 & $<2.76$ & $<0.18$ & $<0.64$ & 2.76 & - & 1 & $1.18\pm0.03$ \\
                    1eRASS $J$043901.1$-$022852 & $<2.85$ & $<0.17$ & $<0.15$ & 2.58 & - & 1 & $1.07\pm0.03$ \\
                    1eRASS $J$053425.5$-$183444 & $<2.87$ & $<0.17$ & $<0.14$ & 2.29 & - & 1 & $1.00\pm0.03$ \\
                    1eRASS $J$085742.5$-$054517 & $<2.87$ & $<0.17$ & $<0.13$ & 1.32 & - & 0 & $1.17\pm0.02$ \\
                    1eRASS $J$091715.4$-$102343 & $<2.85$ & $<0.17$ & $<0.32$ & 2.78 & - & 2 & $0.96\pm0.02$ \\
                    1eRASS $J$094854.3$+$133740 & $<2.75$ & $<0.16$ & $>0.00$ & 3.62 & - & 4 & $1.08\pm0.03$ \\
                    1eRASS $J$102148.4$+$225133 & $<2.89$ & $<0.17$ & $<0.12$ & 1.88 & - & 1 & $1.12\pm0.03$ \\
                    1eRASS $J$113751.5$+$072839 & $<2.90$ & $<0.17$ & $<0.03$ & 0.95 & - & 0 & $0.94\pm0.02$ \\
                    1eRASS $J$124634.6$+$252236 & $<2.87$ & $<0.17$ & $<0.19$ & 1.74 & - & 0 & $1.07\pm0.02$ \\
                    1eRASS $J$133333.8$+$062920 & $<2.86$ & $<0.18$ & $<0.19$ & 2.06 & - & 1 & $1.14\pm0.02$ \\
                    1eRASS $J$133801.7$+$175957 & $<2.86$ & $<0.17$ & $<0.17$ & 1.82 & - & 1 & $1.21\pm0.03$ \\
                    1eRASS $J$140945.2$-$130101 & $2.40_{-0.55}^{+0.55}$ & $<0.17$ & $0.08_{-0.07}^{+0.74}$ & 3.05 & - & 1 & $0.99\pm0.02$ \\
                    1eRASS $J$142000.6$+$095651 & $<2.86$ & $<0.17$ & $<0.19$ & 2.43 & - & 2 & $1.19\pm0.02$ \\
                    1eRASS $J$145552.2$-$030618 & $<2.88$ & $<0.17$ & $<0.14$ & 1.51 & - & 0 & $1.19\pm0.05$ \\
                    \noalign{\smallskip}
                    \hline
                    &   &   & WST/3KK &   &   &   &   \\
                    \hline
                    \noalign{\smallskip}
                    1eRASS $J$051425.6$-$093653 & $<2.88$ & $<0.17$ & $<0.12$ & 0.40 & - & 0 & $0.72\pm0.02$ \\
                    1eRASS $J$115512.7$+$125759 & $<2.89$ & $<0.17$ & $<0.17$ & 2.02 & - & 2 & $1.02\pm0.02$ \\
                    1eRASS $J$121051.6$+$315520 & $<2.89$ & $<0.17$ & $<0.13$ & 0.65 & - & 0 & $0.97\pm0.01$ \\
                    1eRASS $J$140945.2$-$130101 & $<2.87$ & $<0.17$ & $<0.07$ & 0.94 & - & 0 & $0.99\pm0.02$ \\
                    \noalign{\smallskip}            
                    \hline
                    \noalign{\smallskip}
                \end{tabular}
            }
    \end{table*}


\end{appendix}
\end{document}